\documentclass[nofootinbib,superscriptaddress,eqsecnum,tightenlines,11pt]{revtex4-1}
\usepackage{graphicx}
\usepackage{amsmath,amssymb,amsfonts,amsthm,stmaryrd,mathtools,mathtools}
\usepackage{yfonts}
\usepackage{multirow,bigdelim}
\usepackage{dsfont}
\usepackage{color}
\usepackage{graphicx,color,epstopdf,epsfig,psfrag}
\usepackage[dvipsnames]{xcolor}
\usepackage{ mathrsfs }
\usepackage[linktocpage]{hyperref}

\hypersetup{
    colorlinks=true,       
    linkcolor=MidnightBlue,          
    citecolor=BrickRed,        
    filecolor=BrickRed,      
    urlcolor=BrickRed           
}
 
\makeatletter
\newcounter{subsubsubsection}[subsubsection]
\renewcommand\thesubsubsubsection{\thesubsubsection .\@alph\c@subsubsubsection}
\newcommand\subsubsubsection{\@startsection{subsubsubsection}{4}{\z@}%
                                     {-3.25ex\@plus -1ex \@minus -.2ex}%
                                     {1.5ex \@plus .2ex}%
                                     {\centering\normalfont\small\textit}}
\newcommand*\l@subsubsubsection{\@dottedtocline{3}{10.0em}{4.1em}}
\newcommand*{\subsubsubsectionmark}[1]{}
\makeatother

\def\tr{\text{tr}}

\def\SU{\text{SU}}

\def\nn{\nonumber}
\def\q{\qquad}

\def\wG{\widetilde{G}}
\def\wH{\widetilde{H}}
\def\mG{\mathcal{G}}
\def\mR{\mathcal{R}}

\def\uRoman#1{\uRomannumeral{\the\value{#1}}}
\def\be{\begin{equation}}
\def\ee{\end{equation}}
\def\ba{\begin{eqnarray}}
\def\ea{\end{eqnarray}}
\def\bas{\begin{subequations}\begin{eqnarray}}
\def\eas{\end{eqnarray}\end{subequations}}

\renewcommand{\S}{{\mathbb S}}
\renewcommand{\tilde}{\widetilde}

\def\D{\mathfrak{D}}
\def\d{\mathfrak{d}}

\begin{document}

\title{\LARGE Fusion basis for lattice gauge theory and\\ loop quantum gravity}

\author{Clement Delcamp}\email{cdelcampATperimeterinstituteDOTca}
\affiliation{Perimeter Institute for Theoretical Physics,\\ 31 Caroline Street North, Waterloo, Ontario, Canada N2L 2Y5}
\affiliation{Department of Physics $\&$ Astronomy and Guelph-Waterloo Physics Institute \\  University of Waterloo, Waterloo, Ontario N2L 3G1, Canada}
\author{Bianca Dittrich}\email{bdittrichATperimeterinstituteDOTca}
\affiliation{Perimeter Institute for Theoretical Physics,\\ 31 Caroline Street North, Waterloo, Ontario, Canada N2L 2Y5}
\author{Aldo Riello}\email{arielloATperimeterinstituteDOTca}
\affiliation{Perimeter Institute for Theoretical Physics,\\ 31 Caroline Street North, Waterloo, Ontario, Canada N2L 2Y5}

\begin{abstract}

We introduce a new basis for the gauge--invariant Hilbert space of lattice gauge theory and loop quantum gravity in $(2+1)$ dimensions, the fusion basis. In doing so, we shift the focus from the original lattice (or spin--network) structure directly to that of the magnetic (curvature) and electric (torsion) excitations themselves. These excitations are classified by the irreducible representations of the Drinfel'd double of the gauge group, and can be readily ``fused'' together by studying the tensor product of such representations. We will also describe in detail the ribbon operators that create and measure these excitations and make the quasi--local structure of the observable algebra explicit. 
Since the fusion basis allows for both magnetic and electric excitations from the onset, it turns out to be a precious tool for studying the large scale structure and coarse--graining flow of lattice gauge theories and loop quantum gravity. 
This is in neat contrast with the widely used spin--network basis, in which it is much more complicated to account for electric excitations,  i.e. for Gau\ss~constraint violations, emerging at larger scales.  Moreover, since the fusion basis comes equipped with a hierarchical structure, it readily provides the language to design states with sophisticated multi--scale structures.
Another way to employ this hierarchical structure is to encode a notion of subsystems for lattice gauge theories and $(2+1)$ gravity coupled to point particles. In a follow--up work, we have exploited this notion to provide a new definition of entanglement entropy for these theories.

~\\

\end{abstract}

\maketitle

\newpage

\small
\tableofcontents
\normalsize

\section{Introduction}

Yang--Mills theory and general relativity are prime examples of theories with gauge symmetries, which have become indispensable in modern physics. 
The Ashtekar formulation of canonical general relativity  \cite{AshtekarVar} brought the two theories even closer. 
Roughly speaking, this was achieved by including the group of local rotations, as an extra gauge symmetry beside  space--time diffeomorphisms. 
This allowed to incorporate lattice gauge theory techniques in the realm of background independent field theories and led to the development of loop quantum gravity \cite{LQGReviews}.

Lattice gauge theories allow for non--perturbative quantization schemes, which are needed in particular for the understanding of quantum chromodynamics as well as quantum gravity. The success of such schemes relies on a clever choice of  discrete observables \cite{WilsonG} transforming nicely under the gauge symmetries.\footnote{The issue is, however, much more involved  for the space--{\it time} diffeomorphism group \cite{BD-Diff,BD14}.}   These observables are based on holonomies, built out of the gauge connection, and on fluxes, built out of the electric field and---in non--Abelian  gauge theories---out of the connection, too.

The major drawback of gauge formulations is, however, that it still needs the identification of a complete set of mutually independent gauge invariant degrees of freedom and observables.
This is particularly important when it comes to the quantum theory.  
A gauge invariant basis for lattice field theory, allowing a convenient description of the gauge invariant Hilbert space, is the so--called spin--network basis \cite{RovSmol}. Such a basis has found wide applications in both lattice gauge theories and loop quantum gravity. In particular, it solved the problem of over--completeness of the Wilson loop observables, as encoded by the Mandelstam identities, which plagued the early developments of loop quantum gravity,  see e.g. \cite{Loll}.  

The main purpose of the present work is to introduce another basis of the gauge-invariant Hilbert space: the ``fusion basis''.
Among its desirable properties, one of the most important ones is that in this basis coarse--graining of states simplifies considerably with respect to an approach based on the spin--network basis \cite{EteraCG,Clement}. 
 This feature makes it the natural candidate to study  the large scale dynamics of loop quantum gravity in terms of coarse--graining and renormalization \cite{BD14}.

The amenability of the fusion basis to coarse--graining is due to the fact that in non--Abelian gauge theories, effective electric excitations (or torsion excitations, in a gravity context) emerge at large scales even if they are not present at the lattice scale \cite{DG14b}. 
Since these excitations are not present from the onset in the spin--network basis, one needs to devise extension of the ``standard'' framework. See \cite{EteraTags} for proposals.   
In contrast, the fusion basis improves this state of things in a twofold way: on the one hand it allows from the onset for both magnetic (curvature) and electric (torsion) excitations, and on the other it can be designed to have a notion of coarse--graining directly built in its combinatorial structure.

The fusion basis we introduce here is adapted from the theory of topological phases in $(2+1)$ dimensions.
We will therefore restrict here to $(2+1)$ dimensional lattice gauge theories and loop quantum gravity. 
For a strategy to generalize the fusion basis to $(3+1)$ dimensions, see \cite{DelDitt}. 
Furthermore, another simplification we introduce in order to focus on the main ideas without bothering about technical details, is that we will consider only a finite gauge or structure group ${\cal G}$.
We will comment on the application to Lie groups in the discussion section, \ref{discussion}.\\

Let us briefly describe and compare the main features of the spin--network and fusion basis.  The spin--network  basis diagonalizes at each link of the lattice the quadratic Casimir operator built from the electric fluxes. 
These operators are gauge invariant and coincide with the electric contribution to the Yang--Mills Hamiltonian. 
For a non--Abelian structure group, 
additional, gauge-invariant, information on the electric fluxes is encoded at the nodes, in so--called intertwiners.
Thus,
 the spin--network basis provides a polarization of the state space based on the flux observables.%
\footnote{ Flux observables do actually \textit{not} commute in non--Abelian gauge theories. In \cite{BaratinEtAl}, this is made explicit and a polarization is constructed, in which fluxes compose by non--commutative multiplication.}  

The fusion basis, on the other hand, diagonalizes Wilson loop operators, i.e. traces of holonomies associated to closed paths. 
In this sense, the fusion basis provides a polarization dual to the spin--network one.
To avoid over--parametrization, one does, however, not include all possible Wilson loops supported on the lattice, but only a certain hierarchically ordered set. 

A crucial feature of non--Abelian gauge theories is that this set of Wilson loops does {\it not} define a maximal set of commuting observables. In fact, it is necessary to also consider certain flux observables, based again on closed loops, that capture the electric (or torsion) degrees of freedom arising at scales larger than the lattice one.\footnote{We already mentioned this effect when we discussed coarse--graining.}
Maybe surprisingly, these large scale data are not already encoded in the multilevel Wilson loop observables.
The fusion basis is designed to encode both Wilson loop and large scale flux observables in a unified framework.

In fact, it turns out that the fusion basis diagonalizes closed ``ribbon'' operators, which directly classify the magnetic (curvature) and electric (torsion) excitations. 
This notion of excitation has to be understood with respect to some vacuum state.
Here, the relevant one is the so--called BF vacuum.
Taking its name from the BF topological field theory, of which it is a physical state, this vacuum state is a gauge invariant state peaked sharply on vanishing curvature, i.e. on a flat connection. 
It is then not surprising, that the fusion basis framework bares a close relationship with the theory of extended topological quantum field theories on the mathematical side, and with topological phases and their defect excitations on the condensed matter side. 
In particular, BF theory can be described by so-called extended\footnote{Here the attribute `extended' describes the addition of non-gauge-invariant group-representation-space indices,  encoding the choice of local reference frame, which are not present in the `pure' string net models.} string net models \cite{LevinWen, Buersch}.  

The classification of the excitations comes with an interesting mathematical structure, the Drinfel'd double ${\cal D}(\mG)$ of the gauge group ${\cal G}$. For this reason we will introduce and review various basic facts about the Drinfel'd double and its representation theory. This is in fact the fundamental mathematical tool behind the definition of the fusion basis, since the irreducible representations of ${\cal D}(\mG)$ characterize  the excitations of the model, while their tensor product describes their ``fusion''. 

After having introduced the fusion basis and the ribbon operators characterizing it, we will give an overview of various applications.
Firstly, we will discuss how to use the fusion basis to easily design multi--scale states.
It is interesting to compare the tools developed here to the closely related philosophy underlying the introduction of tensor network states, which provide an Ansatz for the ground state of Yang--Mills theories \cite{Ashley}.\footnote{In the context of $(2+1)$dimensional gravity, on the other hand, the BF vacuum already provides the physical state of the theory, i.e. the state invariant under full diffeomorphism symmetry. 
Fusion basis states, then, encode multi--particle states coupled to gravity. Therefore, this basis could be a useful tool to understand their coupled dynamics.}
Secondly, using the muli--scale states, we describe a coarse--graining scheme based on the fusion basis.
 At this point of the discussion, the advantage in using the fusion basis should be obvious: coarse--graining is directly given by the fusion of excitations, which are in turn naturally encoded in the fusion basis itself. 

In a follow up work, we plan to discuss entanglement entropy in non--Abelian lattice gauge theories, a topic which recently attracted increased attention \cite{GaugeEnt}. Specifically, we will make use of the fusion basis to provide a new definition of entanglement entropy for such theories, and to compute it for a certain family of states. \\

Let us conclude this introduction with a note.
In this paper, we will rely on a lot of previous work coming both from the context of topological phases with defects and from extended topological field theories. This material will be translated and adapted to our purposes. The reformulation of lattice gauge theory and loop quantum gravity in terms of extended topological field theory is parallel to \cite{Kir,BalKir,DG16}. The fusion basis has been constructed already for string--net models \cite{LevinWen}. An explicit definition for $\SU(2)_q$, at  $q$ root of unity, was given in \cite{KKR}, which is easily generalizable to modular fusion categories (see also \cite{DG16}). The fusion basis  for more general fusion categories appeared---albeit only implicitly---in \cite{Hu}.

Our characterization of basic excitations is adapted from arguments\footnote{See \cite{Kong} for alternative derivations} in \cite{Ocneanu, Lan} which were developed  also in the context of string--net models. Here, although we make use of the same idea of gluing states, we rephrase it in a context more amenable for applications to lattice gauge theory and loop quantum gravity. For this reason, we develop our arguments for the BF representation \cite{DG14a,BDG15} and work in the holonomy polarization. This will considerably facilitate the interpretation of the excitations generated and measured by ribbon operators in terms of standard gauge theory and loop quantum gravity observables. It will also help us to provide an interpretation of the corresponding operators defined for the Turaev--Viro based representations \cite{DG16}, where the operators are constructed via more abstract arguments within the flux (spin) polarization.

The Drinfel'd double of (finite) groups and their representations have been discussed in \cite{Verlinde,BaisReview}.  Ribbon operators were introduced by Kitaev \cite{Kitaev1} and studied in great detail by Bombin \textit{et al.} \cite{Bombin} in the context of a lattice gauge theory model. Our discussion, however, will rather be based on a lattice-\textit{independent} description of the ribbon operators. While we believe that this can be fruitful in the study of Yang--Mills theory, it is definitely necessary for application to background independent theories, such as loop quantum gravity.
~\\

This paper is organized as follows. In section \ref{sec_BF} we formulate the BF representation in $(2+1)$D. This provides also an interpretation of lattice gauge theories as topological field theories with defects, making the fusion basis available for these cases.
Then, we give the main argument for the appearance of the Drinfel'd double $\mathcal{D}(\mathcal{G})$, in section \ref{sec_drinfeld};
in this section we also review the representation theory of the Drinfel'd double, and fix the relevant notations for the rest of the paper.
Section \ref{sec_fusionbasis} is the core of the paper, where the fusion basis  is introduced.
In section \ref{sec_ribbons}, we introduce the open ribbon operators that generate the fusion basis by acting on the BF vacuum, as well as closed ribbon operators that project onto the fusion basis states.
 Finally, in section \ref{sec_coarsegraining}, we discuss possible applications of the fusion basis, firstly for the multi-scale design of states, and secondly for coarse--graining. 
The paper has also a number of appendices, where technical calculations are relegated and further details on the BF representation are provided.

\section{BF representation in $(2+1)$ dimension\label{sec_BF}}

In lattice gauge theories, observables can be given in terms of holonomies (or Wilson lines), encoding the magnetic degrees of freedom, and fluxes, encoding the electric degrees of freedom. In $(2+1)$ dimension both holonomy and flux observables test the continuum field along a one--dimensional path embedded in the spatial manifold. On a fixed graph (or lattice) one has only  access to a restricted set of such holonomies and fluxes, that is those that can be composed from the elementary holonomies and fluxes associated to the links of the graph itself. In this way different graphs $\Gamma$ lead to different Hilbert spaces ${\cal H}_\Gamma$, hence providing a representation of the holonomies and fluxes based on $\Gamma$.

One can however consider also all possible graphs at once (or at least a suitable set of graphs allowing for infinite refinement) by constructing a so--called inductive limit of the family of Hilbert spaces $\{{\cal H}_\Gamma\}_\Gamma$. This allows for the representation of holonomies and fluxes based on arbitrary paths (or again based on a  suitable set of paths).  Such an inductive limit construction led to the Ashtekar--Lewandowski--Isham (ALI) representation \cite{ALI} of the kinematical\footnote{That is the observables are not completely space--time diffeomorphism invariant.} observable algebra in loop quantum gravity. Here the selection of a  (kinematical) vacuum state is essential, which in the case of the ALI representation is given by a state for which the expectation values vanishes for all operators composed from fluxes. This implies that the resulting Hilbert space supports states which have vanishing flux expectation values almost everywhere. As the fluxes encode the spatial metric the states describe therefore an almost everywhere degenerate geometry.

This was one of the motivations for the construction of an alternative representation based on a different--- actually dual---vacuum, sharply peaked on vanishing curvature \cite{DG14a,DG14b,BDG15}.  This vacuum is a solution of a topological field theory known as BF theory and describes in lattice gauge theory the weak coupling limit.  BF theory plays also an important role in the gravity context: it is itself a formulation of $(2+1)$ dimensional gravity, and moreover, in $(3+1)$ dimensions, it is the starting point for the construction of spin--foam models, a covariant version of loop quantum gravity \cite{AlexReview}.
A quantum deformed version \cite{DG16}, based on the Turaev--Viro topological theory\footnote{This representation is so far only applicable to $(2+1)$ dimensions, for a strategy to generalize to $(3+1)$ dimensions, see \cite{DelDitt}.} \cite{TV}, describing $(2+1)$ dimensional Euclidean gravity with positive cosmological constant, is more directly formulated as an extended topological field theory. Here the notion of defect excitations, supported in $(2+1)$ dimensions on punctures, is essential.   

In this section we will shortly explain the BF representation for loop quantum gravity and a related understanding of lattice gauge theory as an extended topological field theory. The BF representation in \cite{DG14a,DG14b,BDG15} has been based on an inductive limit involving triangulations and their dual lattices. We will review this notion and then lay out an alternative construction, similar to \cite{DG16}, which is nearer to the spirit of extended topological field theory. In the latter case the graphs or lattices have a  less fundamental role.  Instead one  uses punctures (or `defects') which carry the  excitations.  These defect excitations are to be understood as deviations from a vacuum or alternatively violations of constraints, which characterize the vacuum. This vacuum  is here given as the BF vacuum, i.e. a state without curvature (magnetic excitation) or torsion (electric excitation). 

These considerations will also allow to understand lattice gauge theory as an extended topological field theory, that is a topological field theory with a (here fixed) number of defects allowed.

\subsection{Triangulation--based BF representation: review and limitations}

The BF representation is based on a so--called inductive limit of Hilbert spaces. The inductive limit is defined via a family of Hilbert spaces labelled by elements of a partially ordered (and directed) set. Each Hilbert space of this family can be understood to capture a certain subset of the degrees of freedom of the continuum, given by the inductive limit. In this sense a given Hilbert space of this family defines also a discretization. 

In \cite{DG14a,DG14b,BDG15}, such an inductive limit was based on the refinement of triangulations of a given $2D$ hypersurface $\Sigma$.
Specifically, given $\Sigma$ and a triangulation $\Delta$ thereof, the configuration space underlying the Hilbert space ${\cal H}_\Delta$ is given by the moduli space of flat connection on $\Sigma\setminus\Delta_0$, that is ${\cal M}^\text{flat}(\Sigma\setminus\Delta_0)$.
Here $\Delta_0$ is the set of $0$-simplices, i.e. vertices, of the triangulation.

As is well known, ${\cal M}^\text{flat}(\Sigma\setminus\Delta_0)$ can be fully described by considering the set of holonomies\footnote{We use the word `holonomy' for the group--valued path--ordered exponential of a connection along a path between two points on the manifold. It transforms covariantly upon gauge transformations at its starting and ending points, and it is  invariant upon any other gauge transformation. } along the links of a graph $\Gamma$ dual to the triangulation.
Clearly, the flatness conditions ensures that the specific choice of dual graph is irrelevant.
Then, ${\cal H}_\Delta$ is given by the gauge invariant functions of such holonomies, equipped with a specific inner product.
For a well--defined inductive limit, the measure on the underlying gauge group $\mathcal G$ has to be discrete, even if $\cal G$ is a Lie group \cite{DG14a,BDG15}.

It is often convenient to choose a marked point on the manifold, the `root', at which gauge invariance is relaxed.
Fully gauge-invariant functions can be re--obtained via a gauge averaging procedure.
The advantage of having a root is clear if $\cal G$ is a Lie group: the gauge averaging procedure over $\cal G$ equipped with a discrete measure would in general  lead to many subtleties \cite{BDG15}.
Physically, the root can be interpreted as a reference frame internal to the system. 

So far we have described the structure of the Hilbert space ${\cal H}_\Delta$ on a fixed triangulation.
What is missing is the inductive limit construction of the continuum Hilbert space ${\cal H}_\Sigma$.
Consider two triangulations $\Delta$ and $\Delta'$, such that $\Delta'$ is a refininement of $\Delta$, i.e. $\Delta \prec \Delta'$.
Then, the inductive limit is based on the definition of embedding maps
\be
\iota_{\Delta,\Delta'}:\; \mathcal{H}_\Delta\; \hookrightarrow \;{\cal H}_{\Delta'}\,.
\ee
Roughly speaking, in the BF representation, the embedding maps multiply the states in ${\cal H}_\Delta$ with a set of delta--functions---hence the relevance of the discrete  measure on the group---enforcing the triviality of the holonomy around every additional cycle present in $\Gamma'$ but not in $\Gamma$.
Notice that there is one such cycle for every element of $\Delta'_0\setminus\Delta_0$.
This defines ${\cal H}_\Sigma$. 
However, we also need to define operators ${\cal O}$ compatible with the refinement procedure.
This is easily done by requiring,
\ba
{\cal O}_{\Delta'} \,\circ\, \iota_{\Delta, \Delta'} \,=\, \iota_{\Delta, \Delta'}  \, \circ\, {\cal O}_{\Delta}   .
\ea
In \cite{DG14b,BDG15}, such operators have been constructed and fully characterized.
They are of two types.
Firstly, there are holonomy operators along root--based closed cycles of $\Gamma$. 
These operators are labelled by a representation of $\cal G$ and act by multiplication in the obvious way.
Secondly, there are so--called `exponentiated flux operators'.
In $(2+1)D$, they are associated to edges of the triangulation $\Delta$ itself.
They act by translating the holonomies associated to the links of $\Gamma$ dual to the relevant edges of $\Delta$.
Therefore, they act as exponentiated derivate operators, hence their name.\footnote{In the ALI representation of loop quantum gravity, gravitational fluxes act as derivatives on the holonomies.}
Notice that the holonomy translation by the action of the exponentiated fluxes induces curvature defects at some vertices of the triangulation.
In other words, it introduces non--trivial monodromies around cycles of $\Gamma$ dual to some vertices of $\Delta$.
To obtain a state with a curvature defect at an arbitrary position $x\in\Sigma$, one just has to first refine $\Delta$ to $\Delta'$ in such a way that $x\in \Delta'_0$.
Finally, we stress that the operators just described, and properly defined in \cite{DG14b,BDG15}, are either gauge invariant or lead to gauge violations confined at the root.

This last remark is important because in the present work we will allow torsion degrees of freedom to be carried by the vertices of the triangulation. 
Which means that more general gauge--invariance violations will be allowed than in the setting presented above.
Although to avoid technicalities we will do this in the context of a finite group gauge theory, this generalization is conceptually of crucial importance for gravity (which is, of course, based on a Lie group).
This is because, spinning particles induce torsion violation \cite{Sousa,Louapre1}.
The relevant operators, creating this more general type of excitations, have been introduced --- albeit in a slightly different manner with respect to ours --- by Kitaev, in \cite{Kitaev1}.
He called them `ribbon operators'.
In the context of $(2+1)D$ gravity, ribbon operators crucially provide Dirac observables \cite{FreiZap}.
We draw from this further motivation for the present work, in that we want on the one hand to give a lattice--independent definition of ribbon operators, and on the other to use their eigenvalues to fully characterize a basis of the quantum gravity Hilbert space on $\Sigma$.

\subsection{An alternative description of the BF representation}\label{sec_alBF}

Here we present an alternative formulation of the BF-representation.
Its advantages are multiple:
First, its language is closer to that of the Turaev Viro based representation \cite{DG16}.
Second, it translates a range of techniques used in the context of string--net models \cite{LevinWen, Hu} to an holonomy--based formalism.
Finally, it provides a lattice-independent description of the Kitaev model \cite{Kitaev1}, which can in turn be mapped onto an `extended' string--net model \cite{Buersch}.

The basic idea behind this alternative formulation is to replace the triangulation, its vertices and its dual graph, with a less rigid structure provided by punctured surfaces and general graphs on them.
Introducing an equivalence class among graphs allows for a first step towards the continuum limit.
We say `a first step' because in this paper we will work  with the defects' locations, i.e. the punctures, kept fixed.
The second, and last, step to the continuum limit would be to consider the inductive limit in which one allows for the addition of new punctures. 
A possible way to achieve this is sketched in Appendix \ref{app_inductive}.

Let us  provide all the ingredients needed for the construction of this alternative description of the BF representation.\\
 
~\\
{\bf {Finite Group}}\\
As mentioned above, we will work with a finite gauge group ${\cal G}$, with $|{\cal G}|<\infty$ elements. 
Some of our results can be generalized to Lie groups, in both the BF and ALI representations. 
This would, however, require lengthy (measure theoretical) technical discussions.
Here, we rather prefer to emphasize the underlying algebraic structures and the many analogies to the TV representation. 
Indeed, one can understand the $q$--deformation at root of unity characteristic of the TV representation, as in a certain sense turning $\SU(2)$ into a finite (quantum) group $\SU(2)_q$. Spin--foam models with finite groups are  used to study the behaviour of spin--foams under coarse graining \cite{FiniteSF,Clement}, and we hope that the techniques developed here will be useful also in this context.
We denote general elements of $\cal G$ by $G,H,g,u,\dots$ and variations thereof, and the identity element by $e$.
The delta-function on the group is normalized so that $\delta(g,h)=1$ if $g=h$ and vanishes otherwise.\\
 \newpage
~\\
{\bf {Punctured surface}}\\
In our analysis we will for simplicity exclusively work in the case in which $\Sigma$ is the two-sphere $\mathbb{S}$.
Fix $\S$ to have a finite number $|p|$ of marked points, called punctures $\{p\}$.
Define $\S_p$ to be the surface $\S$ with one disc removed around each puncture and  with one point marked on the boundary of each such discs. 
We will call these points  `puncture--nodes'.
This structure is needed to describe torsion defects and later--on to define the gluing of states along punctures. 
Now, consider finite directed graphs embedded into this surface.
The graphs can have `open links', i.e. links ending in a one--valent node, provided this node is a puncture--node.
We require all other nodes to be two- or tri--valent.\footnote{Two--valent nodes are needed only as intermediate steps of the refining procedure.}
This is just a choice, that leads to a triangulation as dual complex and furthermore makes a translation to string nets (via a standard group Fourier transform) more immediate.  This restriction can however be easily dropped.

\begin{figure}[h!]
	\centering
	\begin{minipage}[b]{0.30\textwidth}
		\centering
		\includegraphics[scale = 1]{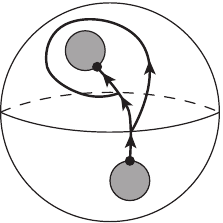}	
	\end{minipage}
	\begin{minipage}[b]{0.30\textwidth}
		\centering
		\includegraphics[scale = 1]{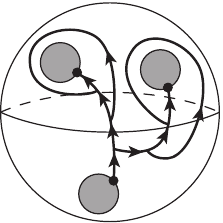}	
	\end{minipage}
	\caption{Examples of minimal graphs on $\S_2$ and $\S_3$, respectively. \label{fig_min}}
\end{figure}

Among all the possible graphs, there is a special subclass of `minimal graphs'.
Minimal graphs on a punctured sphere are defined by the following properties: ({\it i\,}) they capture the first fundamental group of the punctured sphere, $\pi_1(\S_p)$, ({\it ii\,}) they have no contractible faces, that is all their faces enclose a puncture, ({\it iii\,}) they have no two--valent node, and ({\it iv\,}) they have one open link associated to each puncture, see figure \ref{fig_min} for examples.
Given $\S_p$, minimal graphs are by no means unique.
From our definition, it is not difficult to see that a minimal graph on $\S_{p>1}$ must have exactly $L=1 + 2(p-2) + 3(p-1) = 5p - 6$ links, and $N=\tfrac{1}{3}(2L-p)$ internal nodes.

Before moving to the next point, notice that the puncture--nodes play the all--important role of making possible the gluing of surfaces along the boundaries of the punctures' discs.
Following Ocneanu's insight \cite{Ocneanu}, we will show how this operation unveils a wealth of algebraic structure hidden in the theory.

~\\
{\bf {Hilbert Space ${\cal H}_p$}}\\
The configuration space underlying the BF representation is, exactly as before, given by the moduli space of flat connections on $\S$ minus some points (or, equivalently, discs).
In this case, this reads $\mathcal M^\text{flat}(\S_p)$.
This space is now completely characterized by the holonomies along the links of a minimal graph $\Gamma$.
Hence, we define the Hilbert space ${\cal H}_\Gamma$ to be given by the set of gauge invariant functions $\{\psi\}$ on such a space of holonomies:
\be
\psi:\; {\cal G}^L \; \to \; \mathbb{C},
\ee
where $L$ denotes the number of links of the minimal graph. Importantly, we require the gauge group to act only at the internal nodes, and not at the puncture--nodes.  Indeed, imposing gauge invariance at the puncture--nodes would result in the trivialization of the dependence of the state from  the group element associated to the only link ending there. In the following, it will become apparent that avoiding this trivialization is crucial to implement both torsion excitations and a consistent cutting--and--gluing scheme of the states. 

More specifically, a gauge transformation is parametrized by a choice $\{u_n\}_n \in {\cal G}^N$ where $n$ denotes an internal node and $N$ their number.
It acts on a holonomy configuration  $\{g_l\} \in {\cal G}^L$ as
\ba
\{u_n\}_n \triangleright  \{g_l\}_l \,=\, \{u^{-1}_{t(l)} \,g_l \, u_{s(l)} \}_l \, ,
\ea
where $s(l)$ and $t(l)$ denote the source and target nodes of the link $l$, respectively.\footnote{In the equation above, we left understood that $u_{s(l)}\equiv e$ if $s(l)$ is a puncture--node. Similarly for $t(l)$.}
Finally, the inner product in ${\cal H}_\Gamma$ is defined by
\ba\label{innprod1}
\langle \psi_1, \psi_2\rangle  &=& \frac{1}{|{\cal G}|^{L}} \sum_{\{g_l\}} \overline{\psi_1 \{g_l\}} \, \psi_2\{g_l\} .
\ea
To obtain a Hilbert space ${\cal H}_p$ associated directly to $\S_p$, we need to show how to identify various ${\cal H}_{\Gamma'}$ for different choices of (possibly non-minimal) graphs  $\Gamma'$ in $\S_p$ .
We do this by declaring two states based on different graphs as equivalent if they are related by a combination of the four operations we are now going to describe.
The idea is that via a minimal graph one can already characterize $\mathcal{M}^\text{flat}(\S_p)$ completely:
it gives access to the holonomies associated to all the non-contractible cycles (those around the punctures), as well as giving the holonomy (parallel transport) between any couple of punctures.
Since the connection is locally flat, the path underlying each holonomy can be smoothly deformed.
Also, we can refine the graph, provided we ensure that the holonomies associated to the contractible cycles are all trivial, and that gauge invariance is preserved.
As a consequence of gauge invariance, we can freely remove two--valent nodes.
Likewise for a non--minimal graph we can remove links, if the resulting graph still captures $\pi_1(\S_p)$. 
Formally, the operations are:
\begin{itemize}
\item[{\it i})]{\it Link deformation}---A link can be (smoothly) deformed as long as no other link, node or puncture is crossed. Two states $\psi,\psi'$ based on two graphs $\Gamma,\Gamma'$ related by a link deformation are defined to be equivalent if they are described by the same function, i.e. if $\psi(\{g\})=\psi'(\{g\})$ as functions on ${\cal G}^L$.

\item[{\it ii})]{\it Link orientation flip}---After flipping the orientation of a link $l \to l^{-1}$, the state $\psi'$ equivalent to $\psi$ is
\ba
	\label{equ1}
	\psi'( g_{l^{-1}}, \cdots) \,=\, \psi(g^{-1}_l,\cdots)  .
\ea

\item[{\it iii})] {\it Link subdivision/union}---After the subdivision of a link $l \to l_2 \circ l_1$, the state $\psi'(g_{l_1},g_{l_2}, \cdots)$ equivalent to $\psi(g_l, \cdots)$ is
\ba
	\label{equ2}
	\psi'(g_{l_1},g_{l_2}, \cdots)  &=& \psi( g_{l_2} g_{l_1}, \cdots)   .
\ea

\item[{\it iv})]{\it Face removal/addition}---After the addition of a new link $l$, the graph gains a new closed face (that is a contractible cycle) $f$ with holonomy $h_f$ (we are assuming that any link subdivision necessary to the addition of this new link has already been performed). Then the state $\psi'(\{g_{l'}\})$ on the new graph which is equivalent to the original $\psi(\{g_l\})$ is
\ba
	\label{equ3}
	\psi'( \{g_{l'}\}) &=& \sqrt{|{\cal G}|} \,\, \delta(e, h_f) \, \psi( \{g_l\})  .
\ea
where the factor $\sqrt{{\cal G}}$ in (\ref{equ3}) has been introduced to ensure that equivalent states have the same norm. 
\end{itemize}

At this point, it is a simple exercise to show that the inner product is independent of the choice of representative in the equivalence class described above.
This concludes the construction of ${\cal H}_p$.

Notice that the only information that is common to all ${\cal H}_\Gamma$, and therefore that is proper to ${\cal H}_p$ itself, is the embedding of the punctures.
This mirrors the properties of the states in ${\cal H}_p$: excitations are confined to the punctures and the state describe locally--flat gauge--invariant connections away from the punctures. 

To obtain a continuum Hilbert space allowing for excitations at arbitrary positions in $\Sigma$ we have to consider the inductive limit over Hilbert spaces ${\cal H}_p$, where $p$ stands not only for the number of punctures (denoted by $|p|$) but also for their embedding information. 
For a sketch on how to achieve this, we refer to Appendix \ref{app_inductive}.

All the construction presented here can be recast in a spin--network language, essentially by decomposing the states $\psi$ via the Peter-Weyl theorem onto a graph-dependent basis labeled by representation--theoretic data. 
In this formulation one would recover the so--called extended string nets \cite{Buersch}, and the conditions above would be rephrased in a completely algebraic and combinatorial language.

\section{From Ocneanu's tube algebra to the Drinfel'd double\label{sec_drinfeld}}

So far, we have been describing states in a graph-dependent and redundant fashion.
Graph independence is then shown to be recovered thanks to the introduction of appropriate equivalence relations.
It would be, however, much more efficient to characterize the states directly, with no reference to any choice of graph.
To this end we turn our focus on the punctures and the excitations they carry.
Let us start by analyzing the simplest cases, $\S_p$ with $|p|=1,2,3$.

Clearly, $\S_1$ cannot carry excitations, since $\pi_1(\S_1)$ is trivial.
Indeed, a minimal graph on $\S_1$ has one link $l$ surrounding the puncture, and one link $l'$ starting at $n=s(l)=t(l)$ and ending at the puncture--node; 
now, contractibility of $l$ imposes $\psi(g_l,g_{l'}) = \delta(e, g_l) f(g_{l'})$, while gauge invariance at $n$ requires $f$ to be constant.

Thus, the simplest non--trivial case is that of the 2-punctured sphere, $\S_2$, which is topologically just a cylinder. 
The study of states on the cylinder is the subject of this section.
The next--simplest case is $\S_3$. 
The three--punctured sphere is a fundamental object in $2D$ topology.
It goes under the names of `trinions', or--- for obvious topological reasons---`pair-of-pants'.
The analysis of states on $\S_3$ will be the subject of the next section (section \ref{sec_fusionbasis}).

The reason why $\S_3$ is such a fundamental object is because out of it, by the procedure of successive gluing, one can produce any $\S_p$.%
\footnote{ Actually $\S_3$ is enough to build any $\Sigma_p$, although the gluing procedure for the states becomes  more subtle in this case. We will not discuss this in the present paper.}
Therefore, $\S_2$, $\S_3$ and the gluing procedure are all that there is to know.
Let us hence start by discussing cylinders.

\subsection{Characterizing the excitations}

Cylinders play a special role in the characterization of `basic' excitations. 
Simply put, the reason is that cylinders can be glued `around a puncture' without changing the topology of $\S_p$.
Therefore, via the gluing operation, states on a cylinder can also be interpreted as maps acting on ${\cal H}_p$.
By successively gluing cylinders onto one-another, it is straightforward to define a multiplication between cylinder states.
In this way, states on the cylinder define an algebra, called the Ocneanu `tube--algebra'. 

But what is the physical interpretation of this algebra?
By visualizing the cylinder as an annulus of space, one can think of the gluing operation as the addition of `more--space' around an excitation.
Topological excitations relevant to $3D$ gravity should be stable under this operation.
Therefore, they are characterized by idempotents of the tube algebra, or---equivalently---by its indecomposable (representation) modules \cite{Ocneanu, Lan}.
As pointed out by Ocneanu \cite{Ocneanu}, this allows the  interpretation of such idempotents as viable boundary conditions.

In the next subsection, we will show explicitly how the tube algebra is nothing but the Drinfel'd double algebra ${\cal D}({\cal G})$  \cite{DrinfeldDouble}.
Thus, it follows immediately that states on the cylinder are labeled by irreducible representations $\rho$ of ${\cal D}({\cal G})$.
Also, from the above discussion, we see that any puncture on $\S_p$ should be labeled by such a $\rho$.
Hence we conclude that excitations in $3D$ Euclidean gravity can be classified in terms of irreducible representation $\rho$ of ${\cal D}({\cal G})$. 
In the case ${\cal G}=\SU(2)$, the two labels defining such a $\rho$ can be directly interpreted as the mass and the spin of the excitation.

This result is not new, see e.g. \cite{Freidel2, KarimDrinfeld, BonzomEtAl, PranzEtAl}.
However, in previous treatments, it was a found as a consequence of the presence of group--valued constraints (moment maps).%
\footnote{AR thanks Karim Noui and Florian Girelli for a useful discussion at this purpose.}
The gluing of cylinder states seems, however, to provide so far the simplest and most direct argument.

Of course, the next important question is to understand how different `local' states defined at each puncture, eventually fit together.
Not surprisingly, the answer lies in the study of the tensor product of representations, i.e. in the fusion algebra of $\text{Rep}({\cal D}({\cal G}))$. 
This will be the content of section \ref{sec_fusionbasis}.

\subsection{Two--punctured sphere}

A minimal graph $\Gamma_2$ on $\S_2$ possesses four links $\{l_i\}_{i=1,\dots,4}$.
We fix the compositions $l_2^{-1} \circ l_4$ to be a closed loop winding once around the cylinder, and $l_3\circ l_2 \circ l_1 $ to go from the `source' puncture to the `target' puncture.
Finally, label the links of the graph with group elements $\{g_{l_i}  \in \mathcal{G} \}$. For brevity we set $g_{l_i}=g_i$:
 \be
	\begin{array}{c}\includegraphics[scale =1]{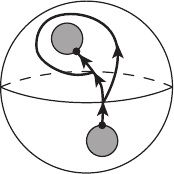}\end{array} =
	\begin{array}{c}\includegraphics[scale =1]{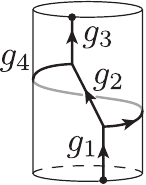}\end{array} =
	\begin{array}{c}\includegraphics[scale =1]{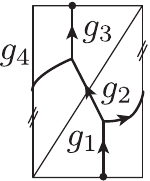}\end{array} .
\ee 

A state in ${\cal H}_{\Gamma_2}$ is given by a gauge invariant state
\be
\psi^{\S_2}(g_1,g_2,g_3,g_4) = \psi^{\S_2}(u g_1, v g_2 u^{-1}, g_3 v^{-1}, v g_4 u^{-1}),
\ee 
for any $u,v \in\cal G$.
Taking advantage of the gauge invariance of $\psi^{\S_2}$, choosing $v = g_3$ and $u= g_3 g_2$, we can fix $g_2 = e = g_3$.
A basis of ${\cal H}_{\Gamma_2}$ is then given by the gauge--fixed states
 \be
\psi^{\S_2}_{G,H}(g_1,g_4)_{|{\rm g.f.}} = |\mathcal{G}|^{3/2} \delta(G,g_1) \delta(H,g_4) \equiv |\mathcal{G}|^{3/2}
	\begin{array}{c}
		\includegraphics[scale =1]{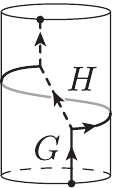}
	\end{array}\label{basisgroup}.
\ee
{ where $|\mG|^{3/2}$ is a normalization factor chosen for later convenience.}
In the above diagram, dashed lines represent gauge fixed group elements, while solid lines carry the group variables.
In fully gauge--covariant form, these basis states read
 \begin{align}
\psi^{\S_2}_{G,H}(g_1,g_2,g_3,g_4)
& = |\mathcal{G}|^{3/2}\delta(G,g_3g_2g_1) \delta(H, g_3 g_4 g^{-1}_2g^{-1}_3).
 \end{align}

This basis can readily proved to be orthogonal:
\begin{align} \nn
	\big\langle \psi^{\S_2}_{G,H}\, , \, \psi^{\S_2}_{\widetilde{G},\widetilde{H}}\big\rangle &= \frac{1}{|\mathcal{G}|^4}\sum_{g_1,\dots,g_4}\overline{\psi^{\S_2}_{G,H}(g_1,\dots,g_4)}\psi^{\S_2}_{\widetilde{G},\widetilde{H}}(g_1,\dots,g_4) 
	\,= |\mathcal{G}|\delta(G,\widetilde{G})\delta(H,\widetilde{H}). \label{innerprodgroup}
\end{align} 
Notice that the states $\psi^{\S_2}_{G,H}$ are not normalized.
The reason for this choice will be made clear later. Henceforth, we will often keep the $\{g_i\}$ implicit, and denote the basis states simply $\psi^{\S_2}_{G,H}$. 

A general element of ${\cal H}_2$ can then be written as
\be
\psi^{\S_2}= \sum_{G,H \in \mathcal{G}}\alpha(G,H) \psi_{G,H}^{\S_2}\,.
\ee
In particular, we define the (unnormalized) $\S_2$ vacuum state to be
\be
	\psi_0^{\S_2}\,=\, \delta(e, g_4 g^{-1}_2)  \, 1(g_1) \, 1(g_3)\, ,
	\label{vacuum}
\ee
where $1(\cdot)$ denotes the constant function evaluating to $1$ on any group element. 
Therefore, $\psi_0^{\S_2}$ is the following linear combination of basis states,
\be
\psi^{\S_2}_0 = |\mG|^{-3/2}\sum_{G\in\mG} \psi^{\S_2}_{G,e} 
\qquad\text{with}\qquad
\langle \psi^{\S_2}_0 , \psi^{\S_2}_0 \rangle =  |{\cal G}|^{-1}.
\ee

\subsection{Ocneanu's tube algebra \label{subsec_ocneanu}}

As we have already mentioned, a fundamental property of punctured manifolds is the possibility of gluing them together.
Let us denote the gluing operation with a star, $\star$.
Then, denoting $\Sigma^g_p$ the Riemann surface of genus $g$ and $p$ punctures,
\be
\Sigma^g_p \star \Sigma^h_q = \Sigma^{g+h}_{p+q-2}.
\ee 
Clearly, gluing a cylinder $\S_2 = \Sigma^0_2$ to any other $\Sigma^g_p$ gives back $\Sigma^g_p$.
At the level of the graphs $\Gamma^g_p$ and $\Gamma^h_q$, the gluing is defined by identifying the marked points associated to the punctures along which the gluing is performed and matching the two edges which end at these marked points. The marked points then become a single two--valent node $n$ in $\Gamma^g_p\star\Gamma^h_q$.
Even if the two original graphs on $\Sigma^g_p$ and $\Sigma^h_q$ were minimal, the resulting graph on $\Sigma^{g+h}_{p+q-2}$ is not.
Indeed, it contains an extra closed face $f$, i.e. a face surrounding no puncture.
Therefore, if we want the gluing to be mirrored at the level of the state spaces, this face must be associated with a trivial holonomy.
This suggests to define first the gluing operation on basis states based on graphs via
\be
\begin{array}{cccccl}
\star:& \;{\cal H}_{\Gamma} \otimes {\cal H}_{\Gamma'} & \hookrightarrow & {\cal F}_{\;\Gamma\star\Gamma'} & \xrightarrow{{\mathbb P}} &{\cal H}_{\Gamma\star\Gamma'}\\
& \psi^\Gamma \otimes \phi^{\Gamma'} & \mapsto & \psi^\Gamma \cdot \phi^{\Gamma'} & \mapsto & \psi^\Gamma \star \phi^{\Gamma'} \simeq \mathbb P( \psi^\Gamma \cdot \phi^{\Gamma'}),
\end{array}
\ee
and then to extend this by linearity to arbitrary states.
{ The $\simeq$ before the last term in the diagram above signals that equivalence relations (Section \ref{sec_alBF}) are generally used at this point.}
Here, ${\cal F}_{\Gamma\star\Gamma'}$ is simply the space of functions on $\Gamma\star\Gamma'$, and the first operation is trivial.
On the other hand $\mathbb{P}$ is a projector, and it is itself the combination of two operations:
\be
\mathbb P = \mathbb P_\text{flat} \circ \mathbb P_\text{gauge}
\ee 
the first being a projection onto gauge invariant states at the newly created two-valent node $n$ of $\Gamma\star\Gamma'$, and the second a projection onto states carrying trivial holonomies around the newly created closed face $f$.

To clarify the above construction, let us consider the important case of gluing two cylinders to one-another.
In this case, $\S_2\star\S_2 = \S_2$, and the gluing defines a multiplication operation on ${\cal H}_2$.
Notice that this is not a standard structure on a Hilbert space.
In particular, thanks to the gluing operation, ${\cal H}_2$ carries a representation of algebra, named Ocneanu's tube algebra.

Consider two minimal states 
\be
\psi^{\S_2}_1(\{g_i\}) , \; \psi^{\S_2}_2(\{g'_i\}) \in {\cal H}_2.
\ee
Then, following the prescriptions above, we obtain
\be
	(\psi^{\S_2}_1\cdot \psi^{\S_2}_2)(\{g_i,g_i'\})= 
	\begin{array}{c}\includegraphics[scale =1]{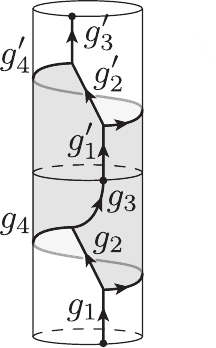}\end{array}
\ee
where the gray face is the new closed face $f$ on which the flatness constraint has to be imposed. 
The node $n$ is the one where the links $l_3$ and $l_1'$ meet.
Gauge invariance at $n$ can be imposed by group averaging:
\begin{align}
	\tilde\Psi(\{g_i,g_i'\})& :=\Big(\mathbb{P}_\text{gauge} (\psi^{\S_2}_1\cdot \psi^{\S_2}_2)\Big)(\{g_i,g_i'\}) \\
	&=\, \frac{1}{|{\cal G}|}\sum_{k \in \mG}  \,\, \psi^{\S_2}_2 ( g'_1 k^{-1},  g_2' ,g_3' ,g_4' ) \psi^{\S_2}_1(g_1,g_2,k g_3 ,g_4 ).
\end{align}
Now, the flatness constraint at the new closed face $f$ is readily imposed as
\be
	\psi_1^{\S_2}\star\psi_2^{\S_2} = \Big(\mathbb{P}_\text{flat}  \tilde\Psi\Big)(\{g_i,g_i'\})\,=\, \delta(e, h_f) \tilde\Psi(\{g_i,g_i'\})
\ee
where $h_f =  g_3 g_4 g_2^{-1} g_3^{-1} g_1'^{-1}  g_4'^{-1} g_2' g_1'$.
The state so constructed is not defined on a minimal graph. 
However, it is in ${\cal H}_{\Gamma\star\Gamma'}$ and hence via the equivalence relation described above, can be identified with a state in ${\cal H}_2$.

Let us be even more specific, by gluing two basis states of ${\cal H}_2$. 
Graphically:
\begin{align} \nn
	\psi_{G',H'}^{\S_2} \star \psi_{G,H}^{\S_2} \, & = \, |\mG|^3
 	\begin{array}{c}\includegraphics[scale =1]{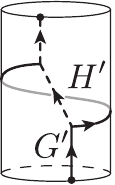}\end{array} \star
	\begin{array}{c}\includegraphics[scale =1]{fig/cylindre2-eps-converted-to.pdf}\end{array}
	\, \simeq \,\mathbb{P}_\text{flat} \circ \mathbb{P}_\text{gauge} \,
	\Bigg( |\mG|^3 \begin{array}{c}\includegraphics[scale =1]{fig/mul1-eps-converted-to.pdf}\end{array} \cdot
	\begin{array}{c}\includegraphics[scale =1]{fig/cylindre2-eps-converted-to.pdf}\end{array}\Bigg) \,
\end{align}
but, on the other hand, the linked cylinder state is explicitly given by
\be \nn
	\begin{array}{c}\includegraphics[scale =1]{fig/mul1-eps-converted-to.pdf}\end{array} \cdot
	\begin{array}{c}\includegraphics[scale =1]{fig/cylindre2-eps-converted-to.pdf}\end{array}
	 = 
	\begin{array}{c}\includegraphics[scale =1]{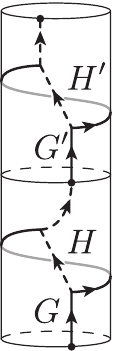}\end{array}=
\ee \\[-5em]
\be
	 = \delta(G,g_3g_2g_1)\delta(H,g_3g_4g_2^{-1}g_3^{-1})
	\delta(G',g_3'g_2'g_1')\delta(H',g_3'g_4'(g_2')^{-1}(g_3')^{-1}) 
	\begin{array}{c}\includegraphics[scale =1]{fig/mul02-eps-converted-to.pdf}\end{array}
\ee
Hence, applying the projectors and rearranging the delta functions, we obtain
\be \nn
	\mathbb{P}_\text{flat}  \circ \mathbb{P}_\text{gauge} \,
	\Bigg( |\mG|^3 \begin{array}{c}\includegraphics[scale =1]{fig/mul1-eps-converted-to.pdf}\end{array} \cdot
	\begin{array}{c}\includegraphics[scale =1]{fig/cylindre2-eps-converted-to.pdf}\end{array}\Bigg)=
\ee
\begin{align}\nn
	=
	|\mG|^3 \frac{1}{|{\cal G}|} \sum_{k\in\mG}\, & \delta(G',g_3'g_2'g_1'k^{-1})\delta(g_2'g_1'g_3g_4 g_2^{-1}g_3^{-1}(g_1')^{-1}(g_4')^{-1})  \\[-4em] \nn
	 \times \, &\delta(G'G, g_3' g_2'g_1'g_3g_2g_1) \,
	\delta(H,(G')^{-1}H'G')\,  \delta(H', g_3'g_4'(g_2')^{-1} (g_3')^{-1})  
	\begin{array}{c}\includegraphics[scale =1]{fig/mul02-eps-converted-to.pdf}\end{array}
\end{align}
Now, we appeal to the equivalence relations of section \ref{sec_BF} to remove the dependence on $g_4$ by removing the corresponding link (and associated face, see Equation \eqref{equ3}). 
We then undo three link subdivisions and declare $l'_1\circ l_3\circ l_2 \circ l_1$ to be the new link $\tilde l_1$.
Hence, we finally obtain the following crucial result%
\footnote{Note that there we could have made different choices to simplify the final form of the state, but all choices have led to the same result.}
\begin{align}	
	\psi^{\S_2}_{{G'},{H'}}\star \psi^{\S_2}_{G,H} 
	\,=\, |\mathcal{G}|^{3/2} \delta(H,(G')^{-1}H'G')
	\begin{array}{c}\includegraphics[scale =1]{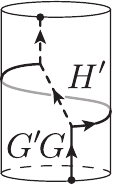}\end{array} \,=\,
	\delta((G')^{-1}H'G',H)\psi^{\S_2}_{{G'}G,{H'}}\,.
	\label{eq_gluecyl}
\end{align}
This multiplication law, together with the usual Hilbert space linear structure, defines the Ocneanu tube algebra.
As a matter of fact, the $\star$ multiplication we have just constructed is exactly the multiplication law of the Drinfel'd double algebra $\mathcal{D}(\mathcal{G})$.
The present construction can be readily generalized to the case where $m\geq1$ links are allowed to end at  the puncture, see appendix \ref{app_mpuncture}.

\subsection{Drinfel'd double of a finite group}

Drinfel'd (or quantum) doubles of a finite group  \cite{DrinfeldDouble} are examples of quasi-triangular Hopf algebras (or quantum groups).
Here we will present the basic properties relevant for our discussion.
More details can be found e.g. in \cite{hopf}.

As a linear space, the Drinfel'd double ${\cal D}({\cal G})$ is isomorphic to
\be\label{tenorD}
\mathcal{D}(\mathcal{G}) \simeq \mathbb{C}\mathcal{G} \otimes \mathcal{F}(\mathcal{G}),
\ee
where $\mathcal F(\mathcal G)$ is the Abelian algebra of linear complex--valued functions on $\mathcal G$, and  $\mathbb{C}\mathcal{G}$ is the complex group algebra. Both $\mathcal F(\mathcal G)$ and $\mathbb{C}\mathcal{G}$ can be made into Hopf algebras and moreover as Hopf algebras are dual to each other.

Due to the isomorphism (\ref{tenorD}) we can state a basis of ${\cal D}(\mG)$ by  $\{G \otimes \delta_{H}  \}_{G,H \in \mathcal{G}}$  where $\delta_H$ is the delta function peaked on $H$ such that $\delta_H({\scriptstyle \bullet}) = \delta(H,{\scriptstyle \bullet})$.
To have a more symmetric notation, we will denote such basis elements as $[G,H] = G\otimes\delta_H $.

The Drinfel'd double being a Hopf algebra, it comes equipped with a series of operations and maps satisfying certain compatibility axioms.
This is a standard construction, and we refer e.g. to \cite{hopf} for the complete list.
Here, we  just provide a brief reminder of some of its structures.
\begin{itemize}
\item[{\it i})] {\it Multiplication}---It is denoted with a $\star$ and is defined by
\begin{align}
\star \;\;  : \;\; \mathcal{D}(\mathcal{G}) \otimes \mathcal{D}(\mathcal{G})\;\;\;\; &\longrightarrow \;\;\mathcal{D}(\mathcal{G}) \\ \nn
	([\widetilde{G},\widetilde{H}],[G,H])\;\; &\longmapsto \;\;[\widetilde{G},\widetilde{H}]\star [G,H] = 
	\delta(\widetilde{H},\widetilde{G}H\widetilde{G}^{-1})[\widetilde{G}G,\widetilde{H}].
\end{align}
It is associative and its identity element is
\be
	\mathbb{I} = \sum_{H \in {\mathcal{G}}}[e,H].
\ee
Note, this multiplication does {\it not} come from the direct product of the algebras $\mathcal F(\mathcal G)$ and $\mathbb{C}\mathcal{G}$, but it rather has a semi--direct product structure where the $\mathbb{C}{\mathcal{G}}$ factor acts on the $\mathcal F(\mathcal G)$ factor.

%
\item[{\it ii})] {\it Comultiplication}---This is needed for defining the action of ${\cal D}(\mG)$ on tensor product representations, and in ${\cal D}(\mG)$ it is non trivial:
\begin{align}
	\Delta \;\;  : \;\; \mathcal{D}(\mathcal{G}) \;\; &\longrightarrow \;\;\mathcal{D}(\mathcal{G}) \otimes \mathcal{D}(\mathcal{G}) \\ \nn
	[G,H]\;\; &\longmapsto \;\;\Delta[G,H] = 
	\sum_{X,Y \in \mG \atop XY =H}[G,X]\otimes [G,Y].
\end{align}
Importantly, this operation is co-associative, i.e.
\be
	(\Delta \otimes \text{id})\circ \Delta  = (\text{id} \otimes \Delta)\circ \Delta,
\ee
where we introduced the identity map on the double, $\text{id}({\scriptstyle \bullet}) = {\scriptstyle \bullet}$. 
\item[{\it ii})] {\it Antipode}---We will not make essential use of it in this paper, but since it is what generalizes the inverse element to the context of quasitriangular Hopf algebras, it is worth recalling its definition
\be\label{eq_antipode}
	S([G,H]) = [G^{-1},G^{-1}H^{-1}G].
\ee
The defining relation for the antipode, which qualifies it as the generalization of the notion of inverse, is
\be
	(\text{id}\star S)\circ\Delta = (S \star \text{id})\circ \Delta = \epsilon ,
\ee
where $\epsilon$ is a linear map on ${\cal D}(\mG)$, known as the counit, defined by $\epsilon([G,H]) = \delta(H,e)\mathbb I $.
Notice the role played by both the multiplication and the comultiplication in the above identity.
\end{itemize}

As our notation suggests,  the $\star$--multiplication corresponds to the gluing product introduced in the previous section. 
The $\star$-product of two basis states of the form $\psi^{\S_2}_{G,H}$ is a different basis state (of the same form).
However, if we want to follow the idea that physical excitations must be stable under the operation of gluing cylinders on the top of them, we are motivated to look for an alternative basis, labeled by $\rho$, such that
\be
	\psi_{\rho'}\star \psi_{\rho} \propto \delta(\rho,\rho')\psi_{\rho}.
	\label{idempo}
\ee
Mathematically speaking, we look for a basis of ${\cal D}(\mG)$ that is idempotent under the  $\star$--multiplication.
Since the Drinfel'd double algebra ${\cal D}(\mG)$ is semisimple, the idempotent states we are looking for are directly provided by the irreducible representations of the Drinfel'd algebra.
By this, we mean that the $\rho$--index above labels irreducible representations of the Drinfel'd algebra.

\subsection{Irreducible representations of ${\cal D}(\mG)$ \label{sec_irreps}}

Irreducible representations $\{\rho\}$ of the Drinfel'd double ${\cal D}(\mG)$ are constructed as induced representations \cite{Verlinde,BaisReview}.
This is possible because the multiplication operation
\be
[\tilde G, \tilde H] \star [G,H] = \delta(  \tilde H, \tilde G H \tilde G^{-1}) [\tilde G G, \tilde H]
\ee
can be roughly read as `$\tilde G$ multiplies $G$, while acting on $H$'.

Since $\tilde G$ acts on $H$ by conjugation, a fundamental ingredient to build the $\{\rho\}$ is the set of conjugacy classes of $\mG$.
The property of being conjugated to each other is an equivalence relation:
\be
H\sim H'\quad\text{iff}\quad \exists \, G\in\mG ,\; H' = G H G^{-1}.
\ee
Therefore, the group $\mG$ is partitioned by the set of its conjugacy classes $\mG = \bigsqcup C$ (to not burden the notation even further we do not introduce an index labeling different conjugacy classes).
Denote the elements of $C$ by $c_i$ with $i=1,\dots,|C|$. 
And say that $c_1 \in C\subset\mG$ is the `representative' of the conjugacy class $C$.

Now, having fixed a representative, we can define $N_C$ to be the stabilizer (`little group') of $c_1$, i.e.
\be
N_C = \{ G\in \mG | G c_1 G^{-1} = c_1 \}.
\ee
We now want to define `standard' transformations that bring the representative $c_1$ to any other element of $C$.
Clearly, there is no canonical choice in $\mG$.
And therefore we need to make a set of choices, we will call $\{q_i\} = Q_C$:
\be
c_i = q_i c_1 q_i^{-1}.
\ee
Note that $q_1$ can always be taken to be the identity, $q_1 = e$.
Note also that
\be
Q_C \simeq \mG/N_C
\ee
and therefore $|Q_C| = |C|$ exactly as desired.

We also introduce a `label function' $k_C$ that associates to any $c_i\in C$ their `number' withing $C$,
\be
k_C(c_i) = i.
\ee

The last ingredient for constructing $\{\rho\}$ is the set of unitary irreducible representations (`irreps') of $N_C$, $\{ R\}$.
Denote the matrix elements of $\widehat G\in N_C$ in the representation $R$, by
\be
D^R_{M' M} (\widehat G)
\ee
where $M, M'$ are the `magnetic indices' which take $d_R$ different values, $d_R$ being the dimension of the irrep $R$.
The corresponding characters are
\be
\chi^R(\widehat G) = \tr( D^R(\widehat G) ).
\ee
A useful consequence of Schur's orthogonality relations is that the (Kronecker) delta function on the group can be decomposed on the characters as
\be
\delta(e, {\scriptstyle \bullet}) = \frac{1}{|N_C|} \sum_{R} d_R \chi^R({\scriptstyle \bullet}) \q \label{delta_stabilizer} .
\ee

The idea is then to separate the action of each element $G\in\mG$ into its action within $C$ and $N_C$.
For this we unfortunately need to introduce some extra notation.
Fixing a $G\in\mG$ and a label $i$ in $C$, we define the index $i'$ via $i' = k_C(G c_i G^{-1})$, i.e.
\be
c_{i'} = G c_i G^{-1}.
\ee
Now, this allows us to construct out of $G$ and $i$ an element $\widehat G_i \in N_C$ as
\be
\widehat G_i = q^{-1}_{i'} G q_i.
\ee
Indeed, this follows by the comparison of the first and last terms of the following series of equalities
\be
q_{i'} c_1 q^{-1}_{i'} = c_{i'} = G c_i G^{-1} = G q_i c_1 q_i^{-1} G^{-1}.
\ee

At this point we have all is needed to introduce the irreducible representations $\{\rho\}$ of ${\cal D}(\mG)$. 
These are labeled by a conjugacy class $C$ and an irrep $R$ of $N_C$: $\rho=(C,R)$.
The relevant vector space on which $\rho$ acts is given by the linear (complex) span of the following vectors (in a ket--bra notation):
\be
V^{(C,R)} = \text{Span}\Big\{| c_i, M \rangle  \; \Big|\; i=1,\dots,C; \; M=1,\dots, d_R \Big\}.
\ee
The action of the Drinfel'd double ${\cal D}(\mG)$ algebra on $V^{(C,R)}$ is thus defined by the following action on the above basis:
\be
D^{C,R}([G,H]) | c_i, M\rangle = \delta(H,G c_i G^{-1} ) \sum_{M'} D^R_{M M'} (\widehat G_i) |  G c_i G^{-1} , M \rangle.
\ee
Equivalently, the matrix elements of $D^{C,R}([G,H])$ are 
\be
D^{C,R}_{i'M',iM}([G,H]) \,=\, \delta(H, c_{i'})\, \delta( c_{i'} , G c_i G^{-1}) \, D^R_{M'M}( \widehat G_i).
\ee
This is then extended linearly to the whole vector space.

We can interpret such a definition as follows. First  one acts with the adjoint action of $G$ on $c_i \in C$. However $G$ contains also a part in $N_C$, and this part acts via $D^R$ on the $R$ representation index. Then one projects out the $H$ component $|H,\cdot \rangle$.

The representations $D^{C,R}$ can also be extended  to elements of the Drinfel'd double of the more general form
\be
\alpha =\sum_{G,H\in\mG} \alpha_{G,H} \;G\otimes \delta_H ,
\ee
by linearity, i.e.
\begin{align}
D^{C,R}(\alpha) 
& = \sum_{G,H} \alpha_{G,H} \delta(H, c_{i'})\, \delta( c_{i'} , G c_i G^{-1}) \, D^R_{M'M}( \widehat G_i)\nn\\
&  =\sum_G \alpha_{G, c_{i'}}\, \delta( c_{i'} , G c_i G^{-1}) \, D^R_{M'M}( \widehat G_i) .
\end{align}

Often, it will be not necessary to have a grasp on the precise value of the magnetic indices.
For this reason we introduce the short--hand notation 
\be
	D^{\rho}_{I',I}({\scriptstyle \bullet}) \equiv  D^{C,R}_{i'M',iM}({\scriptstyle \bullet}) ,
\ee
with $\rho \equiv C,R$, $I' \equiv i'M'$ and $I \equiv iM$.

\subsubsection{Some properties of the irreducible representations of ${\cal D}(\mG)$\label{sec_relationsdouble}}

The dimension of the representation $\rho=(C,R)$ is given by
\be
d_\rho = d_{C,R}=d_R \cdot|C|.
\ee
The following consequence of Schur's relation will be useful
\be
	\sum_R d_R^2 = |N_C|.
\ee
Using the two equations above, it is easy to check that $\text{dim}({\cal D}(\mG)) = |\mG|^2 =  \sum_{C,R} d^2_{C,R}$.
From the contragradient representations of $N_C$, we deduce an expression for the complex conjugate of the matrix elements
\begin{align} \nn
	\overline{D^{C,R}_{i'M',iM}([G,H])} &= \delta(H,c_i')\delta(c_i,G^{-1}c_i'G)D^R_{MM'}(q_i^{-1}G^{-1}q_{i'}) \\
	&= D^{C,R}_{iM,i'M'}([G^{-1},G^{-1}HG]) \label{contra}.
\end{align}
Note that the element of ${\cal D}(\mG)$ appearing in the last term fails to be the antipode $S([G,H])$ of $[G,H]$.
Therefore, the above formula does not define the representation dual to $(C,R)$. 
Anyway, such a notion of dual representation will not be necessary for our discussion.

Finally, the characters of the irreducible representations labeled by $(C,R)$ are defined as $\chi^{C,R}( {\scriptstyle \bullet}) \equiv \text{tr}( D^{C,R}( {\scriptstyle \bullet}))$ and they satisfy the property
\be
	\chi^{C,R}([G,H]) =
	\begin{dcases}
		&\chi^R(\widehat G_{k_C(H)})\;\;\text{if}\;\; H \in C \;\;\text{and}\;\; G^{-1}H^{-1}GH = e \\
		&0  \qquad\qquad\quad\, \text{otherwise} \\
	\end{dcases}
\ee
where recall that $k_C(c_i)$ is the label function $k_C(c_i)=i$.

Using this remark, it is easy to check the following character orthogonality relation \cite{Verlinde}
\be
	\frac{1}{|\mathcal{G}|}\sum_{G,H } \chi^{\rho_1}([G,H])\overline{\chi^{\rho_2}([G,H])} = \delta_{\rho_1,\rho_2}. 
	\label{orthoChar}
\ee

Other important relations which directly descend from the definitions above are the following (proofs are relegated to appendix \ref{app_irrepsandsoon}). 
First of all the fact that $D^{\rho}([G,H])$ actually defines a representation of the Drinfel'd double star product:
\begin{align} 
	D^{\rho}_{I' I}([\widetilde{G},\widetilde{H}]\star [G,H]) \, = \,\sum_{I''} D^{\rho}_{I' I''}([\wG,\wH])D^{\rho}_{I'' I}([G,H])
		\label{defining}
\end{align}
Then, come the orthogonality relations
\be
	\frac{1}{|\mathcal{G}|}\sum_{G,H }D^{\rho_1}_{I'_1 I_1}\big( [G,H]\big)\overline{D^{\rho_2}_{I_2',I_2}\big( [G,H]\big)}
	= \frac{\delta_{\rho_1,\rho_2}}{d_{\rho_1}}\delta_{I_1',I_2'}\delta_{I_1I_2},
	\label{orthoMat}
\ee
as well as the completeness relations
\begin{align}
	\sum_{\rho}\sum_{I',I} d_{\rho}\; D^{\rho}_{I' I}([G,H])\overline{D^{\rho}_{I' I}([\widetilde{G},\widetilde{H}])}
	 = |\mathcal{G}| \delta(G,\widetilde{G})\delta(H,\widetilde{H}) .
 	 \label{completeness}
\end{align}

\subsubsection{Diagonalizing the star-product}

We have now all the ingredients needed to diagonalize the star product in the sense of Equation \eqref{idempo}.
Consider the following change of basis in ${\cal H}_2$:
\begin{align}
\begin{dcases}
 \psi^{\S_2}_{\rho,I'I}  = \frac{1}{|\mG|} \sum_{G,H} \sqrt{d_\rho}\;D^\rho_{I'I}([G,H]) \; \psi^{\S_2}_{G,H}  \\
 \psi^{\S_2}_{G,H} =  \sum_{\rho}\sum_{I',I} \sqrt{d_\rho}\;  \overline{D^\rho_{I'I}([G,H])} \;  \psi^{\S_2}_{\rho,I'I}
\end{dcases}
\end{align}
Then, the new basis $\{ \psi^{\rho}_{I'I} \}$ diagonalizes the star-product: 
\be
 \psi^{\S_2}_{\rho_2,I'_2 I_2} \star  \psi^{\S_2}_{\rho_1,I'_1 I_1 } = \frac{\delta_{\rho_1,\rho_2}}{\sqrt{d_{\rho_1}}}\; \delta_{I_2 , I_1'}\;  \psi^{\S_2}_{\rho_1,I'_2I_1}.
 \label{comultrivia}
\ee
This crucial relation is proven in Appendix \ref{app_diagprod}.
Recall that according to the discussion at the end of the previous section, the importance of such a basis is that it is labeled by the physically `stable' properties of the punctures. 
In other words $\rho=(C,R)$ can be interpreted as the `charge' carried by the puncture.

With a little stretch of the formalism, we could in principle consider $\mG=\SU(2)$.
Then, interpreting the punctures as point particles, $C$ would correspond to the mass of the particle, as measured by the (curvature) conical defect it induces, and $R\in\mathbb N$ would correspond to its spin, i.e. the torsion defect.

\subsubsection{Tensor products of representations and the Clebsch--Gordan series\label{sec_CG}}

In order to consider the tensor product of two representations, it is necessary to make use of the comultiplication $\Delta$ to `redistribute' the Drinfel'd double elements to the various factors:%
\footnote{In the case of the tensor product of representations of a group, the comultiplication is in principle needed as well. However, in this case, it is trivial ($\Delta(g)=g\otimes g$, for $g$ a group element) and passes therefore unnoticed.}
\be
	(D^{C_1,R_1} \otimes D^{C_2,R_2})(\Delta[G,H]) = \sum_{X,Y \in \mathcal{G} \atop XY=H}(D^{C_1,R_1} \otimes D^{C_2,R_2})([G,X]\otimes [G,Y]).
\ee
The fact that the tensor product of representations can be itself decomposed onto irreducible representations, leads to the notion of fusion category.
More precisely, the fusion structure relies on the existence of `fusion rules' of the form
\be
 (C_1,R_1) \otimes (C_2,R_2) = \bigoplus_{C_3,R_3}N^{C_1C_2C_3}_{R_1R_2R_3} \, (C_3,R_3).
\ee 
 with the fusion coefficients $N^{C_1C_2C_3}_{R_1R_2R_3}$ being integers.
If $N^{C_1C_2C_3}_{R_1R_2R_3} \in \{0,1\}$, the fusion category is said to be multiplicity free.
Henceforth, we assume { for notational convenience} that the fusion category of ${\cal D}(\mG)$ is multiplicity free,{ thus avoiding extra multiplicity indices. The subsequent derivations, however, could be easily generalized}. 
A non-zero fusion coefficient signifies the presence of a non-trivial recoupling channel, which translates into the existence of an invariant subspace in the tensor product of the corresponding three representation spaces. 
Using the orthonormality of the characters, we obtain the following expression for the fusion rules \cite{Verlinde}
\begin{align}
	N^{C_1C_2C_3}_{R_1R_2R_3} &= \frac{1}{|\mathcal{G}|}\sum_{G,H \in \mathcal{G}}\text{tr}\big((D^{C_1,R_1} \otimes D^{C_2,R_2})(\Delta[G,H]) \big)
					\overline{\chi^{C_3,R_3}([G,H])} \\
	&=  \frac{1}{|\mathcal{G}|}\sum_{G \in \mathcal{G}}\sum_{H' \in C_3 \atop H \in C_1}\chi^{C_1,R_1}([G,H]) \chi^{C_2,R_2}([G,H^{-1}H'])
					\overline{\chi^{C_3,R_3}([G,H'])} .\label{fusion}
\end{align}

Now, we look for a relation between the matrix elements of the representations $\rho_1 \otimes\rho_2 = (C_1,R_1)\otimes(C_2,R_2)$ and the matrix elements of its irreducible components $\rho_3$.
From our hypothesis, there exists a unitary map $\mathcal{U}^{[\rho_1,\rho_2] }: \oplus_{\rho_3 \in \rho_1\otimes \rho_2} V_{\rho_3}\rightarrow V_{\rho_1} \otimes V_{\rho_2}$ which satisfies the relation
\be\label{CG0}
	D^{\rho_1}_{I_1',I_1} \otimes D^{\rho_2}_{I_2',I_2}(\Delta[G,H])
	=  \sum_{\rho_3}\sum_{I_3, I_3'}
	\mathcal{U}^{[\rho_1,\rho_2]}_{I_1'I_2',\rho_3I_3'}\,
	D^{\rho_3}_{I_3'I_3}([G,H])\,
	({\mathcal{U}^{[\rho_1,\rho_2] \dagger}) _{\rho_3I_3,I_1I_2}},
\ee
where the matrix indices are given by the composed labels $I_1'I_2'$ and $\rho_3I_3'$. 
The map $\mathcal{U}$ corresponds to the analogue of the Clebsch-Gordan coefficients for the Drinfel'd double and therefore we will make use of the following notation
\be
	\mathcal{C}^{\rho_1 \rho_2\rho_3}_{I_1 I_2 I_3} = 
	\mathcal{U}^{[\rho_1 ,\rho_2]}_{I_1 I_2,\rho_3I_3}.
\ee
{ Note that relaxing the multiplicity-free assumption of the fusion category would lead to extra indices for the Clebsch-Gordan coefficients}.
As for the fusion rules, we can use the orthogonality of the representation matrices, to obtain
\begin{align}
 	\frac{1}{|\mathcal{G}|}\sum_{G \atop H_1,H_2 }D^{\rho_1}_{I_1'I_1}([G,H_1]) D^{\rho_2}_{I_2'I_2}([G,H_2]) 
	\overline{D^{\rho_3}_{I_3'I_3}([G,H_1 H_2])}
	 = \frac{1}{d_{\rho_3}}\,
	\mathcal{C}^{\rho_1 \rho_2\rho_3}_{I_1' I_2' I_3'}\;
	\overline{\mathcal{C}^{\rho_1 \rho_2\rho_3}_{I_1 I_2 I_3}} \label{defCGcoeffs},
\end{align}
from which one can compute explicitly the values of the Clebsch-Gordan coefficients (notice that there is an ambiguous overall phase, exactly as in the $\SU(2)$ Clebsch--Gordan coefficient).

From the unitarity of $\mathcal{U}$, it follows the following orthogonality and completeness relation for the Clebsch--Gordan coefficients:
\be
	\sum_{I_1,I_2}
	\mathcal{C}^{\rho_1 \rho_2\rho}_{I_1 I_2 I} \cdot
	\overline{\mathcal{C}^{\rho_1 \rho_2\rho'}_{I_1 I_2 I'}}
	 = \delta_{\rho,\rho'}\delta_{I'I} \label{orthoW3J},
\ee
and
\begin{align}
	\sum_{\rho}\sum_I
\mathcal{C}^{\rho_1 \rho_2\rho}_{I_1' I_2' I} \cdot
\overline{\mathcal{C}^{\rho_1 \rho_2\rho}_{I_1 I_2 I}} 
= \delta_{I_1'I_1}\delta_{I_2',I_2} .
	\label{resolutionCC}
\end{align}
(One can easily prove this equation---see Appendix \ref{app_complete}---using the completeness of the $D^\rho_{I'I}$.)

Furthermore from the defining equation (\ref{CG0}) one can derive the following invariance property of the Clebsch-Gordan coefficients
\ba\label{CGinvariance}
\sum_{H_1,H_2} \sum_{I_1,I_2,I_3} 
D^{\rho_1}_{I_1'I_1}([G,H_1])   D^{\rho_2}_{I_2'I_2}([G,H_2]) \overline{ D^{\rho_3}_{I'_3 I_3}([G, H_1H_2])  }
\,\,\mathcal{C}^{\rho_1 \rho_2\rho_3}_{I_1 I_2 I_3} &=&
\mathcal{C}^{\rho_1 \rho_2\rho_3}_{I'_1 I'_2 I'_3}.
\ea
This is shown in Appendix \ref{app_invar}. 
Note that in this formula the summations over $H_1$ and $H_2$ (not present in the analogue formula for a group instead of a double) have the following origin: one sum over $H_1$ and $H_2$ restricted to $H_1H_2=H$ fixed, comes from the coproduct $\Delta(H)$, while the sum over all possible $H$ is there because we are considering the Drinfel'd double identity element $\mathbb{I}=\sum_{H} [e,H]$.

We can now use this invariance property to show that the Clebsch-Gordan coefficients automatically implement both flatness and gauge invariance. Consider the following contraction of the Clebsch-Gordan coefficients
\begin{align}\label{CGconstraint0}
	&\sum_{I_1,I_2,I_3} 
	D^{\rho_1}_{I_1'I_1}([G_1,H_1])   D^{\rho_2}_{I_2'I_2}([G_2,H_2]) 
	\,\mathcal{C}^{\rho_1 \rho_2\rho_3}_{I_1 I_2 I_3}  \, D^{\rho_3}_{I_3 I'_3}([G_3, H_3])   \nn\\ &\q =
	\sum_{\tilde H_1, \tilde H_2}\sum_{I_1,I_2,I_3} \sum_{I''_1,I''_2,I''_3} 
	D^{\rho_1}_{I_1'I_1}([G_1,H_1])   D^{\rho_2}_{I_2'I_2}([G_2,H_2])  D^{\rho_3}_{I_3 I'_3}([G_3, H_3])  \nn\\
	&\q\q \q\q \times  D^{\rho_1}_{I_1I''_1}([G,\tilde H_1])   D^{\rho_2}_{I_2I''_2}([G,\tilde H_2]) \overline{ D^{\rho_3}_{I_3 I''_3}([G, \tilde H_1\tilde H_2])  }
	\,\mathcal{C}^{\rho_1 \rho_2\rho_3}_{I''_1 I''_2 I''_3} \,,
\end{align}
where on the RHS we used equation \eqref{CGinvariance}, which holds for any $G \in {\cal G}$. The summations over $I_1,I_2,I_3$ implement the $\star$-multiplication in the Drinfel'd double, leading to three delta functions. Two of these can be solved for $\tilde H_1$ and $\tilde H_2$, thus resolving also the sum over these group elements. We finally obtain
\begin{align}\label{CGconstraint}
	&\sum_{I_1,I_2,I_3} 
	D^{\rho_1}_{I_1'I_1}([G_1,H_1])   D^{\rho_2}_{I_2'I_2}([G_2,H_2]) 
	\,\mathcal{C}^{\rho_1 \rho_2\rho_3}_{I_1 I_2 I_3}  \, D^{\rho_3}_{I_3 I'_3}([G_3, H_3])   \nn\\ & \q =
	\sum_{I_1,I_2,I_3} 
	\delta(H_3, G_1^{-1} H_1 G_1 G_2^{-1} H_2 G_2) \nn\\
	&\q \q \q \times 
	D^{\rho_1}_{I_1'I_1}([G_1G,H_1])   D^{\rho_2}_{I_2'I_2}([G_2G,H_2]) 
	\,\mathcal{C}^{\rho_1 \rho_2\rho_3}_{I_1 I_2 I_3}  \, D^{\rho_3}_{I_3 I'_3}([G^{-1} G_3,  G^{-1} H_3 G]),
\end{align}
with $G\in\mG$ an arbitrary group element.
When we will use the Clebsch-Gordan coefficients to construct the fusion basis, the ``extra'' delta function on the RHS of equation \eqref{CGconstraint} will ensure flatness of the state, while the fact that equation \eqref{CGconstraint} holds for arbitrary $G$ shows the gauge invariance of the construction. 

This concludes the set of preliminaries that we needed before getting to the core of the paper.

\section{The fusion basis}\label{sec_fusionbasis}

In this section, we make use of the notions introduced previously to construct a new basis for the Hilbert space ${\cal H}_{p}$.
The idea is to label the punctures by its physical charges $\rho=(C,R)$, and to use the recoupling theory of ${\cal D}(\mG)$ to `put these charges together' into singlet states on $\S_p$. 
The result of this construction is a basis with a direct physical interpretation, which mathematically resembles a spin--network basis where $\mG$ has been replace by $\mathcal{D}(\mathcal{G})$.
Thanks to the use of recoupling theory at the level of the defect charges, this basis will also trivialize the notion of merging---or coarse--graining---defects.
Heuristically, we can imagine the merging of defects by replacing two punctures by a single one defined by a disc containing the two puncture--discs  to be merged:
\begin{align}
	\begin{array}{c}\includegraphics[scale =1]{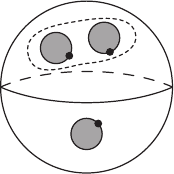}\end{array} \rightarrow
	\begin{array}{c}\includegraphics[scale =1]{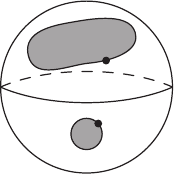}\end{array}
\end{align} 
In practice, such a merging will be realized by performing a fusion of the corresponding irreducible representations.
For this reason, we will refer to this basis as the fusion basis and label the corresponding basis states with an $\mathfrak{f}$.
Since any surface $\S_p$ can be decomposed into trinions (a.k.a. `pair of pants'), it is only necessary to define the states $\psi^{\S_2}_{\mathfrak{f}}$ and $\psi^{\S_3}_{\mathfrak{f}}$,  as well as a procedure  to glue them to one another.

\subsection{The two-punctured sphere}

In the case of $\S_2$, we have already introduced the fusion basis (although without naming it this way).
Consider the $\{\psi^{\S_2}_{G,H}\}$ basis of ${\cal H}_2$, 
\be
\psi^{\S_2}_{G,H} (g_1, \dots, g_4) = |\mG|^{3/2} \delta(G, g_3 g_2 g_1) \delta(H, g_3 g_4 g_2^{-1} g_3^{-1}).
\ee
Then, the following change of basis defines the fusion basis $\{\psi^{\S_2}_{\frak f} [\rho, I'  I] \}$:
\begin{align}
\begin{dcases}
 \psi^{\S_2}_{\frak f}[\rho,I'I]  = \frac{1}{|\mG|} \sum_{G,H} \sqrt{d_\rho}\;D^\rho_{I'I}([G,H]) \; \psi^{\S_2}_{G,H}  \label{fusbasis}\\
 \psi^{\S_2}_{G,H} =  \sum_{\rho}\sum_{I',I} \sqrt{d_\rho}\;  \overline{D^\rho_{I'I}([G,H])} \;  \psi^{\S_2}_{\frak f}[\rho,I'I]\\
\end{dcases}
\end{align}
With the above normalizations, the fusion basis can be shown to be orthonormal in ${\cal H}_2$:
\begin{align} \nn
	&\big\langle \psi^{\S_2}_{\mathfrak{f}}[\rho,I'I]\, , \,  \psi^{\S_2}_{\mathfrak{f}}[\widetilde{\rho},\tilde{I}'\tilde{I}]\big\rangle \\
	&\q = \frac{1}{|\mathcal{G}|^6}\sqrt{d_{\rho}d_{\widetilde{\rho}}}\sum_{g_1,\dots,g_4}
	\sum_{G,H \atop \widetilde{G},\widetilde{H}}\overline{\psi^{\S_2}_{G,H}(g_1,\dots,g_4)}\psi^{\S_2}_{\widetilde{G},\widetilde{H}}(g_1,\dots,g_4)
	\overline{D^{\rho}_{I'I}([G,H])}D^{\widetilde{\rho}}_{\tilde{I}'\tilde{I}}([\widetilde{G},\widetilde{H}])\nn  \\
	&\q = \delta_{\rho,\widetilde{\rho}}\delta_{I\tilde{I}}\delta_{I'\tilde{I}'}
	\label{orthofus2}
\end{align}
The calculation above uses the explicit form of $\psi^{\S_2}_{G,H}$ and the orthogonality relation \eqref{orthoMat}.

Similarly, one can explicitly show that the basis is complete in ${\cal H}_2$: 
\begin{align}
	&\sum_{\rho}\sum_{I'I}\psi^{\S_2}_{\mathfrak{f}}[\rho,I'I](\{g\}) \overline{\psi^{\S_2}_{\mathfrak{f}}[\rho,I',I](\{\widetilde{g}\})} \nn \\
	& \q = |\mathcal{G}|^2 \delta(g_3g_2g_1,\widetilde{g}_3\widetilde{g}_2\widetilde{g}_1)\delta(g_3g_4g_2^{-1}g_3^{-1},\widetilde{g}_3\widetilde{g}_4\widetilde{g}_2^{-1}\widetilde{g}_3^{-1}),
\end{align}
where once more use was made of the explicit form of $\psi^{\S_2}_{G,H}$, as well as of the completeness \eqref{completeness}.\footnote{ A generalization of this construction to cylinders with multiple marked points at the punctures is provided in appendix \ref{app_mpuncture}.}

\subsection{The three-punctured sphere}

Using the Clebsch-Gordan coefficients which play the role of intetwiners between the irreducible representations of $\mathcal{D}(\mathcal{G})$, one can now construct the fusion basis states for the 3-punctured sphere.
Once again, we start from the basis in the $[G,H]$--picture.
After gauge--fixing, a basis of ${\cal H}_{\S_3}$ is given by 
\begin{align}\label{3pstate1}
\psi^{\S_3}_{G_1,H_1;G_2,H_2}(\{g\},\{g'\})|_\text{g.f.} \, = \, |\mG|^3 \cdot
	 \begin{array}{c}\includegraphics[scale =1]{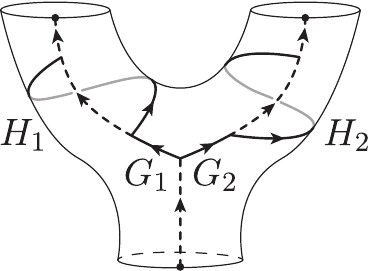}\end{array}  = 
	\psi^{\S_2}_{G_1,H_1}(\{g\})|_\text{g.f.}
	\psi^{\S_2}_{G_2,H_2}(\{g'\})|_\text{g.f.}.
\end{align} 
where we borrowed the notation of Equation \eqref{basisgroup}.
The definition on a non--gauge--fixed state is readily recovered by reintroducing the other group elements and averaging over the gauge action at the five internal nodes.
Now, we perform the transformation to the $[\rho, I'I]$--picture on each of the $\S_2$ factors:
\begin{align}
\psi^{\S_3}_{\rho_2,I'_2 I_2 ;\rho_1, I'_1 I_1}
\, = \,\frac{1}{|\mG|^2} \sum_{G_1,H_1 \atop G_2, H_2} 
\sqrt{d_{\rho_2} d_{\rho_1} }\; 
 {D^{\rho_2}_{I'_2 I_2 }([G_2,H_2 ])} \;
  {D^{\rho_1}_{I'_1 I_1}([G_1 ,H_1 ])}\;
\psi^{\S_3}_{G_1,H_1;G_2,H_2}.
\end{align} 
where we have suppressed from our notation the dependence on $\{g,g'\}$.
In this basis, the charges at the ``top'' punctures 1 and 2 are fixed to $\rho_1$ and $\rho_2$.
However, little can be said for what concerns the charge of the bottom puncture.
Moreover, we are left with `magnetic' indices of ${\cal D}(\mG)$, $I_1$ and $I_2$, associated with the ``bottom'' of the two cylinders we are gluing, which---from the trinion perspective---'sit' in the very middle of the graph.
What is needed, is therefore a unitary transformation that maps two magnetic indices into one representation label (the charge of the third puncture) and one magnetic index (now sitting at the bottom of the graph).
This is exactly the job of the Clebsch--Gordan coefficients constructed in section \ref{sec_CG}.
Hence, we finally define the fusion basis of ${\cal H}_{\S_3}$ as
\begin{align}\label{deffus48}
\begin{dcases}
\psi^{\S_3}_{\frak f}
	 { \scriptsize\begin{bmatrix} \rho_1,I_1'\\ \rho_2,I_2'\\ \rho_3,I_3\end{bmatrix}}
&\,= \,\frac{1}{|\mG|^2} \sum_{I_1, I_2} \sum_{G_1,H_1 \atop G_2, H_2} 
 \sqrt{d_{\rho_2} d_{\rho_1} }\;
 \mathcal{C}^{\rho_1 \rho_2\rho_3}_{I_1 I_2 I_3}\;
 {D^{\rho_2}_{I'_2 I_2 }([G_2,H_2 ])} \;
  {D^{\rho_1}_{I'_1 I_1}([G_1 ,H_1 ])}\;
\psi^{\S_3}_{G_1,H_1;G_2,H_2}\\
\psi^{\S_3}_{G_1,H_1;G_2,H_2}
&\,= \, \sum_{\rho_3, I_3} \sum_{\rho_1,I_1 I'_1 \atop \rho_2, I_2  I'_2 } 
 \sqrt{d_{\rho_2} d_{\rho_1} }\;
 \overline{\mathcal{C}^{\rho_1 \rho_2\rho_3}_{I_1 I_2 I_3}}\;
 \overline{{D^{\rho_2}_{I'_2 I_2 }([G_2,H_2 ])}} \;
  \overline{{D^{\rho_1}_{I'_1 I_1}([G_1 ,H_1 ])}}\;
\psi^{\S_3}_{\frak f} { \scriptsize\begin{bmatrix} \rho_1,I_1'\\ \rho_2,I_2'\\ \rho_3,I_3\end{bmatrix}}
\end{dcases}
\end{align} 
To prove the consistency of the two formulas above, it is sufficient to use the orthogonality and completeness of both $ \mathcal{C}^{\rho_1 \rho_2\rho_3}_{I_1 I_2 I_3}$ and ${D^{\rho}_{I' I}([G ,H  ])}$. That is equations \eqref{orthoMat}, \eqref{completeness}, \eqref{orthoW3J} and \eqref{resolutionCC}.

The  orthonormality of the basis,
\begin{align}\nn
\Bigg\langle \psi^{\S_3}_{\mathfrak{f}}
	{ \scriptsize\begin{bmatrix}
		\rho_1,I_1' \\ \rho_2,I_2' \\ \rho_3,I_3
	\end{bmatrix}},
	\psi^{\S_3}_{\mathfrak{f}}
	{ \scriptsize\begin{bmatrix}
		\widetilde{\rho}_1\tilde{I}_1' \\ \widetilde{\rho}_2,\tilde{I}_2' \\ \widetilde{\rho}_3,\tilde{I}_3
	\end{bmatrix}} \Bigg\rangle 
 = \, \delta_{\rho_1,\widetilde{\rho}_1}\delta_{\rho_2,\widetilde{\rho}_2}\delta_{\rho_3,\widetilde{\rho}_3}\delta_{I_1',\tilde{I}_1'}\delta_{I_2',\tilde{I}_2'}\delta_{I_3,\tilde{I}_3},
\end{align} 
is proved in Appendix \ref{app_orthostates}, while its completeness follows from the completeness of the basis $\{\psi^{\S_3}_{G_1,H_1; G_2, H_2}\}$ and the change of basis above.

\subsection{States on $\S_p$}\label{fusionB}

 Here we define the fusion basis states $\psi_{\mathfrak{f}}^{\S_p}$ for the $p$-punctured sphere by generalizing the construction followed for the case of the three--punctured sphere. In particular, this means that the $\S_2$ factors associated with $p-1$ punctures are transformed to the $[\rho',I'I]$-picture and then a fusion tree is constructed by contracting Clebsch-Gordan coefficients together. In the following subsection, we will present how such states can be recovered by gluing states defined on three-punctured spheres as outlined at the beginning of this section. 

To make the construction more transparent, we introduce a more synthetic graphical notation. Since all the operations defined on the fusion basis states can be performed at the level of the representations, it is not necessary to look at the group variables $\{g\}$ in details.
Therefore we will represent the fusion basis state $\psi^{\S_3}_{\mathfrak{f}}$ as follows 
\begin{align}
	\psi^{\S_3}_{\mathfrak{f}}{ \scriptsize\begin{bmatrix}
		\rho_1, I_1' \\ \rho_2,I_2' \\ \rho_3,I_3
	\end{bmatrix}}=
	 \begin{array}{c}\includegraphics[scale =1]{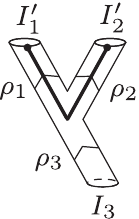}\end{array}
	\label{newrepS3}  . 
\end{align}
The tube structure provides the following combinatorial information:
\be
	 \begin{array}{c}\includegraphics[scale =1]{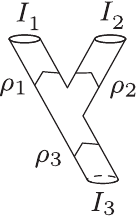}\end{array} = 
	\mathcal{C}^{\rho_1,\rho_2,\rho_{3}}_{I_1,I_2,I_3}
\ee
while the bold edge signals where the group variables are inserted (see equation \eqref{3pstate1}), i.e. in a sense where we consider the degrees of freedom to be.
By this we mean that the lower cylinder in (\ref{newrepS3}) ``does not carry any degree of freedom'' because the flatness constraint tells us that a complete knowledge of the state is encoded in the knowledge of the upper tubes only. Nevertheless, it is possible to use the expression for the flatness constraint in order to rewrite the states on the three-punctured sphere in a more symmetric form (see equation \eqref{3punct-sym}), in which each tube is associated to a representation matrix.

To obtain the states on $\S_p$, we first perform the transformation to the $[\rho',I'I]$-picture using equation \eqref{fusbasis} on each of the $p-1$ upper tubes respectively associated to $p-1$ punctures. The upper tubes are then connected to each other two by two via Clebsch-Gordan coefficients so as to form a fusion tree. Using the graphical notation introduced above, the resulting states are given by
\begin{align} \nn
	\psi_{\mathfrak{f}}^{\S_p} \big[ \{\rho_i\}_{i=1}^{2p-3},\{I'_k\}_{k=1}^{p-1},I_p \big]&=
	 \begin{array}{c}\includegraphics[scale =1]{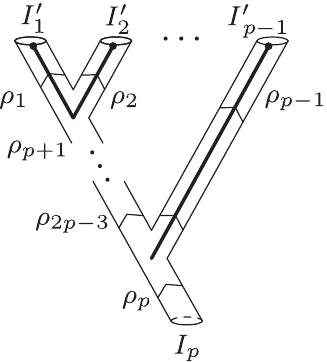}\end{array} \\	\nn
	&= \frac{1}{|\mG|^{p-1}}\sum_{\{I\}}\sum_{\{G_k,H_k\}_{k=1}^{p-1}}
	\mathcal{C}^{\rho_1\rho_2\rho_{(p+1)}}_{I_1I_2I_{(p+1)}} 				 
	\mathcal{C}^{\rho_{(2p-3)}\rho_{(p-1)}\rho_p}_{I_{(2p-3)}I_{(p-1)}I_p} 
	\; \psi^{\S_p}_{\{G,H\}}\\
	& \q \q \q  \times \prod_{k=1}^{p-1}\sqrt{d_{\rho_k}}D^{\rho_k}_{I_k'I_k}([G_k,H_k])
	\prod_{i=p+1}^{2p-4}\mathcal{C}^{\rho_i\rho_{(i-p+2)}\rho_{(i+1)}}_{I_iI_{(i-p+2)}I_{(i+1)}}\, .
\end{align}
The subindex $i\in\{1,\dots,2p-3\}$ labels the edges of the fusion tree. The subindex $k=\{1,\dots,p\}$ labels the $p$ punctures, which are in one--to--one correspondence with the leaves of the fusion tree. The orthonormality of these states is proven in Appendix \ref{app_orthostates}.

It is important to recall that the CG coefficients of $\mathcal{D}({\cal G})$, $\mathcal{C}^{\rho_1\rho_2\rho_{3}}_{I_1 I_2 I_3}$, are not symmetric in all its indices, and $\{\rho_3,I_3\}$ actually play a distinguished role (see Equation \eqref{defCGcoeffs}). 
Therefore the above graphs are directed. 
Different choices of root trees defining the states above are related by a change of basis, as it is most easily seen by going back to a group representation.

Notice that this is just the simplest example of fusion tree.
More refined construction can be built in a similar way, possibly with the idea in mind of reproducing the multi--scale design underlying the tensor network states. For this we refer to section \ref{sec_coarsegraining}

\subsection{Fusion basis via gluing}

As we mentioned in the beginning of this section, every punctured-sphere can be decomposed into trinions so that the fusion basis state for $\S_p$ boils down to a ``gluing'' of states defined on three-punctured spheres.

To start with we want to represent the fusion basis for the three--punctured sphere in a more symmetric manner.  To this end we use the equivalence relations in section \ref{sec_alBF} and  express the sate (\ref{3pstate1}) on an extended graph
\begin{align}\label{extgr1}
	\psi^{\S_3}_{G_1,H_1;G_2,H_2}(\{g\},\{g'\},\{g''\})|_\text{g.f.} &= |\mG|^{7/2}  \sum_{G_3,H_3}
	 \begin{array}{c}\includegraphics[scale =1]{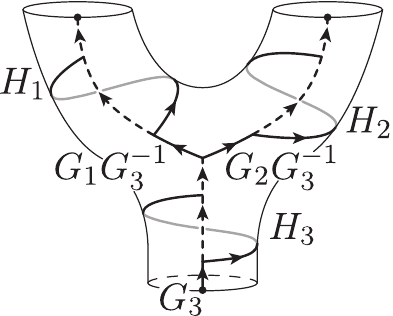}\end{array} \\
	&=  |\mG|^{7/2} 
	\sum_{G_3,H_3} \delta(g_4,H_1) \delta(g_1,G_1G_3^{-1}) \delta(g'_4,H_2) \delta(g'_1,G_2G_3^{-1})  \nn\\ \nn
	&\q \q \times  \delta(g''_4,H_3) \delta(g''_1, G_3) \,
	\delta(G_3^{-1}H_3G_3, G_1^{-1} H_1 G_1 G_2^{-1} H_2 G_2) \, .
\end{align}

Now the following identity (equation \eqref{CGconstraint2}) can be obtained from equation (\ref{CGconstraint}), which spells out the relation of the Clebsch-Gordan coefficients to the flatness and Gau\ss~constraints, by first setting $G_3=e$ and then summing over $H_3 \in {\cal G}$, that is by evaluating it on the Drinfel'd double identity $\mathbb{I} = \sum_{H_3} [e,H_3]$:
\begin{align}\label{CGconstraint2}
	&\sum_{I_1,I_2,I_3} 
	D^{\rho_1}_{I_1'I_1}([G_1,H_1])   D^{\rho_2}_{I_2'I_2}([G_2,H_2]) 
	\,\mathcal{C}^{\rho_1 \rho_2\rho_3}_{I_1 I_2 I'_3}  \\ \nn & \q =
	\sum_{I_1,I_2,I_3} 
	D^{\rho_1}_{I_1'I_1}([G_1G^{-1},H_1])   D^{\rho_2}_{I_2'I_2}([G_2G^{-1},H_2]) 
	\,\mathcal{C}^{\rho_1 \rho_2\rho_3}_{I_1 I_2 I_3}  D^{\rho_3}_{I_3 I'_3}([G ,  G G_1^{-1} H_1 G_1 G_2^{-1} H_2 G_2 G^{-1}]).
\end{align}
Using equation (\ref{extgr1})  for the expression of $\psi^{\S_3}_{G_1,H_1;G_2,H_2}$ on an extended graph, we can express the fusion state as
\begin{align}
\psi^{\S_3}_{\frak f}
	 { \scriptsize\begin{bmatrix} \rho_1,I_1'\\ \rho_2,I_2'\\ \rho_3,I'_3\end{bmatrix}}_{|{\rm g.f}}
	&= \frac{1}{|\mG|^2} \sum_{I_1, I_2} \sum_{G_1,H_1 \atop G_2, H_2} 
	\sqrt{d_{\rho_2} d_{\rho_1} }\;
	\mathcal{C}^{\rho_1 \rho_2\rho_3}_{I_1 I_2 I'_3}\;
	{D^{\rho_2}_{I'_2 I_2 }([G_2,H_2 ])} \;
 	{D^{\rho_1}_{I'_1 I_1}([G_1 ,H_1 ])}\;
	{\psi^{\S_3}_{G_1,H_1;G_2,H_2}}_{|{\rm g.f}} \nn\\
	&=
	|\mG|^{3/2} \!\!\!\sum_{I_1, I_2,I_3}\!\!\! \sum_{G_1,H_1 \atop G_2, H_2,G_3,H_3}\!\!\!\!\!\!
	\sqrt{d_{\rho_2} d_{\rho_1} }\;
	\mathcal{C}^{\rho_1 \rho_2\rho_3}_{I_1 I_2 I_3}\;
	{D^{\rho_2}_{I'_2 I_2 }([G_2G_3^{-1},H_2 ])} \;
	{D^{\rho_1}_{I'_1 I_1}([G_1 G_3^{-1} ,H_1 ])}\;  \nn\\  
	& \q\q\q\q \times {D^{\rho_3}_{I_3 I'_3}([G_3 ,H_3 ])}\; 
	\delta(g_4,H_1) \delta(g_1,G_1G_3^{-1}) \delta(g'_4,H_2) \delta(g'_1,G_2G_3^{-1})  \nn\\
	&\q\q\q\q \times  \delta(g''_4,H_3) \delta(g''_1, G_3) \,\,
	\delta(G_3^{-1}H_3G_3, G_1^{-1} H_1 G_1 G_2^{-1} H_2 G_2)\,. \nn\\
\end{align}
We first translate the summation variables $G_1 \rightarrow G_1 G_3$ and $G_2 \rightarrow G_2 G_3$, then we apply identity (\ref{CGconstraint}) again (this time with $G=e$), hence ``reabsorbing'' the delta function into the Clebsch-Gordan coefficient. In this way, we finally arrive at the following representation of the fusion state on ${\mathbb S}_3$:
\ba
	\label{3punct-sym}
	\psi^{\S_3}_{\frak f}
	 { \scriptsize\begin{bmatrix} \rho_1,I_1'\\ \rho_2,I_2'\\ \rho_3,I'_3\end{bmatrix}}_{|{\rm g.f}}
	&=&
	|\mG|^{3/2} \!\!\!\sum_{I_1, I_2,I_3}\!\!\! \sum_{G_1,H_1 \atop G_2, H_2,G_3,H_3}\!\!\!\!\!\!
	\sqrt{d_{\rho_2} d_{\rho_1} }\;
	\mathcal{C}^{\rho_1 \rho_2\rho_3}_{I_1 I_2 I_3}\;
	{D^{\rho_2}_{I'_2 I_2 }([G_2,H_2 ])} \;
 	{D^{\rho_1}_{I'_1 I_1}([G_1 ,H_1 ])}\;
	{D^{\rho_3}_{I_3 I'_3}([G_3 ,H_3 ])}\; \nn\\
 && \q\q\q \q \times \delta(g_4,H_1) \delta(g_1,G_1) \delta(g'_4,H_2) \delta(g'_1,G_2)   \delta(g''_4,H_3) \delta(g''_1, G_3) \; .
\ea
We have thus obtained a more symmetric representation of the fusion state on the three--punctured sphere. Note however that the dimension factors $d_\rho$ are still not equally distributed, which is due to an asymmetry in the Clebsch-Gordan coefficients.

The fusion basis state on the three--punctured sphere is now expressed such that each leg carries a state that is locally equivalent to a cylinder state. We know how these states behave under gluing  and thus we can now proceed to build a fusion state on the e.g. four--punctured sphere by gluing two three--punctured sphere fusion states: 
\begin{align}
	\psi^{\S_4}_{\frak f}[ \{\rho_i\}_{i=1}^5, \{I'\}_{i=1}^3,I_4]
	&=  \begin{array}{c}\includegraphics[scale =1]{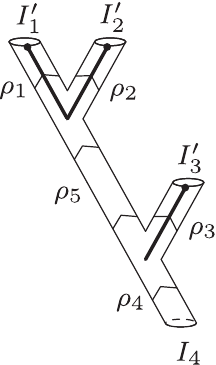}\end{array} = 
	 \, \sqrt{d_{\rho_5}} \,
	 \begin{array}{c}\includegraphics[scale =1]{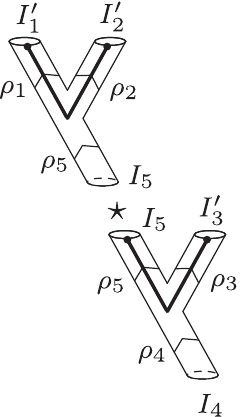}\end{array}  \\
	&=\, \sqrt{d_{\rho_5}} \,\,
	\psi^{\S_3}_{\frak f}
	 { \scriptsize\begin{bmatrix} \rho_1,I_1'\\ \rho_2,I_2'\\ \rho_5,I_5\end{bmatrix}} \star \psi^{\S_3}_{\frak f}
	 { \scriptsize\begin{bmatrix} \rho_5,I_5 \\ \rho_3,I_3'\\ \rho_4,I_4\end{bmatrix}} \nn\\
	&=	 \frac{1}{ \sqrt{d_{\rho_5}}} \sum_{I_5} 
	 \psi^{\S_3}_{\frak f}
	 { \scriptsize\begin{bmatrix} \rho_1,I_1'\\ \rho_2,I_2'\\ \rho_5,I_5\end{bmatrix}} \star \psi^{\S_3}_{\frak f}
	 { \scriptsize\begin{bmatrix} \rho_5,I_5 \\ \rho_3,I_3'\\ \rho_4,I_4\end{bmatrix}}  \,,
\end{align}
where  in the first line we used that for the gluing of cylinder states the ``glued'' indices $I$ have to coincide but drop out in the final result (see equation \eqref{comultrivia}), while in the last line we summed over this index and included the corresponding normalization factor.


\subsection{Gauge invariant projections of the fusion basis}\label{Pfusion}

The fusion basis  $\psi^{\mathbb{S}_p}_{\mathfrak{f}}$ describes both curvature and torsion excitations at the punctures. Often we are interested in having only curvature excitations, that is states that are also gauge invariant at the punctures.  We can obtain such states by applying the Gau\ss~constraint projector ${\mathbb P}_{\rm gauge}$ to the punctures. 

Let us for example consider a fusion basis state on a cylinder
\ba
\psi^{\mathbb{S}_2}_{\mathfrak{f}}[CR,i'M',iM](g_1,\cdots,g_4)&=&
|{\cal G}|^{1/2} \sum_{G,H} \sqrt{d_{C,R}}\,\, \delta(H,c_{i'}) \, \delta(c_{i'}, Gc_iG^{-1}) \, D^R_{M'M}(q^{-1}_{i'}Gq_i)\, \nn\\&&
\q\q\times \delta(G,g_3g_2g_1) \delta(H,g_3g_4g_2^{-1}g_3^{-1})
\ea
and apply ${\mathbb P}_{\rm gauge}$ to the source node of the link carrying $g_1$, i.e. at the source puncture of the cylinder state. 
\ba
{\mathbb P}^{(s)}_{\rm gauge}\,\,\psi^{\mathbb{S}_2}_{\mathfrak{f}}[CR,i'M',iM]&=&
|{\cal G}|^{-1/2} \sum_{h}\sum_{G,H} \sqrt{d_{C,R}}\,\, \delta(H,c_{i'}) \, \delta(c_{i'}, Gc_iG^{-1}) \, D^R_{M'M}(q^{-1}_{i'}Gq_i)\, \nn\\&&
\q\q\q \times \delta(G,g_3g_2g_1h) \delta(H,g_3g_4g_2^{-1}g_3^{-1})  .
\ea
One finds (see appendix \ref{PfusionA})
\ba
{\mathbb P}^{(s)}_{\rm gauge}\,\,\psi^{\mathbb{S}_2}_{\mathfrak{f}}[CR,i'M',iM]&=&
 \delta_{R,0 } \delta_{M'0} \delta_{M0}  \frac{1}{|Q_C|}\sum_{j}\psi^{\mathbb{S}_2}_{\mathfrak{f}}[C0,i'0,j0]   .
\ea
Likewise, applying ${\mathbb P}_{\rm gauge}$ to the target puncture we find
\ba
{\mathbb P}^{(t)}_{\rm gauge}\,\,\psi^{\mathbb{S}_2}_{\mathfrak{f}}[CR,iM,i'M']&=&
 \delta_{R,0 } \delta_{M'0} \delta_{M0}  \frac{1}{|Q_C|}\sum_{j'}\psi^{\mathbb{S}_2}_{\mathfrak{f}}[C0,j'0,i0]   .
\ea
Note that the gauge averaged states have now norm equal to $1/|Q_C|$, to get normalized state we should multiply by $\sqrt{|Q_C|}$.

This generalizes to the fusion basis for $p$--punctured spheres: applying a gauge averaging at a puncture $p$ forces the corresponding labels $R_p$ and $M_p$ to be trivial and leads to an averaging over the $i_p$ index.

\section{Ribbon operators \label{sec_ribbons} }

In the previous section we have introduced the fusion basis, which gives immediate access to the excitation structure of a state.
We are now going to construct operators that generate and measure these excitations.
For reasons that will be clear soon, these operators are called `ribbon operators'.
They come in two families: open ribbon operators, which generate excitations, and closed ribbon operators, which measure them.
In particular, we will see that we can define operators that are diagonal in the fusion basis.

\subsection{ Open ribbon operators \label{oribbon}}

Choosing as our configuration space group holonomies, that describe locally flat connections, we have at our disposal two types of operators.
On the one hand, multiplication operators, known as holonomy or Wilson path (loop) operators; 
on the other hand, translation operators, which translate an argument of the wave function either on the left or on the right.

Wilson path operators, $W^f_\gamma$, multiply wave functions by an $f:{\cal G}\to\mathbb C$, 
\ba
(W^f_\gamma \psi ) (g_1, \ldots, g_L) \,=\, f( h_\gamma) \psi(g_1, \ldots, g_L),
\ea 
where $h_\gamma = g_{l_N} \cdots g_{l_1}$ is the holonomy associated to the path $\gamma= l_N \circ \cdots \circ l_1$ (clearly, care must be taken with respect to the orientation of the links).
Being a multiplication operator, $W^f_\gamma$ preserves any flatness constraints, which are multiplication operators themselves. 
Gauge invariance (i.e. Gau\ss~constraints) is preserved only if $\gamma$ is a loop and $f$ a class function. 

Translation operators $T_k[H]$ act by finite translations, and can therefore be thought of as the exponentiated version of momenta.
In loop quantum gravity momenta are known as fluxes, hence the name of `exponentiated flux'.
On a group translations can act either on the left or on the right. 
We choose to work with left multiplication:\footnote{Right translation can be implemented by $T_k[g_k H g_k^{-1}]$.}
 \ba
 (T_k[H]\psi)( \ldots, g_k, \ldots)  \,=\, \psi(\ldots, H^{-1} g_k, \ldots) .
\ea
Note, however, that $T_k[H]$ in general violates all flatness constraints involving the group element $g_k$ carried by the link $l_k$, as well as the Gau\ss~constraints at the target node of $l_k$, $n=t(l_k)$. 
Thus, this operator in general takes a state out of its definition Hilbert space, ${\cal H}_{\S_p}$, and is therefore not viable as it is.
{ Hence, we need to} adjust the definitions of the above operators to correct this issue.
Before doing this, however, we need to understand the structure of constraint violations their action induces.

A translation of---say---the group element associated to  the link $l_1$ will change the holonomies of the two faces---say---$f_1$ and $f_2$ which are adjacent to $l_1$. 
Now, by changing in a precise way also the holonomy associated to another link---say---$l_2 \in f_2$, it is clear that we can re-gain flatness at $f_2$. 
Nevertheless, this comes in general at the cost of changing the holonomy of a third face $f_3$, and so on.
The argument can be used to push around Gau\ss~constraint violations as well. To do so, we can first parallel transport the argument $g_k$ which is about to be translated from its target node to another node $n$ along a path $\gamma$. Once the translation is performed, we then transport back the translated holonomy. The resulting operators are denoted by $T_{k,\gamma}[H]$ and their action reads
\be
	(T_{k,\gamma}[H]\psi)(\ldots,g_k,\ldots) = \psi (\ldots, h_{\gamma}^{-1}H^{-1}h_{\gamma}g_k, \ldots)
\ee
with $h_{\gamma}$ the holonomy along the path $\gamma$.  Notice how $h_\gamma$ involves an implicit dependence on all the group elements $g_l$ corresponding to links $l \in \gamma$.

What we actually learn from this discussion is that curvature excitations and Gau\ss~constraint violations are always generated in pairs. Now, recall that---by construction---punctures are locations in $\S_p$ where constraint violations are allowed.
Therefore, we are led to considering operators that generate pairs of excitations whose positions coincide with a pair of punctures.%
\footnote{Operators generating curvature defects at the end of a certain path have been defined in \cite{DG14a,DG14b,BDG15} as (integrated) exponentiated flux operators.}
Also, Wilson path operators $W^f_\gamma$, which are associated to open paths $\gamma$, generate defects in pairs.
In this case the defects are Gau\ss~constraint violations that appear at the two ends of the Wilson path. 
Again, such violations are allowed if the Wilson path starts and ends at punctures.

\subsubsection{Kitaev's ribbon operators}

Kitaev, in \cite{Kitaev1}, combined translation operators and Wilson path operators into so--called ribbon (or dyonic) operators. 
He also showed that ribbon operators carry an algebraic structure given by the Drinfel'd double of the underlying (discrete) gauge group. 

We first define ribbon operators on $\S_2$ (with a minimal graph), and generalize to more general punctured surfaces in a second moment.
We will show that ribbon operators generate the basis $\{\psi^{\S_2}_{G,H}\}$ (Equation \eqref{basisgroup}) of ${\cal H}_2$, thus revealing already a connection to the Drinfel'd double ${\cal D}(\mG)$.  \\

~\\
{\bf {Kitaev's ribbon on $\S_2$}}\\
Kitaev's ribbon operator on $\S_2$ is defined as the combination of a translation and a Wilson path operator.

The translation operator acts at the link $l_4$, going `around' the cylinder.  
The translating element is parallel transported to the target puncture, i.e. the target node of $l_3$.
We write this 
\ba
(T_{4,3}[H] \psi^{\S_2})(g_1,\ldots,g_4) \,=\, \psi^{\S_2}(g_1,\ldots, g_3^{-1} H^{-1} g_3g_4 )  .
\ea
After the action of $T_{4,3}[H]$ the inner vertices remain gauge invariant. 
The Wilson path operator involves the `longitudinal' holonomy in between the two punctures, and it is characterized by a function $f$ {which acts by multpilication. A basis for these operators is provided by delta functions $\{\delta(G,\cdot)\}_{G\in\mG}$}:
\ba
(W_{321}[G]\psi^{\S_2})(g_1,\ldots,g_4)\,=\,  \delta(G,g_3g_2g_1) \;\psi^{\S_2}(g_1,\ldots,g_4)  .
\ea
With these ingredients, we define on $\S_2$ the Kitaev's ribbon operator ${\cal R}[G,H]$ to be:
\be
{\cal R}[G,H] =  W_{321}[G]\circ T_{4,3}[H].
\ee

Acting on the cylinder (global) vacuum state,
\ba
	\psi^{\S_2}_0(g_1,\ldots,g_4)\,:=\,\delta( e, g_4g^{-1}_2)1(g_1)1(g_3)
\ea
with $1(\cdot)$ the constant function of value $1\in\mathbb C$, we see that $T[H]$ and $W[G]$ generate the whole basis $\{\psi^{\S_2}_{G,H}\}$ of ${\cal H}_2$ (Equation \eqref{basisgroup}): 
\begin{align}
	({\cal R}[G,H] \psi^{\S_2}_0)(g_1,\ldots,g_4)\,&= \,\left( W_{321}[G]T_{4,3}[H] \psi^{\S_2}_0\right)(g_1,\ldots,g_4)\\
	 \, &= \,\begin{array}{c}\includegraphics[scale =1]{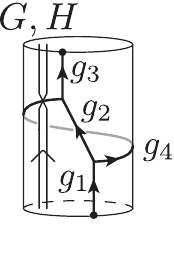}\end{array} \,=\, 
	\begin{array}{c}\includegraphics[scale =1]{fig/cylindre2-eps-converted-to.pdf}\end{array} \, = \,|\mathcal G|^{-3/2}\, \psi^{\S_2}_{G, H}  .
\end{align}
Hence, ribbon operators ${\cal R}[G,H]$ generate all possible pairs of excitations at the punctures of $\S_2$. 

 Let us briefly mention the fact that reversing the direction of the ribbon operator involves the antipode in $\mathcal{D}(\mG)$:
\be
	(\mathcal{R}_{\text{rev}}[G,H]\psi_0^{\S_2})(g_1,\ldots,g_4) = (\mathcal{R}(S[G,H])\psi_0^{\S_2})(g_1,\ldots,g_4)  \, ,
\ee
where $S([G,H])=[G^{-1},G^{-1} H^{-1} G]$ is the antipode of $[G,H]$ defined in equation \eqref{eq_antipode}.
\\

\begin{figure}[h!]
	\includegraphics[scale =1]{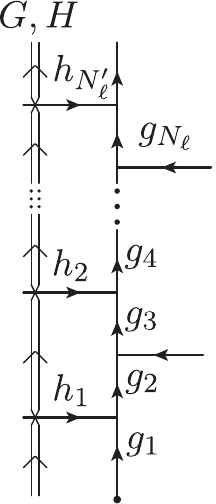}
	\caption{Action of the open ribbon operator.}  \label{Rib1}
\end{figure}

~\\
{\bf {Kitaev's ribbon on $\S_p$}}\\
The above considerations can be generalized, to ribbon operators on $\mathbb{S}_p$ which start and end at two punctures.
Consider two punctures connected by a directed link $\ell$ , possibly composed of several elementary links $\ell=l_{N_\ell} \circ \cdots \circ l_1$ with associated group elements $g_{N_\ell} \cdots g_1$, from which several links are departing to the right  and to the left with respect to the orientation of $\ell$. 
If necessary, we change orientations so that edges departing to the left are ingoing to $\ell$, see Figure \ref{Rib1}. 
Note that the graph underlying the state under consideration can be always brought into this form using the equivalences of Section \ref{sec_alBF}. 

We can now define the action of a ribbon operator acting from the left. 
To this end, draw a ribbon to the left of the link $\ell$, connecting the two punctures.
It will be (over--)crossed by all the links departing to the left of $\ell$. 
We denote the group elements associated to these links $h_1, \cdots,h_{N'_\ell}$ as in Figure \ref{Rib1}. 
We also denote by $g'_l$ the ordered products of the $\{g_l\}$ from the target of $h_l$ to the target puncture of $\ell$.

The (left) ribbon operator along $\ell$, ${\cal R}_\ell[G,H]$, is then defined by
\ba
&&({\cal R}_\ell[G,H] \psi)(g_1,\cdots,g_{N_\ell},h_1,\cdots ,h_{N'_L},\cdots)\nn\\ 
&&\q=\delta(G, g_{N_\ell} \cdots g_1) \psi(g_1,\cdots,g_{N_\ell},  (g'_{1})^{-1}  H^{-1}  g'_{1} h_{1} \,  \cdots ,  ,\, (g'_{N'_\ell})^{-1} H^{-1} g'_{N'_\ell} h_{N'_\ell},\cdots )\, . \,\,
\ea
As before the action of the ribbon operator splits into two parts: 
a Wilson path operator part which fixes to $G$ the holonomy from the source to the target punctures of $\ell$, and a translation operator part which translates by $H^{-1}$ and from the left the (anti-clockwise) holonomy around the target puncture of $\ell$:
\ba
(g'_{N'_\ell}) h_{N'_\ell}\cdots  (g'_{N'_\ell})^{-1} \;\; \mapsto \;\;  \,  H^{-1} \,(g'_{N'_\ell}) h_{N'_\ell}\cdots  (g'_{N'_\ell})^{-1}  .
\ea
At the same time the (clockwise) holonomy around the source puncture of $\ell$ is changed by 
\ba\label{RHsource}
(g'_0)^{-1} h_1 \cdots  g'_0 \;\;\mapsto\;\; 
&&(g'_0)^{-1}        (g'_1)^{-1}         \cdots (g'_{N'_\ell})^{-1}   H^{-1} g'_{N'_\ell} \cdots g'_1 h_1 \cdots  g'_0 \nn\\  
&& =    G^{-1} H^{-1} G \, \, (g'_0)^{-1} h_1 \cdots  g'_0   .
\ea

Note that the face holonomies stay trivial for any closed face. 
To ensure this the prescription of how the group elements $h_l$ are translated is essential: for any closed face being affected, there are always two group elements  $h_l$ and $(h_{l+1})^{-1}$ translated in an opposite manner, so that the net effect is leaving the face holonomy trivial. 

In fact we can imagine that we slide the ribbon operator from one face (lying left to the link $\ell$) to the next face, keeping the upper end fixed at the target puncture.
By this `sliding'  curvature and torsion excitations are moved from one face to the next, until one reaches the source puncture.

\subsubsection{Charge ribbon operators}

As we have seen the ribbon operators $\mR[G,H]$ generate the basis $\{\psi^{\S_2}_{G,H}\}$ of the two--punctured sphere (Equation \eqref{basisgroup}).
Then, the same transformation that allowed us to introduce the fusion basis can be used to define ribbon operators generating the basis $\{\psi^{\S_2}_{\frak f}[\rho,I' I]\}$ (Equation \eqref{fusbasis}).
This is just a Fourier--Peter--Weyl transform performed from the functions on the Drinfel'd double to functions on its representation labels:%
\footnote{The factor $d_\rho$ is not evenly distributed across the following two formulas in order to have equation \eqref{brexit?} to hold as it is, with no extra dimensional factors. }
\begin{align}
	\begin{dcases}
	&\mR[\rho,I' I] = \frac{d_{\rho}}{|\mathcal{G}|}\sum_{G,H}\mR[G,H] D^{\rho}_{I' I}([G,H])\\
	&\mR[G,H] = \sum_{\rho}\sum_{I', I} \mR[\rho,I' I] \overline{D^{\rho}_{I' I}([G,H])} 
	\end{dcases}.
\end{align}
%
And
\be
	\psi_{\mathfrak{f}}^{\S_2}[\rho,I' I] = \frac{|\mG|^{3/2}}{\sqrt{d_{\rho}}} \,\,(\mathcal{R}[\rho,I' I]\psi_0)(g_1,\cdots,g_4) \equiv \frac{|\mG|^{3/2}}{\sqrt{d_\rho}}
	\begin{array}{c}\includegraphics[scale =1]{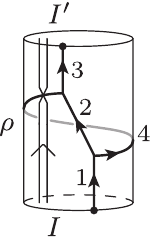}\end{array}.
\ee
The fusion basis had projective (idempotence) properties under the gluing operation defining the $\star$--product for cylinder states. 
This qualified its labels as physical charges carried by the punctures.
For this reason, we refer to $\mathcal{R}[\rho,I' I]$ as the $\rho$--charge ribbon operator.

In calculations, the following expression of  $\mathcal{R}[\rho,I' I]$ is sometimes more useful
\begin{align}
	{\cal R}[C,R;i'M',i,M]&= \frac{{d_{C,R}}}{|\mG|} \sum_{G,H} \delta(c_i,G^{-1}c_{i'}G) \, \delta(H,c_{i'}) D^R_{M'M}(q_{i'}^{-1} Gq_i)\, {\cal R}[G,H] \nn\\
	&=\frac{d_R}{|N_C|} \sum_{n\in N_C} D^R_{M'M}(n) \, {\cal R}[q_{i'}nq_i^{-1},\,c_{i'}] \, .
\end{align}

It is straightforward to extend the definition of the charge ribbon operators to $\mathbb{S}_p$.
It is indeed enough to transform the $[G,H]$ labels of ${\cal R}_\ell[G,H]$ to $[\rho, I'I]$.

\subsection{Properties of ribbon operators}

~\\
{\bf {Deformation invariance of ribbons}}\\
The action of the ribbon operator ${\cal R}_\ell[G,H]$ between two punctures $p_1$ and $p_2$ along $\ell$ changes the quasi--local charges at the punctures.
This action, however, does not depend on the precise path $\ell$.
Indeed, one can check that the action is invariant under isotopic deformations of the path (with regard to other punctures).
On the one hand, only the holonomies around the punctures are changed by the ribbon. 
This translation is determined by the parameter $H$ and the parallel transport along $\ell$ from $p_1$ to $p_2$.
On the other hand, the state is multiplied by a delta--function, which fixes the holonomy from puncture $p_1$ to puncture $p_2$  along $\ell$.  
And since we are dealing with locally flat states, only the isotopy class of $\ell$ matters, for both the parallel transport and the evaluation of the holonomies.
This is the reason why the action of ${\cal R}_\ell[G,H]$ is invariant under isotopic deformations of $\ell$. \\

Ribbon operators can be combined in different ways.
We can glue two ribbons by their extremities and in this way define a lengthwise product.
Or we can consider the operator product of two ribbons associated with the same path, which we call lateral product, obtaining a linear combination of ribbon operators.
Again, these operations can be described by the structure of the Drinfel'd double of the group \cite{Kitaev1}. \\

~\\
{\bf {Lengthwise product}}\\
To combine ribbons lengthwise, we consider a ribbon ${\cal R}_{\ell_1}[G_1,H_1]$  extending from a source puncture $p_1$ to a target puncture $p_2$, as well as a second ribbon ${\cal R}_{\ell_2}[G_2,H_2]$  extending from the (now) source puncture $p_2$ to a target puncture $p_3$. 
We assume that $p_2$ does not carry any excitation, i.e. Wilson loops around the puncture give trivial results, and the wave function has a trivial dependence on the holonomy associated to the link arriving at the puncture.%
\footnote{Later, we will define closed ribbon operators that project onto wave functions with prescribed charges at a given puncture.}

We then demand that the lengthwise product should be such that it does not induce any excitation at the `middle' puncture $p_2$.
And hence that this product in fact coincides with some (not self--crossing) ribbon operator along $\ell = \ell_2\circ\ell_1$, directly going from $p_1$ to $p_3$. 
To achieve this we will project onto the flatness and Gau\ss~constraints at the puncture $p_2$. 
This construction is analogous to the gluing of ribbons for the $\SU(2)_k$ case described in \cite{DG16}.
Moreover, as it will be apparent, this construction parallels the gluing of cylinder states.

If the links $\ell_1$ and $\ell_2$ are consistently oriented, to preserve the flatness at $p_2$ we need to require
\ba
H_1 \stackrel{!}{=}  G^{-1}_2 H_2 G_2  .
\ea
On the other hand, to avoid torsion excitations at $p_2$, we have to apply a group averaging at $p_2$ to the resulting state.
This operation eliminates the delta-function $\delta(G_1, g_{\ell_1})$ (here, $ g_{\ell_1}$ is the holonomy along $\ell_1$), which results from the action of ${\cal R}_{\ell_1}[G_1,H_1]$, but keeps the delta-function $\delta(G_2 G_1, g_{\ell_2} g_{\ell_1})$ fixing the holonomy along the combined path $\ell = \ell_2\circ\ell_1$.

The resulting action of the procedure we just described is---as expected---equivalent (modulo normalizations) to that of a single ribbon operator acting along $\ell=\ell_2\circ\ell_1$ and modifying the charge structure at $p_1$ and $p_3$:

\ba\label{8.16}
|\mG|\,  {\mathbb P}^{p_2}_\text{gauge}  {\mathbb P}^{p_2}_\text{flat} \,\, {\cal R}_{\ell_2}[G_2,H_2]  {\cal R}_{\ell_1}[G_1,H_1]  =
\delta(H_1, G_2^{-1} H_2 G_2) \, {\cal R}_{\ell_2\circ\ell_1}[G_2 G_1, H_2]   .
\label{eq_ribbonslengthprod}
\ea
Appendix \ref{gluingribbons} exemplifies the gluing of two ribbons for states on the three--punctured sphere. 

We now consider the lengthwise product of charge ribbon operators ${\cal R}_{\ell_2}[\rho, I' I]$ and ${\cal R}_{\ell_1}[\rho, I' I]$.
Using (\ref{8.16}) one finds (see Appendix \ref{GCR})
\ba
|{\cal G}|\,{\mathbb P}^{p_2}_\text{gauge}  {\mathbb P}^{p_2}_\text{flat} \, {\cal R}_{\ell_2}[\rho_2, I'_2 I_2]   {\cal R}_{\ell_1}[\rho_1, I'_1 I_1]
= \delta_{\rho_2,\rho_1}  \delta_{I_2, I_1'}  \,   {\cal R}_{\ell_2\circ\ell_1}[\rho_2, I'_2 I_1] .
\label{eq_chargeribbonslengthprod}
\ea
Note that the resulting ribbon does not involve the indices $I_2=I'_1$ at the `middle' puncture $p_2$. 
Thus, for the gluing of two charged ribbons, we can also define that the magnetic indices of the ribbons meeting at the puncture have to be contracted. 
This would introduce an extra factor $d_{\rho_1}= |C_1| d_{R_1} $ in the final result. 

Comparison with equations \eqref{eq_gluecyl} and \eqref{comultrivia} immediately shows that there is a direct relation between the gluing of cylinders and the lenghtwise multiplication of open ribbon operators.
This means that the composition of ribbons agrees with the multiplication of the ${\cal D}(\mG)$ algebra.
To make this completely explicit, we introduce a $\star$--product notation for the left--hand side of equations \eqref{eq_ribbonslengthprod} and \eqref{eq_chargeribbonslengthprod}:
\be
	\mR_{\ell_2}[G_2,H_2] \star \mR_{\ell_1}[G_1,H_1] = \delta(H_1, G_2^{-1}H_2G_2)\mR_{\ell_2\circ \ell_1}[G_2G_1,H_2]
\ee
and
\be
	\mR_{\ell_2}[\rho_2,I_2',I_2]\star \mR_{\ell_1}[\rho_1,I_1',I_1] = \delta_{\rho_1,\rho_2} \delta_{I_2,I_1'} \,  \mR_{\ell_2\circ \ell_1}[\rho_2, I_2',I_1].
	\label{brexit?}
\ee

~\\
{\bf {Lateral product}}\\
We now consider the operator product of two ribbons based on the same  path $\ell$, which we name lateral product. Due to the deformation invariance of the ribbons this is equivalent to having the product of two ribbons that are based on paths parallel to each other, and which start as well as end at the same punctures.  
Hence, we can drop in this section the path label, from ${\cal R}[G_i,H_i],\; i=1,2$.
%
%
It is straightforward to verify that the lateral product of two ribbons is a third ribbon operator (of course based on the same path):
\ba\label{opprod}
	{\cal R}[G_2,H_2] \,  {\cal R}[G_1,H_1] \,=\, \delta(G_1,G_2)     \, {\cal R}[G_1, H_2 H_1]   .
 \ea 
To prove the previous formula one can e.g. consider the consecutive action of two ribbons on the (global) vacuum state on $\S_2$:
\begin{align}
	&({\cal R}[G_2,H_2]  {\cal R}[G_1,H_1]\psi_0^{\S_2})(g_1,\cdots,g_4)  =
	\begin{array}{c}\includegraphics[scale =1]{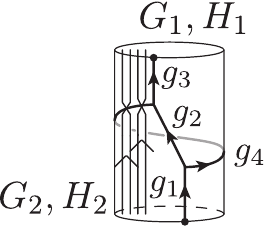}\end{array} \nn\\
	&\q =\delta(G_1,g_3g_2g_1) ({\cal R}[G_2,H_2]  \psi_0^{\S_2})(g_1,\cdots,g_3^{-1}H_1^{-1}g_3g_4) \nn \\ &\q =
	\delta(G_1,g_3g_2g_1)\delta(G_2,g_3g_2g_1)\,\,\psi_0^{\S_2}(g_1,\cdots,g_3^{-1}H_1^{-1}H_{2}^{-1}g_3g_4) = 
	\begin{array}{c}\includegraphics[scale =1]{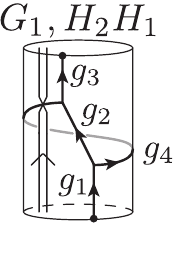}\end{array} \nn\\
	&\q=  \delta(G_1,G_2)     \, ( {\cal R}[G_1, H_2 H_1] \psi_0^{\S_2})(g_1,\cdots,g_4).
\end{align}

We can also consider the lateral product of two charge ribbons based, i.e. the operator product of two charge ribbons based on the same path. 
Here, the two ribbons generate two basic excitations at the same puncture. 
We therefore expect that the resulting excitation should arise from a fusion of the two basic excitations. 
In fact, the lateral product involves the tensor product of the Drinfel'd double representations (and their dual): 
\begin{align}\label{OpF} \nn
	&\mathcal{R}[\rho_2,I_2' I_2] \, \mathcal{R}[\rho_1,I_1' I_1] \\ \nn
	& \q  = \frac{{d_{\rho_1}d_{\rho_2}}}{|\mG|^2}\sum_{G_1,H_1 \atop G_2,H_2} D^{\rho_2}_{I_2' I_2}([G_2,H_2]) 
	D^{\rho_1}_{I_1' I_1}([G_1,H_1]) \mathcal{R}[G_2,H_2]\, \mathcal{R}[G_1,H_1] \\ \nn
	&\q \stackrel{ (\ref{opprod})}{=} \frac{{d_{\rho_1}d_{\rho_2}}}{|\mG|^2}\sum_{G\atop H_1,H_2} D^{\rho_2}_{I_2' I_2}([G,H_2])
	D^{\rho_1}_{I_1' I_1}([G,H_1]) \mathcal{R}[G,H_2H_1] \\ \nn
	& \q = \frac{{d_{\rho_1}d_{\rho_2}}}{|\mG|^2}\sum_{G\atop H_1,H_2}\sum_{\rho_3}\sum_{I_3 I_3'} D^{\rho_2}_{I_2' I_2}([G,H_2])
	D^{\rho_1}_{I_1' I_1}([G,H_1])\overline{D^{\rho_3}_{I_3' I_3}([G,H_2H_1])} \mathcal{R}[\rho_3,I_3' I_3] \\ 
	& \q =  \frac{ 1}{|\mG|}\sum_{\rho_3}\sum_{I_3 I_3'}\frac{d_{\rho_2}d_{\rho_1}}{d_{\rho_3}}\;
	{\mathcal{C}^{\rho_2 \rho_1 \rho_3}_{I_2' I_1' I_3'}}	\;
	\overline{\mathcal{C}^{\rho_2 \rho_1 \rho_3}_{I_2 I_1 I_3}} \,\, \mathcal{R}[\rho_3,I_3' I_3].
\end{align}

 Note that  also the lateral product of two ribbons reflects an algebraic structure of the Drinfel'd double, namely its co-multiplication  $\Delta([G,H]) = \sum_{H_2,H_1}\delta(H_2H_1,H) [G,H_2]\otimes[H,H_1]$.  Similarly the lateral product allows us to write a given ribbon as a sum over all possible pairs of ribbon operators whose product is the desired one:
 \ba
 {\cal R}[G, H] \,=\, \frac{1}{|{\cal G}|} \sum_{H_2,H_1} \delta(H_2H_1,H) \, {\cal R}[G,H_2] \,  {\cal R}[G,H_1]   \, .
 \ea

\subsection{Closed ribbons}

By gluing the ends of an open ribbon, starting and ending at the same puncture, we obtain a closed ribbon.
Closed ribbons do not generate excitations, they just measure the excitation content of the region they enclose.
In the context of BF theory on a surface with fixed punctures (or higher genus), closed ribbon operators provide a complete basis of Dirac observables.
This is because closed ribbons {are defined in such a way to} commute with the flatness and Gau\ss~constraints.
And the fusion basis constructed in section \ref{sec_fusionbasis} diagonalizes the (charge) closed ribbon operators. 

\begin{figure}[h!]
	\includegraphics[scale =1]{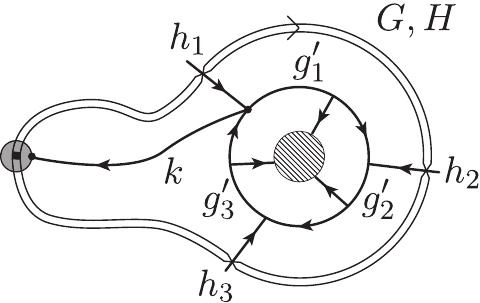} \q \q \q \q 
	\includegraphics[scale =1]{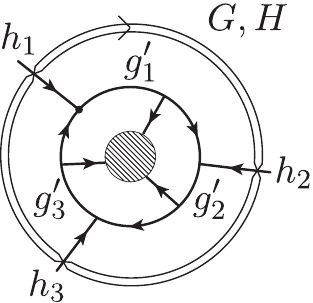}
	\caption{ The construction of a closed ribbon operator. The left panel shows the auxiliary puncture and an auxiliary link with holonomy $k$ going to this puncture. This holonomy plays no role in the final action of the closed ribbon operator as described in (\ref{pcr1}). }\label{figcr}
\end{figure}

To explicitly construct a closed ribbon operator, we start with an open one as in section \ref{oribbon}. 
It might be necessary to introduce an auxiliary puncture, at which the open ribbon starts and ends. 
By applying the refining operations detailed in section \ref{sec_alBF}, we can always consider this puncture connected to the graph underlying the state under consideration via a link carrying a holonomy $k$ (see Figure \ref{figcr}).
The refined state would then be constant in $k$, i.e. not depend on this holonomy.
The ribbon crosses $L$ links with associated group elements $h_1, \cdots, h_L$ which are incoming to a closed (circular) combination of links with associated holonomy $g'_L \cdots g'_2 g'_1$. 
We also define $g'_{Ll}:=g'_Lg'_{L-1} \cdots g'_l$, i.e. $g'_{Ll}$ is the parallel transport from the target node carrying $h_l$ to the target node carrying $h_1$. 
Note that $g'_{L1}:=g'_Lg'_{L-1}\cdots g'_1$ is given by the holonomy going around the cycle defined by the ribbon. 
The (open) ribbon operator, then acts as
\begin{align}
	&\left({\cal R}[G,H] \psi \right)(k, g'_1,\cdots, h_1,\cdots, \cdots) \nn
	\\ &\q = \delta(G, k g'_{L1}k^{-1}) \, \psi(g'_1,\cdots,  (g'_{L1} )^{-1} k^{-1} H^{-1} k g'_{L1} \, h_1 ,\,\,  
	\cdots,  (g'_{L} )^{-1} k^{-1} H^{-1} k g'_{L} \, h_L, \cdots).
\end{align}
We know that the ribbon will preserve both the flatness and Gau\ss~constraints for every faces, with the only exception given by ({\it i}) the flatness constraints for the face containing the auxiliary puncture, since this face contains the holonomy combination $h_1^{-1} g'_L h_{L}$, and ({\it ii}) the Gau\ss~constraint at the target node of the link carrying the holonomy $k$.

To deal with the flatness violation,%
\footnote{Even if the face we are considering here is a priori not closed, we can apply the refinement operations detailed in section \ref{sec_alBF}, so that this face is closed. After applying the closed ribbon operator we can go back to the coarser graph again, applying a coarse graining transformation, to reach a state based on the initial graph.}
we first notice that the holonomy combination $h_1^{-1} g'_L h_{L}$ is shifted to
\ba
h_1^{-1} g'_L h_{L} \; \rightarrow\;  h_1^{-1} (g'_{L1})^{-1} k^{-1}Hk g'_{L1} g'_L (g'_{L})^{-1} k^{-1} H^{-1}k g'_L h_L\,=\, h_1^{-1} k^{-1} G^{-1} H G H^{-1} k g'_L h_L  \, .\;\;\;\,\,\,\;
\ea
Therefore, to avoid a curvature excitation at the auxiliary puncture, we need to demand $GHG^{-1}H^{-1}=e$, which can be taken care of by introducing an extra delta--function factor $\delta(GHG^{-1}H^{-1},e)$. 

Then, we have to ensure gauge invariance at the target node of the link carrying $k$.  
This is achieved by applying the gauge averaging projector for this node.
Using that the initial state is gauge invariant this results in
\begin{align}\label{pcr1}
& \left( {\mathbb P}^n_\text{gauge} \circ {\mathbb P}^n_\text{flat} {\cal R}[G,H] \psi\right) (k,g'_1,\cdots, l'_1,\cdots, \cdots) \nn\\ 
&\q=  \delta(GHG^{-1}H^{-1},e)
\frac{1}{|{\cal G}|}  \sum_h \delta(G,  h  kg'_{L1}k^{-1} h^{-1})
\psi(g'_1,\cdots,  (g'_{L1} )^{-1} k^{-1} h^{-1} H^{-1} h k g'_{L1} l_1 , 
\cdots, \cdots) \q\nn\\
&\q= \delta(GHG^{-1}H^{-1},e)  \frac{1}{|{\cal G}|}  \sum_h  \left({\cal R}[ h G h^{-1} , h H h^{-1}]   \, \psi\right)(g'_1,\cdots, l'_1,\cdots, \cdots)  .
\end{align}

Note that due to the group averaging  the dependence on the (auxiliary) holonomy $k$ disappears.
Furthermore the (projected) closed ribbon operator does not depend anymore on the choice of face, among the faces crossed by the ribbon, at which the auxiliary puncture was inserted.  
Analogously to the open ribbons, the closed ribbons path dependence is limited to its isotopy class. 

Note also that, due to the projections onto flatness and Gau\ss~ constraints, not all information contained in the pair $(G,H)$ is actually relevant.
To see this we first rewrite $G$ using the notation of section \ref{sec_irreps} for the description of the Drinfel'd Double representations. 
This way we obtain, $G=q_i c_1 q_i^{-1}$, where $c_1$ is a representative of the conjugacy class $C$ of $G$ and $q_i \in  Q_C= {\cal G}/N_C$ with $ N_C$ the stabilizer group of $c_1$.  
Now, due to the delta function $\delta(GHG^{-1}H^{-1},e)$ in \eqref{pcr1} we see that $H$ must be of the form $H=q_i \tilde n q_i^{-1}$  for some $\tilde n \in N_C$.  
Therefore, using the fact that $G$ and $H$ must commute, we have
\ba\label{clr2}
\sum_{h \in {\cal G}} {\cal R}[h G h^{-1}, h H h^{-1}]  
&=& \sum_{h \in {\cal G}} {\cal R}[h  q^{-1}_i  \, q_i c_1 q_i^{-1}  \, q_i  h^{-1}, h  q^{-1}_i    q_i \tilde n q_i^{-1}    q_i  h^{-1}]  \nn\\
&=& \sum_{q_j \in Q_C} \sum_{n\in N_C}    {\cal R}[  q_j c_1 q_j^{-1} ,\, q_j n \tilde n n^{-1} q_j^{-1}] \nn\\
&=& |N_D|  \sum_{q_j \in Q_C} \sum_{ d \in D}   {\cal R}[  q_j c_1 q_j^{-1} ,\, q_j d  q_j^{-1}] 
\label{clr2}
\ea  
where in the first step we shifted the summation argument by $q_i$, and in the second step we split the summation over $h \in {\cal G}$ into a one over $q_j \in Q_C$ and $n \in N_C$ (here, we use that each group element has a unique representation of the form $h=q_j n$). 
In the third step, we split again the summation over $N_C$ into one over the stabilizer group $N_D \subset N_C$ and a conjugacy class $D$ of the group $N_C$.  

Thus the group averaging over ribbons ${\cal R}[G,H]$ (with $G$ and $H$ commuting) does only depend on the conjugacy class $C$ of ${\cal G}$  (such that $G \in C$) and a conjugacy class $D$ of ${N_C}$ (such that  $H$ is conjugated to an element of $D$).

Hence, we define closed ribbon operators as
\ba\label{defK}
{\cal K}[C,D] := \sum_{q_j \in Q_C} \sum_{ d \in D}   {\cal R}[  q_j c_1 q_j^{-1} ,\, q_j d  q_j^{-1}]   ,
\ea
where $C$ is a conjugacy class of $\mG$ and $D$ is a conjugacy class of $N_C$, the stabilizer group of $c_1 \in C$.

We constructed closed ribbon operators from gluing open ribbons. 
We arrive at the same definition as in \cite{Bombin}, where the closed ribbons ${\cal K}[C,D]$  are defined (via the third line of \ref{clr2}) based on more abstract reasoning.

\subsubsection{ Closed charge ribbon operators} \label{clchri}

In the case in which we consider punctured spheres only, the closed ribbon operators measure the excitation content of the region enclosed by the ribbon.%
\footnote{On higher genus surfaces, the closed ribbons could wind around non--contractible cycles.}
We are now going to construct closed ribbons with  projective properties, which allow to project onto a region with a certain charge content. 
In this case, what is needed, is the projective property with respect to the lateral product, rather then the (lengthwise) $\star$--product.

Using \eqref{opprod} of the ${\cal R}[G,H]$, we can deduce the lateral product for the closed ribbons:
\begin{align}
	{\cal K}[C_2,D_2]   \, {\cal K}[C_1,D_1] &= 
	\sum_{q_i \in Q_{C_2} \atop q_j \in Q_{C_1}}\sum_{d_2 \in D_2 \atop d_1 \in D_1}\delta(q_i c^{(2)}_1q_i^{-1},q_j c^{(1)}_1 q_j^{-1})
\,\,	\mathcal{R}[q_i c^{(2)}_1 q_i^{-1},q_i d_2 q_i^{-1} q_j d_1 q_j^{-1}] \\
	& = \delta_{C_2,C_1} \sum_{q \in Q_{C_2}} \sum_{d_1 \in D_1, d_2 \in D_2}
	{\cal R}[ q c^{(2)}_1 q^{-1},  q d_2 d_1 q^{-1}].
	\label{opprodK}
\end{align}
Defining coefficients %
 $N^{D_3}_{D_2D_1}$ via\footnote{ The set $\{\sum_{d \in D} d\}_D$, where $D$ is an index labelling the conjugacy classes of $N_C$,  gives a basis of (group algebra) elements commuting with all $n \in N_C$. Thus also the product of $\sum_{d_2 \in D_2} d_2$ with $\sum_{d_1 \in D_1} d_1$ commutes with $n\in N_C$ and can be expanded in this basis.}
 \be
 \sum_{d_1\in D_1,d_2\in D_2} d_2 d_1 \,=\, \sum_{D_3} N^{D_3}_{D_2D_1} \sum_{d_3 \in D_3} d_3
 \ee
  (both sides are to be understood as elements of the group algebra $\mathbb{C}[N_{C_2}]$) we arrive at
\ba
{\cal K}[C_2,D_2]  \, {\cal K}[C_1,D_1] &=& \delta_{C_2,C_1} \sum_{D_3}  N^{D_3}_{D_2D_1} \, {\cal K}[C_2,D_3]   .
\ea
Therefore, the closed ribbon ${\cal K}[C,D]$ are already projective in $C$, but not so in $D$. 
To reach fully projective closed ribbons under the lateral product, we define the charge closed ribbons via the formula
\ba\label{defcl2}
{\cal K}[C,R] &:=& \frac{d_R}{|N_C|} \sum_{D} \chi^R(D) \, {\cal K}[C,D]\,,
\ea
 where $R$ is an irrep of the stabilizer group $N_C$ (see also \cite{Bombin}---although, there slightly different conventions are used).
The inverse transformation is given by 
\be
{\cal K}[C,D]=   \frac{|N_C|}{|N_D|} \, \sum_R \frac{1}{d_R} \overline{\chi^R}(D) \,\, {\cal K}[C,R]  .
\ee

Now, it is straightforward to check (see Appendix \ref{appOP2}) that the lateral product of two charge closed ribbons is simply
\ba
{\cal K}[C , R]\,  {\cal K}[C' , R' ]  = \delta_{C , C'} \delta_{R, R'} {\cal K}[C , R].
\ea
Hence, the charged closed ribbons ${\cal K}[C,R]$ define a family of orthogonal projectors.
We are now going to show that they do actually project precisely on the fusion basis states. \\

~\\
{\bf {Diagonalization of closed ribbon operators}}\\
We consider the action of a closed charge ribbon ${\cal K}[C,R]$ applied to a fusion basis state on the cylinder. 
(We will later generalize to fusion basis states on $\S_p$.)
Using a minimal graph, the fusion basis state can be  expressed in the holonomy representation as
\ba\label{basiscylR}
	\psi^{\S_2}_{\frak f}[CR;i'M',iM]&=&|\mG|^{1/2}
\sqrt{d_{C,R}} \sum_{n\in N_C}  \delta( q_{i'}nq_{i}^{-1},g_3g_2g_1)\, \delta(c_{i'}, g_3g_4 g_2^{-1}g_3^{-1}) D^R_{M'M}(n)  .
\ea

\begin{figure}[h!]
	\centering
	\includegraphics[scale = 1]{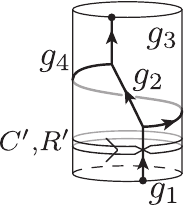}	
	\caption{The closed ribbon operator applied to a state on a cylinder.}
	\label{fig:clrc}
\end{figure}

We apply a closed ribbon ${\cal K}[C',R']$ that goes  anti-clockwise around the cycle with holonomy $g_2^{-1}g_4$ and crosses only the link with holonomy $g_1$,  as in Figure \ref{fig:clrc}.

\noindent Then, the action of the closed ribbon ${\cal K}[C',R']$ on  $\psi^{\S_2}_{\frak f}[CR;i'M',iM]$ can be readily evaluated in the holonomy basis (see Appendix \ref{app:clrc}).
We expect that the closed ribbon does not change the charge content of the states.
And indeed, the fusion basis states are eigenstates of the closed ribbon operator:
\ba
	{\cal K}[C',R']\, \,\psi[CR;i'M',iM] &=&\delta_{C,C'} \delta_{R,R'} \,\psi^{\S_2}_{\frak f}[CR;i'M',iM].
\ea
Or, more succinctly 
\ba
	{\cal K}[\rho']\, \,\psi[\rho, I' I] &=&\delta_{\rho,\rho'} \,\psi^{\S_2}_{\frak f}[\rho, I'  I].
\ea
Thus the  closed charge ribbon operator ${\cal K}[C',R']$ projects onto the basis states $\psi^{\S_2}_{\frak f}[CR;i'M',iM]$. 

This result can be immediately generalized to the fusion basis states on $\S_p$. 
In this case, we consider a closed ribbon going around one leg of the trinion decomposition of $\S_p$ underlying the fusion basis.  
We can then choose the graph on this trinion to be the same as in figure \ref{fig:clrc}.
Hence, the action of the closed charge ribbon can be evaluated in the same way as there.  
Again, the closed charge ribbon ${\cal K}[C,R]$ will project onto fusion basis states with charge labels $(C,R)$ for the trinion leg in question.

\subsection{ An alternative  closed ribbon operator}

In the previous section we started with a ribbon ${\cal R}[G,H]$ based on a closed path, and then projected onto its flatness and gauge--invariance preserving component.
We saw that the resulting operators only depend on the conjugacy class $C$ of $G$ and a conjugacy class $D$ in the stabilizer group $N_C$. 
Alternatively, we could also start with the charge ribbons ${\cal R}[\rho; I' I]$, again based on a closed path, and project these. 
This provides an alternative basis of closed ribbon operators.
We are going to discuss these here, as these ribbons mimic the closed ribbons discussed in \cite{DG16} for the quantum group case $\SU(2)_{\rm k}$, where the group representation is not available.  
We will in particular see that the two types of closed ribbons are in a certain sense dual to each other: they are related by a specific transform that can be interpreted as Fourier transform within ${\cal D}(\mG)$ \cite{KoornFT}.
 
Recall the following expression of the charge ribbon operators
\be
	{\cal R}[C,R;i'M',iM]= \frac{d_R}{|N_C|} \sum_{n\in N_C} D^R_{M'M}(n) \, {\cal R}[q_{i'}nq_i^{-1},\,c_{i'}]  .
\ee
Aiming at the definition of a closed ribbon operator, we sum over the indices $i=i'$ and $M=M'$:
\ba\label{clA1}
\sum_{i,M}{\cal R}[C,R;iM,i,M]&=& \frac{d_R}{|N_C|} \sum_{n\in N_C}  \chi^R(n)  \sum_{q_i \in Q_C} {\cal R}[q_inq_i^{-1},q_i c_1 q^{-1}_i]  .
\ea
As $c_i=q_ic_1 q_i^{-1}$ and $n$ is in the stabilizer group of $c_1$ we see that  $G=q_inq_i^{-1}$ and $H=c_i$ do commute, and hence the flatness constraints are already satisfied.
The contraction of the ribbon as defined in \eqref{clA1} is also invariant under the group averaging projector:
\ba
{\mathbb P}_\text{gauge} \sum_{i,M}{\cal R}[C,R;iM,i,M]
&=& \frac{d_R}{|N_C||\cal G|} \sum_{h \in {\cal G}}\sum_{n\in N_C}  \chi^R(n)  \sum_{q_i \in Q_C} {\cal R}[h q_inq_i^{-1} h^{-1}, hq_i c_1 q^{-1}_ih^{-1}]  \nn\\
&=&
 \frac{d_R}{|N_C|^2} \sum_{n\in N_C}  \chi^R(n)   \sum_{q_j \in Q_C} \sum_{n' \in N_C}    {\cal R}[q_j n' n (n')^{-1}q_j^{-1}, q_j c_1 q_j^{-1}] \nn\\
&=&  \frac{d_R}{|N_C|} \sum_{\tilde n \in N_C} \sum_{q_j \in Q_C} \chi^R(\tilde n)\,\, {\cal R}[q_j \tilde n q_j^{-1}, q_j c_1 q_j^{-1}]\nn\\
&=&  \sum_{i,M}{\cal R}[C,R;iM,i,M]  .
\ea
In the above calculation, we first shifted the summation over $h$ by $q_i^{-1}$, making the sum over $q_i \in Q_i$ superfluous. Then, we split again $h$ as $h=q_j n'$ and redefined the variable $n$ to $\tilde n=n' n (n')^{-1}$, hence making the sum over $n'$ superfluous. 

This shows that the following is a viable definition of an operator on ${\cal H}_{p}$, since it preserves both the flatness and Gau\ss~constraints:
\be
\tilde {\cal K} [C,R] := \sum_{i,M}{\cal R}[C,R;iM,i,M],
\ee
or, equivalently,
\be
\tilde {\cal K} [\rho] := \sum_{I}{\cal R}[\rho;I I].
\ee

In particular, the above formulas show that $\tilde{\cal K}[C,R]$, when expressed in terms of ${\cal R}[G,H]$, has essentially  the same form as the ribbon operators ${\cal K}[C,R]$ defined at Equations \eqref{defK} and \eqref{defcl2}.
The only difference is that the role of the entries in the ribbon operator ${\cal R}[G,H]$ is exchanged. 
Indeed, the transformation between the two types of closed ribbon operators reveals why this is the case. 

To express $\tilde {\cal K}[C,R]$ in terms of $ {\cal K}[C',R']$ operators, we write
\ba
\tilde {\cal K} [C,R]&=&{\mathbb P}_\text{gauge} \sum_{i,M}{\cal R}[C,R;iM,i,M]\nn\\&=&
 \frac{d_R}{|N_C|} \sum_{n\in N_C}  \chi^R(n)  \sum_{q_i \in Q_C} \, {\mathbb P}_\text{gauge}  {\cal R}[q_inq_i^{-1},q_i c_1 q^{-1}_i] \nn\\
 &\stackrel{\eqref{clr2}}{=}& \frac{d_R}{|N_C|} \sum_{n\in N_C}  \chi^R(n) \frac{    |N_{D(n,c_1)}| }{|N_C|}    \,\, {\cal K}[ C_n,D_{n,c_1}] \nn\\
 &\stackrel{\eqref{defcl2}}{=}& \frac{d_R}{|N_C|} \sum_{n\in N_C}   \chi^R(n)     \sum_{R'} \frac{1}{d_{R'}}\overline{\chi^{R'}( D_{n,c_1}) }\,\,{\cal K}[ C_n,R'] ,
\ea
where in the third line, we used the definition (\ref{clr2}) of the ribbon operators ${\cal K}[C,D]$, and where we made use of the following notation:
 $C_n$ stands for the conjugacy class of $n$ in ${\cal G}$ and $D_{n,c_1}$ for the conjugacy class in $N_{C_n}$, which includes the element
 \ba\label{3.39}
  {}_nq_{k}^{-1}\, c_1 \, {}_nq_{k}  \q \text{where}   \q n= {}_nq_{k} \, {}_nc_1\, {}_nq_{k}^{-1} \q \text{and} \q {}_nq_{k} \in Q_{C_n} ,\,\,{}_nc_1 \in C_n  .
\ea
Therefore, we conclude that $\tilde{\cal K}[C,R]$ is a linear combination of operators ${\cal K}[C',R']$. 
This can be summarized with the formula,
\ba
\tilde{\cal K}[C,R] \,=\, \sum_{C',R'}  {\cal S}_{CR,C'R'}  \, {\cal K}[C',R'],
\ea
where
\ba
{\cal S}_{CR,C'R'} &=&  \frac{d_R}{ d_{R'} \, |N_C|} \sum_{n\in N_C}  \chi^R(n)   \overline{\chi^{R'}( D_{n,c_1})} \, \delta_{C',C_n}     .
\ea
This matrix turns out to be related to the so--called S--matrix of the Drinfel'd double of the group.
This is defined as \cite{Pasquier,Verlinde} 
\ba\label{SmatrixV}
{\bf S}_{CR,C'R'} \,=\, \frac{1}{|\cal G|}  \sum_{h_i  \in C, h'_j \in C'}  \delta( h_i h'_j, \, h'_j h_i) \, \, \overline{\chi^R( q_i^{-1}h'_jq_i)} \,\, \overline{ \chi^{R'} ( (q'_j)^{-1} h_i q'_j) } .
\ea
where $h_i:= q_i c_1 q_i^{-1}$ and $h'_j:= q_j c'_1 q_j^{-1}$, with $c_1 \in C, \, c'_1 \in C'$ and $q_i \in Q_C, \, q'_j \in Q_{C'}$. 
As it is shown in Appendix \ref{SmatrixR}, the S--matrix ${\bf S}_{CR,C'R'}$ can be rewritten as
\ba\label{3.45}
{\bf S}_{CR,C'R'} &=& \frac{1}{|\cal G|}  \sum_{h_i  \in C}  \sum_{n \in N_C} \delta_{C',C_n}   \, \, \overline{\chi^R( n) }\,\, \overline{ \chi^{R'} ( D_{n,c_1})} \nn\\
 &=& \frac{1}{|N_C|}  \,  \sum_{n \in N_C} \delta_{C',C_n}   \, \, \overline{\chi^R ( n)} \,\, \overline{\chi^{R'}(D_{n,c_1}) }\,,
\ea
and thus 
\ba
{\cal S}_{CR,C'R'}\,=\, \frac{d_R}{d_{R'}} \, {\bf S}_{CR^*,C'R'}\,,
\ea
where $R^*$ denotes the contragredient representation to $R$. 

Using this result, it is straightforward to deduce the action of $\tilde{\cal K}[C,R]$ on the fusion basis. In the conventions of figure \ref{fig:clrc}, and with the usual short--hand notation:
\ba
\tilde {\cal K}[\rho]\, \,\psi[\rho, I' I] &=& {\cal S}_{\rho,\rho'} \,\,\psi[\rho', I' I] ,
\ea
i.e. the fusion basis states are also eigenstates of $\tilde{\cal K}[C,R]$, but this time with eigenvalues determined by the entries of the S--matrix.

The relation between the two basis of closed ribbon operators and the fusion basis on the cylinder can be understood as follows.
The label $C$ in  $\psi^{\S_2}_{\frak f}[C R;i'M',iM]$ denotes the conjugacy class of the holonomy ($H$) around the cylinder, whereas the representation label $R$ encodes information about the functional dependence of the wave function on the `longitudinal' holonomy ($G$) along the cylinder. 
Going back to the construction of the  closed charge ribbon ${\cal K}[C,R]$ (Equations \eqref{clr2} and \eqref{defcl2}), we see that $C$ is again the conjugacy class of the holonomy around the cylinder and $R$ captures information about the holonomy along the cylinder. 
This explains why ${\cal K}[C,R]$ projects onto wave functions $\psi^{\S_2}_{\frak f}[C,R; I', I]$.

In turn, if we consider a closed ribbon $\tilde{\cal K}[C,R]$ 
going around the cylinder, $C$ now captures information about the `longitudinal' holonomy along the cylinder (the one crossed by the ribbon), whereas $R$ encodes information about the holonomy going around the cylinder. 

In fact, on the torus $\mathbb T$---obtained e.g. by gluing the two punctures of the cylinder---we can consider closed ribbons associated to the two cycles generating the torus fundemental group.
We can then define two different basis of ${\cal H}_{\mathbb T}$ diagonalizing the two different closed ribbons. 
The transformation between these two basis is given by the S--matrix.
This is for the same reason why the S--matrix appears in the transformation between ${\cal K}$ and $\tilde {\cal K}$: it exchanges the role of the longitudinal and transverse holonomies.
But, on the torus, the role of longitudinal and transverse holonomy is the same.
Hence, the complete duality in this case.

At the level of the Drinfel'd double, ${\cal D}(\mG) = {\cal F}(\mG)^\ast  \otimes {\cal F}(\mG) $, the S--matrix defines a transform exchanging the role of ${\cal F}(\mG)$ and its dual    $  {\cal F}(\mG)^\ast  \simeq {\mathbb C}{\mathcal G}$ \cite{KoornFT}.
In particular, this translates into the fact that the role of multiplication and co-mutiplication are also exchanged in a proper sense.
This is why, in the analysis above, we have seen both the $\star$--multiplication and the co-multiplication structures appearing naturally in the context of lateral products.

The S--matrix can also be defined through the eigenvalues for the operator defined by two interwoven closed ${\tilde K}$--ribbons (Figure \ref{figS}):
\begin{figure}[h]
	\includegraphics[scale =1]{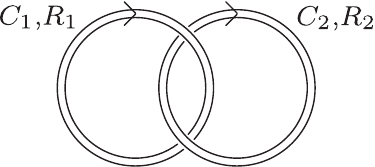} 
	\caption{Two braided closed ribbon operators. These can be constructed by gluing open ribbon operators, which act in a certain order, e.g. under--crossing pieces of ribbons act before the over--crossing pieces of ribbons.}
	\label{figS}
\end{figure}
To define this interwoven operator one needs to build the closed ribbons by `gluing' open ribbons, after having applied the latter to the state in the appropriate order. See \cite{DG16} for details.

\subsection{Back to the fusion basis}

\begin{figure}[h]
	\includegraphics[scale =1]{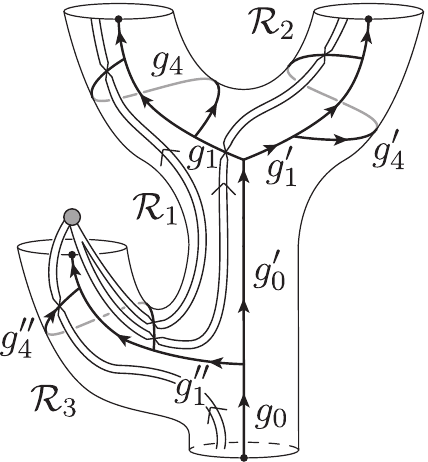} 
	\caption{Construction of the fusion basis states on the three-punctured sphere using charge ribbon operators. An auxiliary puncture is introduced at which the three ribbons are fused via a Clebsch-Gordan coefficient. \label{fig3S}}
\end{figure}

We have previously shown that the charge ribbon operators generate the fusion basis on the cylinder
\be
	\psi_{\mathfrak{f}}^{\mathbb{S}_2}[\rho,I',I] = \frac{|\mG|^{3/2}}{\sqrt{d_{\rho}}} (\mathcal{R}[\rho,I',I]\psi^{\mathbb{S}_2}_0)(g_1,\cdots,g_4) \equiv\frac{|\mG|^{3/2}}{\sqrt{d_{\rho}}}
	\begin{array}{c}\includegraphics[scale =1]{fig/inner01-eps-converted-to.pdf}\end{array}.
\ee
This statement can be generalized to spheres ${\mathbb S}_p$ with more punctures.  Consider for instance the three--punctured sphere ${\mathbb S}_3$. We wish to obtain a fusion basis state by using the charge ribbon operators ${\cal R}_1[\rho_1, I_1',I_1],\, {\cal R}_2[\rho_2, I_2',I_2]$ and ${\cal R}_3[\rho_3, I_3',I_3]$ for each of the legs of the three--punctured sphere. However the three ribbons need to be fused (or glued) at an auxiliary puncture, we therefore need to consider a four--punctured sphere, see figure \ref{fig3S}. Moreover, we need to contract the free indices arriving at the auxiliary puncture with a Clebsch-Gordan coefficient.

By construction, the fusion procedure at the auxiliary puncture includes a projection of this puncture to vanishing electric and magnetic (curvature and torsion) charge. This allows us to understand the resulting state as a state on the three--punctured sphere again. As shown in appendix \ref{RibbonsTOFusion} this gives
\begin{align}
&\sum_{I_1',I_2',I_3'}\,    \bigg(\Big( \mathcal{R}_1[\rho_1,I_1',I_1] \, \mathcal{R}_2[\rho_2,I_2',I_2] \,  \mathcal{C}^{\rho_1\rho_2\rho_3}_{I_1 I_2 I_3}\Big) \, \star\, {\cal R}_3[\rho_3,I_3,I'_3] \bigg) \,\psi_0^{{\mathbb S}_3} = \frac{1}{|{\cal G}|^3 } \,d_{\rho_3}  \sqrt{ d_{\rho_1}d_{\rho_2} } \, \,\, \psi^{\S_3}_{\frak f}
	 { \scriptsize\begin{bmatrix} \rho_1,I_1'\\ \rho_2,I_2'\\ \rho_3,I'_3\end{bmatrix}} .
\end{align}
This construction can be easily generalized to spheres with more punctures. 


\section{Applications: Multi-scale design of states and coarse graining \label{sec_coarsegraining}}

 We finally come to applications of the fusion basis and the related ribbon operators. Here we will discuss applications that make use of the multi--scale control the fusion basis offers.  In a follow-up work we will discuss a new notion of entanglement entropy for non--Abelian lattice gauge theories that can be defined with the help of the fusion basis \cite{Toappear}.

\subsection{Multi-scale design of states}\label{multiscale}

In the previous sections we constructed the fusion basis as well as ribbon operators which either generate it (open ribbon operators ${\cal R}$) or  project onto it (closed ribbon operators ${\cal K}$).
Crucially, the fusion basis is quite different from e.g. a spin--network basis, since it allows a direct access to observables at different scales. 
In fact, the $\{\rho\}$ labels of a fusion basis state $\psi_{\mathfrak{f}}^{{\mathbb S}_p}[ \{\rho\}, \{I\}]$ correspond to the $\rho$ labels appearing in the set of closed ribbon operators $\{{\cal K}[\rho]\}$, which project onto the fusion basis state.
In turn, such closed ribbon operators go around different number of punctures.
This number provides us with a notion of `scale', which we can associate to the closed ribbon operator ${\cal K}[\rho]$, and hence to the label $\rho$ itself.
In the case of gravity the geometry is encoded in the states.
Thus this notion of `scale' is not a priori associated to a notion of length or metric.
It is rather an auxiliary notion, from which one can however deduce a length scale \textit{once a choice of state is given},  see e.g. \cite{Ditt12b,BD14}.

\begin{figure}[h]
	\includegraphics[scale =1]{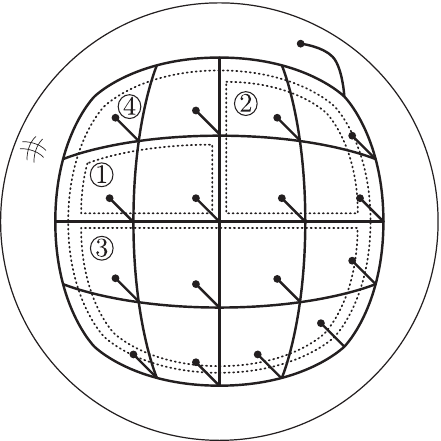} 
	\caption{We can embed a regular square lattice on the sphere by closing it with an exterior `big plaquette'. Each plaquette, including the  exterior one, can carry an elementary curvature excitation. The solid lines represent the embedded graph while the dashed lines represent a possible scheme for coarse--graining the lattice. \label{fig_latt}}
\end{figure}

Note that we are not forced to follow the `linear' construction of the fusion basis as indicated in section \ref{fusionB}: we can be more flexible.
Take for example a regular square lattice%
\footnote{This lattice (or graph) would of course have four--valent nodes. Nevertheless, the techniques developed in this paper can be straightforwardly applied to graphs with nodes of valence higher than three. Alternatively, four--valent nodes can always be expanded into three--valent ones in a regular manner.} 
with $N \times N$ plaquettes, with $N=2^K$ for some $K\in\mathbb{N}$. To obtain the topology of a punctured sphere%
\footnote{Alternatively, we can allow for a punctured torus topology, for which one can also define a fusion basis. This implements periodic boundary conditions.}
we close off this lattice with one `big plaquette', see figure \ref{fig_latt}.  This corresponds to the choice of free boundary conditions for the original lattice.
The  `big plaquette' has 4 two--valent and $(4N-4)$ three--valent nodes along its boundary.

Each plaquette can then carry an elementary curvature (i.e. magnetic) excitations. 
The Gau\ss~constraint violations (i.e. torsion or electric excitations) would usually sit on the nodes of the lattice.%
\footnote{In lattice gauge theories, as well as in loop quantum gravity, one restricts quite often  attention to gauge invariant states, which would make the introduction of open links unnecessary. As we will see, coarse graining of non--Abelian gauge theories does however introduce torsion excitations, and one might want thus to include such cases in the discussion. On the other hand it is straightforward to restrict to a basis which is gauge invariant, by setting the torsion excitations for the initial plaquettes  to zero.}
However, for the regular square lattice it is immediate (and natural) to move also these excitations onto the plaquettes.
To do this, we restore gauge invariance at the two--,  three-- and four--valent nodes, but add an open link  to all the four--valent nodes, pointing in one direction, toward the center of---say---the top-left plaquette. 
   Thus it is only the two--valent and  three--valent nodes on the boundary of the `large' plaquette, for which we ignore the Gau\ss~constraint violations. 
To also allow for a Gau\ss~ constraint violation at the `big  plaquette' we introduce an open link at one of its corners.

We are now in the setting discussed in this paper: each plaquette can be identified with a puncture, which can carry both curvature and torsion excitations. 
To define a fusion basis we have to decide on an ordering in which the punctures or plaquettes are fused to larger ones.  
To reach a homogeneous definition, we can  first fuse pairs of plaquettes in $x$--direction and then pairs of plaquettes in $y$--direction (leaving the `big plaquette' untouched). 
This coarse--graining procedure can be repeated until we remain with just two plaquettes, which represent the two--punctured sphere.

The corresponding fusion basis diagonalizes closed ribbon operators, ${\cal K}_1$ that go around single plaquettes, operators ${\cal K}_2$ that go around pairs (in $x$--direction) of plaquettes, operators ${\cal K}_4$ that enclose quadruples of plaquettes, and so on. 
Correspondingly, the different scales of the basis states are described by sets of representation labels $\{\rho^i_k\}$, where $k=0,1,2, \dots, K$ indicates the scale given by the number $2^k$ of plaquettes surrounded. 

Hence, we see that the fusion basis is ideal to design states with a prescribed multi-scale behavior of observables. 
We expect that this will help to design low--energy states for Yang--Mills (lattice) theory, by merging our tools with the techniques developed for this purpose in the context of tensor network states or MERA (multi--scale entanglement renormalization Ansatz, \cite{VidalM}), see \cite{Ashley} and also \cite{Luca} for some recent developments.

 The advantage of using the fusion basis is that it comes with multi--scale observables, that are automatically diagonalized by the fusion basis itself.

The fusion basis can also be useful in  covariant (space--time) approaches to renormalization and coarse graining of lattice gauge theories and spin--foams \cite{FiniteSF, Clement,BS}. 
Here, the partition function associated to a space--time building block can be represented by a state on the boundary of this block \cite{Ditt12b,DittStein13,DecTNW}. 
Using the fusion basis to represent this state would allow to keep control in particular over the torsion excitations, which are generated by coarse graining in non--Abelian gauge theories, and which are rather difficult to handle in the spin--network basis (see next section, and especially \cite{Clement}).


\subsection{Coarse--graining in lattice gauge theories and loop quantum gravity}

\subsubsection{Coarse--graining in terms of density matrices}\label{densitym}

We will discuss here the coarse--graining of gauge theory and loop quantum gravity states, explain the intricacies of this procedure, and motivate the use of the fusion basis to define coarse--graining. 
We will work in the context of a fixed (initial)  graph, or lattice, thus the discussion in this section is independent of the question on which representation (ALI versus BF) we use. 
In the context of loop quantum gravity coarse graining has been discussed in \cite{RovelliQuanta, EteraCG, EteraTags, DG14b,BDG15}. 

To start with, one considers an (initial) graph $\Gamma$ and associates to it the Hilbert space ${\cal H}_\Gamma$ of functions $\psi \in {\cal F}({\cal G}^L)$ of the graph connection. 
Here, $L$ denotes the number of links, while the inner product in ${\cal H}_\Gamma$ is given by \eqref{innprod1}.

Coarse graining in a canonical framework is usually discussed using density matrices, which we will here denote by $\D$ (instead of $\rho$ which we reserved for the representation labels of the Drinfel'd double ${\cal D}({\cal G})$). 
Pure states are then represented in the holonomy basis by
\ba\label{pureD}
\D = \Big|\psi \Big\rangle\Big\langle \psi\Big| = \sum_{g,\tilde g} \mathfrak{d}\Big[\{ g\}, \{\tilde g\}\Big]\Big| \{ g\}\Big\rangle\Big\langle\{ \tilde g\}\Big|,
\ea
where
\ba
 \mathfrak{d}\Big[\{ g\}, \{\tilde g\}\Big]\,=\, \psi( \{ g\})\,\, \overline{  \psi( \{ \tilde g\}) } \, .
\ea

The coarse--graining of a density matrix is defined as follows. 
First, choose a splitting of the holonomies $\{ g\}$ attached to the links of the graph $\Gamma$ under consideration, into two sets of finer $\{ g^f\}$ and coarser $\{ g^c\}$ holonomies. 
Starting from a density matrix $\D$ for the initial system, a coarser density matrix can then be defined by summing over the finer degrees of freedom,
\ba\label{cdensity}
\mathfrak{d}^c( \{g^c\}, \{ \tilde g^c\}) \,=\, \frac{1}{|{\cal G}|^{L_f}} \sum_{g^f}  \, \mathfrak{d}( \{g^c\}, \{g^f\}, \{\tilde g^c\}, \{g^f\}) \,,
\ea
where $L_f$ denotes the number of finer links, i.e. those links carrying `finer' holonomies. 

In general, however, the graph under consideration will `break apart' once the finer links are removed.
To avoid this, we can first (unitarily) transform the state onto a lattice where all the finer links one wishes to integrate out are given  by loops, see e.g.  \cite{EteraCG,Ashley}. (In \cite{Ashley} such transformations are called `controlled rotation unitary gates'.) 
This way, removing these loops leaves us with a connected coarser lattice. But this coarse--graining procedure has at least two major drawbacks. 

($a$) Despite providing a certain control over the coarser and finer variables in terms of the holonomies, it completely lacks  control over their conjugated variables, i.e. the (electric) fluxes. 
This is an important issue, especially in the context of loop quantum gravity, where the fluxes encode the metrical information of the (spatial) geometry. 
From this perspective, one would rather be tempted by defining a coarse--graining procedure in terms of both holonomies and flux variables.

($b$) Moreover, the coarse density matrix is in general gauge invariant only under diagonal transformations, that is under those gauge transformations which agree in their action on the $\{g^c\}$ and $\{\tilde g^c\}$ variables. 
This is the  case even if the finer density matrix was invariant under arbitrary gauge transformations at every single node. 
This full invariance holds in particular for pure density matrices constructed as in \eqref{pureD} from gauge invariant states. 
Note that this issue arises only in \textit{non}--Abelian gauge theories.
Indeed, in Abelian gauge theories this issue does not appear if the finer variables are associated to loops. 
The reason being that gauge transformations  act by adjoint action on holonomies associated to loops and are therefore trivial for Abelian structure groups. 
In other terms, in this latter case one has a simple procedure to coarse--grain gauge--invariant variables.
On the contrary, for non--Abelian gauge theories, the spin--network basis for density matrices is not stable under coarse graining, and is therefore quite inconvenient for this purpose.

Of course, one could consider an extension to non--gauge invariant spin--network states as proposed in \cite{EteraCG,EteraTags}, however, the main appeal of the spin--network basis is the straightforward implementation of gauge invariance. 
 
A neat solution to issue ($b$) would consist in providing a basis which allows for a coarse--graining in terms of gauge--invariant variables. Here, one not only needs a maximally commuting subset of observables (which specifies a choice of basis), but also their conjugated observables, which brings us back to issue ($a$).

Before discussing a proposal for such a procedure in the next section, let us mention another possibility based on density matrix factorization.
This consists in finding a transformation which decouples the finer holonomy variables (which we assume to be based on loops) from the rest of the state. 
That is, after the transformation, the density matrix takes the product form
\ba
\mathfrak{d}\Big[ \{g^c\}, \{g^f\}, \{\tilde g^c\}, \{g^f\} \Big] \,=\,  \mathfrak{d}^c \Big[ \{g^c\},  \{\tilde g^c\}\Big]  \times \mathfrak{d}^\text{loops}\Big[\{g^f\},\{g^f\}\Big] .
\ea
Upon coarse--graining, this would simply yield the density matrix $\D^c$. 
In this case $\D^c$ would be fully gauge invariant---provided this is the case for the initial density matrix $\D$. 
However, such a transformation which allows us to cast density matrices into a product form, clearly depends on the initial states. 
Therefore, the coarse--graining itself would not be controlled by a choice of coarser and finer observables, but rather by the form of the initial states. 
We mention this possibility here, because this type of decoupling of finer and coarser variables underlies the MERA approach \cite{VidalM}.

\subsubsection{Coarse--graining based on the splitting of the  observable algebra}

Let us now discuss a coarse--graining procedure in which the splitting of the observable algebra into coarser and finer variables is central.  Here one can consider the kinematical, that is gauge covariant  observable algebra, or the algebra of almost\footnote{since we will use a root that provides a global reference system} gauge invariant observables. Such splittings of the observable algebra are also important for the construction of the continuum Hilbert spaces by an inductive limit \cite{DG14b} or projective techniques \cite{Lanery}. 

We will use a phase space description, and to this end assume that ${\cal G}$ is a compact semi--simple Lie group. 
In this case%
\footnote{See e.g. \cite{DG14b} for a more detailed review of the phase space structures.}
the phase space associated to a graph is given by pairs $(g_l,X_l)$ for each link $l$ of the graph. $X_l\in\text{Lie} ({\cal G})$ are the Lie algebra valued (electric) fluxes. 
We will often express them in the basis $\tau^i$ as $X_l=\sum_i X^i_l \tau^i$. 
The phase space carries the {canonical} symplectic structure
\ba\label{PB1}
\{X^i_l, X^j_{l'}\}\,=\,\delta_{l,l'}f^{ijk}X^k_l \, , \q \{X^i_l,g_{l'}\} \,=\, \delta_{l,l'} g_l \tau^i - \delta_{l^{-1},l'} \tau^i g_{l'} \q\text{and}\q \{g_l,g_{l'}\}=0 .
\ea
Here, it is understood that we associate to an inverted link $l^{-1}$ an hololonomy $g_{l^{-1}}=g^{-1}_l$ and a flux $X_{l^{-1}}=-g_l X_l g_l^{-1}$.
Both fluxes and holonomies transform under gauge transformations, which are parametrized by $\{u_n\in{\cal G}\}_n$, with $n$ labeling the nodes of the graph: 
\ba
g_l \rightarrow  u_{t(l)} g_l u^{-1}_{s(t)} \q \text{and}\q  X_l \rightarrow u_{s(l)} X_l u^{-1}_{s(l)} .
\ea
Here, $s(l)$ and $t(l)$ denote the source and target nodes of $l$, respectively. 

A coarse--graining procedure based on gauge--covariant observables can be achieved in two steps, as follows. 
(1) First, find a canonical transformation such that the new variables split into coarser and finer sets of variables. 
Crucially, the sets of coarser and finer variables must commute with each other.
Also, one should take care of preserving the form of the symplectic structure given by (\ref{PB1}), since this is at the basis of the interpretation of the variables in terms of holonomies and fluxes.
(2) Then, as before, one can simply use a polarization of the wave functions in the new holonomy variables and integrate out the finer holonomy variables, while keeping the coarser holonomies fixed, as in (\ref{cdensity}). 
The coarser holonomies and the coarser fluxes give (conjugated) observables characterizing the coarser states. 

Therefore, this procedure is not different from the one described at the end of section \ref{densitym}, but rather an amendment thereof. This amendment, which basically prescribes in more detail how to split the holonomies into coarser and finer sets, allows us, to gain control over the coarse--graining of the fluxes as well.

Now, one can ask what kind of transformations would preserve the symplectic structure (\ref{PB1}) hence keeping the interpretation of the variables as holonomies and fluxes intact. 
Examples for such transformations are constructed in detail in \cite{DG14b,BDG15,Lanery2}.  
We review the construction in \cite{DG14b,BDG15} shortly, as it is closely related to ribbon operators. 

Holonomies are easy to treat, since we can simply consider compositions $g_{l'}= g_{l_n} \cdots  g_{l_1}$ that result in `new' holonomies $g_{l'}$ attached to `new' links $l'=l_n\circ\dots\circ l_1$ built out of the corresponding links on the initial graph. 
For the fluxes, we can consider combinations of the following type (see \cite{DG14b} for more detailed definitions)
\ba\label{cFl}
X_{l'}:= \sum_{l \in {\bf S}(l')}   g^{-1}_{s(l) s(l')}  X_l \, g_{ s(l) s(l')} .
\ea
Here, ${\bf S}(l')$ is a set of links, so that the dual of these links form a connected path made out of edges of the triangulation (or---more generally---out of edges of the dual complex to the graph under consideration). 
This connected path should be interpreted to be dual to a `new' link $l'$.
The holonomies $g_{s(l) s(l')}$ denote the parallel transport from a node $s(l')$, which will be the source node of the new link $l'$ to the source node $s(l)$ of the link $l$.
In this way, we sum up the fluxes in one and the same reference system, provided by $s(l')$. 
An explicit procedure to find phase space splittings into coarser and finer variables based on such transformations can be found in \cite{DG14b}.
\begin{figure}[h]
	\includegraphics[scale =1]{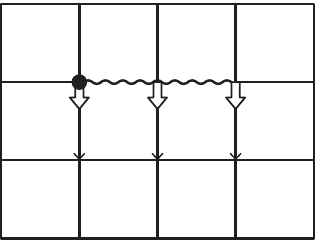}  \q \q \q  \q 
	\includegraphics[scale =1]{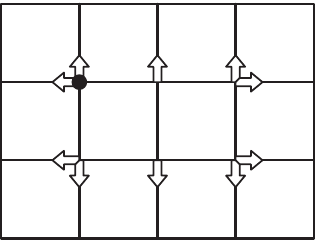} 
	\caption{Each link of the graph is associated with an holonomy and the thick arrow represents the corresponding conjugated flux. The black dot represents the root at which  gauge invariance is relaxed. The left panel shows an example of construction of a coarser flux. Fluxes are first parallel transported to the root following the path defined by the wiggly line and then added to form a coarser flux. The right panel shows an example of closed path. As before, fluxes must be first parallel-transported to a common frame before being added to each other. In presence of curvature, this sum will not be zero meaning that the Gau{\ss} constraint is violated. }
\end{figure}

In a loop quantum gravity context, the fluxes are $\mathfrak{su}(2)$--valued and encode the edge vectors---for the edges dual to the links---in a reference frame associated to the nodes of the graph. 
This interpretation also holds for the coarse--grained fluxes (\ref{cFl}) and thus justifies such a coarse--graining prescription.  

The exponentiated action of the  coarser or `integrated' fluxes, as defined in \eqref{cFl}, agrees with the translation part of the open ribbon operators as discussed in section \ref{oribbon}, see \cite{BDG15}. 
The holonomy (or multiplication) part of the ribbon operator is also constructed via a composition of holonomies, following the same prescription we employed in this section. 
Hence, ribbon operators use a coarse--graining of fluxes and holonomies analogous to the one described here.
There is nevertheless an important difference. The holonomy part and the translational part of a ribbon operator commute with each other: the translational part corresponds to a flux integrated along a path of edges in the triangulation, and the holonomy part  is associated to a path in the dual lattice. 
For the ribbon these two paths are parallel (in fact could be seen as the boundaries of the ribbon), whereas a holonomy and its conjugated flux are based on a link and edge respectively, that are transversal to each other. 

We have so far discussed the gauge covariant phase space. 
The coarse--graining procedure described above would give us control over the coarser holonomy and flux variables, but would suffer also from a violation of (full) gauge invariance for the coarse density matrix. 
In the canonical formalism, gauge invariance is encoded in Gau\ss~constraints associated to each node $n$,
\ba\label{Gauss2}
G_n\,=\, \sum_{ l: s(l)=n} X_l \,+\, \sum_{l: t(l)=n} X_{l^{-1}} .
\ea
Geometrically, these constraints demand the closure%
\footnote{Sometimes, Gau\ss~constraints are renamed `closure constraints'.}
into a polygon of the edges dual to the links ending or starting at $n$. 

With such an interpretation, one might expect that the Gau\ss~constraint are preserved under coarse--graining, as the coarser Gau\ss~constraints would demand that the coarser edges of the coarser triangulation close, too. 
This is, however, generally not the case \cite{DG14b}. The reason is the following.
The coarser fluxes have to be parallel transported to a common frame. 
Now, if this parallel transport has to go through a region with curvature, the coarser Gau\ss~constraint will in general not hold. 
 Hence, we effectively obtain torsion, defined as a violation of the Gau\ss~constraint, due to the presence of curvature. 
Such an effect, which was named curvature--induced torsion in \cite{DG14b}, is strictly related to the need of deforming the Gau\ss~constraint in phase spaces describing piecewise homogeneously--curved  (instead of piecewise flat) geometries \cite{NewRegge, BonzomEtAl, Flo2,EteraCharles} (see also \cite{HHR-AHP16}, for an analysis in four dimensions).
In terms of defect excitations discussed in this paper,  torsion excitations interpreted as spinning particles can arise from the fusion of two spinless defects, since two particles can have orbital angular momentum. 

As mentioned, we could attempt to use gauge invariant variables for the coarse--graining, which would avoid loosing gauge invariance. 
At the phase space level, this would also mean that the Gau\ss~constraints become redundant. 
In fact, implementing gauge invariance in phase space means to consider only the constraint hypersurface where the Gau\ss~constraint is satisfied, and at the same time only gauge invariant functions on such phase space.

But it is a very involved task to come up with a phase space description involving only completely gauge invariant observables. 
It is much easier to work with an almost gauge invariant set--up.
This consists in choosing a node, called root $r$, as global reference frame, to which all fluxes and holonomies are parallel transported.
In other words, in considering only   root--based holonomies and fluxes.
The resulting variables are invariant under all gauge transformations, except for those at the root, which act by adjoint action.

A further, necessary, step consists in identifying an independent set of observables.
Indeed, the fluxes are related by the Gau\ss~constraints (\ref{Gauss2}), while holonomies along loops must satisfy certain algebraic relations.
Such an independent set of variables can be obtained by choosing a rooted spanning--tree of the underlying graph $\Gamma$. 
This defines leaves $\ell$, i.e. links which are not part of the tree.
Each leaf $\ell$ defines a unique closed loop, which starts and ends at the root and which contains only links of the tree and the one leaf $\ell$. 
Hence, each leaf defines an holonomy variable $h_\ell$. 
Furthermore, for what concerns the fluxes, we can consider the leaf's flux  $X_\ell$ parallel transported to the root: ${\bf X}_\ell =   g^{-1}_{s(\ell)  r}     X_\ell  \, g_{s(\ell)  r}$.
The set of variables $\{(h_\ell, {\bf X}_\ell)\}_\ell$, with $\ell$ running through all leaves, gives a complete parametrization of the almost gauge invariant phase space.%
\footnote{The leaf--associated loops allow the reconstruction of all other (root based) loops by construction. Furthermore we are only left with the fluxes associated to the leaves. The fluxes associated to the remaining links can be reconstructed using the Gau\ss~constraints (see \cite{DG14b} for the procedure). }
And they do so by essentially preserving the form of the symplectic structure of the gauge covariant phase space:
\ba
\{{\bf X}^i_\ell, {\bf X}^j_{\ell'}\}\,=\,\delta_{l,l'}f^{ijk}{\bf X}^k_\ell \, , \q \{{\bf X}^i_\ell,h_{\ell'}\} \,=\, \delta_{\ell,\ell'} h_{\ell} \tau^i - \delta_{\ell^{-1},\ell'} \tau^i h_{\ell'} \q\text{and}\q \{h_\ell,h_{\ell'}\}=0 .
\ea

The coarse--graining procedure can now be run analoguesly to the gauge covariant case.
In  particular, there is a well defined sense in which both the graph and the tree are coarse--grained to a coarser graph and tree. 
Based on such a choice of coarser graph and tree, one can perform a split into coarser and finer variables, as needed for coarse--graining.
All this is discussed in detail in \cite{BDG15}. 

Notice that within this procedure, one is working with an ``almost'' gauge--invariant state space, which after coarse--graining still captures the ``almost'' gauge--invariant observables. 
Therefore, we have in this way exhibited a structure which is stable under the coarse--graining procedure. 
This comes, however, at a price: whereas all the initial fluxes could have been reconstructed via the Gau\ss~constraint, this is not the case at the level of the coarser fluxes. 
Of course, one could use some (ad hoc) Gau\ss~constraints of the form (\ref{Gauss2}) to define fluxes associated to the links of the coarse tree.
But these fluxes would not correspond to the fluxes one obtains via coarse--graining from the finer fluxes. 
 In this sense, one looses {important} information, which in the case of loop quantum gravity encodes the coarser spatial metric.
Once again, the underlying reason is that curvature can lead to torsion on a coarse--grained level.  
This is naturally taken into account in a coarse--graining scheme based on the fusion basis.
We now turn to describing such a scheme.

\subsubsection{Coarse--graining in the fusion basis}

 As discussed in section \ref{multiscale}, the fusion basis diagonalizes operators which can be naturally interpreted as describing different coarse--graining scales. 
Therefore, the fusion basis $\psi_{\mathfrak{f}}[\{\rho_j\},\{I_k\}]$ comes equipped with a natural coarse--graining scheme, in which one sums directly over its ${\cal D}(\mG)$ representation labels.

Let us first review, what kind of observables these representation labels are related to.  
As discussed in section \ref{clchri}, the fusion basis diagonalizes the closed charge ribbon operators associated to the fusion tree structure.
More generally, closed ribbon operators ${\cal K}_i[\rho']$ project onto states $\psi_{\mathfrak{f}}[\{\rho_j\},\{I_k\}]$ for which $\rho_i=\rho'$, where $i$ and $j$ label the branches in the fusion tree associated to the fusion basis, whereas $k$ labels its one--valent nodes (i.e. its endpoints).
 
We want now to compare the coarse--graining procedure provided by the fusion basis and closed ribbon operators, to the one provided by the holonomy polarization in the almost gauge invariant set--up.  
To this end, we assume that we work with a $p$--punctured sphere $\S_p$ and a minimal graph, but do not have torsion excitations at the punctures, i.e. we have gauge invariant wave functions. 

The holonomy polarization uses a basis which can be symbolically written as $\psi[\{G_\ell\}](\bullet)= \prod_\ell \delta(G_\ell, \bullet)$, where $\ell$ runs over the leaves of a spanning tree in the graph $\Gamma$, and the number of leaves is given by $|\ell|=p-1$.  
The operators diagonalizing this basis are given by root--based Wilson loop operators $W^f$.
These Wilson loops need \textit{not} be restricted to class functions, i.e. functions of the trace of the loop holonomy. 
 
On the other hand, the closed ribbon operators ${\cal K}_j[C_j,R_j]$ are fully gauge invariant observables. 
In particular, the label $C_j$ measures the conjugacy class (or trace) of the Wilson loop along the closed ribbons, instead of the full loop holonomy. 
While with the holonomy basis we describe holonomies only around a fundamental set of $(p-1)$ cycles, the  subindex $j$ of the fusion basis $\psi_{\mathfrak{f}}[\{\rho_j=(C_j,R_j)\},\{I_k=(M_k,i_k)\}]$ runs over  $(2p-3)$ values and we have as many (not completely independent)\footnote{The closed charge ribbon operators result \textit{not} being completely independent, since their possible results are restricted by the coupling rules. E.g. for an Abelian theory, all the coarser closed ribbon operators are determined by the finest closed ribbons around $(p-1)$ punctures.} closed ribbon operators.

In addition to the holonomy information, the closed ribbon operators ${\cal K}_j[C_j,R_j]$ encode flux observables within the labels $\{R_j\}$. 
More precisely, these are integrated fluxes associated to a closed path. 
Note that in the almost gauge invariant phase space, discussed in the previous section, we only had fluxes associated to open paths,%
\footnote{Before coarse--graining one can obtain fluxes associated to closed paths by combining the open paths fluxes and by using the Gau\ss~constraints to reconstruct the fluxes associated to the tree links. The  Gau\ss~constraints are however not anymore valid for `coarse grained' nodes.}
provided we assume that the graph does not include loops (i.e. links with the same source and target node).  

 Another way to talk about the $\{R_j\}$ labels is to say that they measure torsion (electric charge). 
Indeed, if we assume a gauge invariant state we immediately find (see section \ref{Pfusion}) that $R_j$ is equal to the trivial representations for those ribbons ${\cal K}_j[C_j,R_j]$ that go directly around the punctures.
Notice that there are exactly $p$ of such punctures.
But for non-Abelian groups, we can well have non-trivial labels $R_j$ for the remaining $(p-3)$ closed ribbons. 
These observables are crucial to keep track of how the Gau\ss~constraint gets deformed under coarse--graining.

As explained in section \ref{Pfusion}, a basis for fully gauge invariant states satisfying the Gau\ss~constraints at all punctures can be obtained from the fusion basis $\psi_{\mathfrak{f}}[\{\rho_j\},\{I_k\}]$ by setting the appropriate indices $R_k$ equal to the trivial representation, denoted by $0$. Correspondingly, the relative $M_k$ indices in the $I_k=(M_k, i_k)$ multi-index are also trivialized. 
Furthermore, we also sum over the $\{i_k\}$ labels 
\ba
\psi^{\rm g.i.}_{\mathfrak{f}}[\{(C_{m},R_{j})\}, \{C_k\}]\,=\,  \left(\prod_k  \frac{1}{\sqrt{ |Q_{C_k}|}}\right)  \,\, \sum_{\{i_k\}} \psi_{\mathfrak{f}}[\{(C_{m},R_{m})\}, \{C_{k},0\}; \{i_k,0\}].
\ea
In this formula, we have split the index $j$ running over the edges of the fusion tree into two sets $k$ and $m$, labelling the punctures (or leaves of the fusion tree) and the remaining edges of the fusion tree, respectively.

Note that if one allows for states violating the Gau\ss~constraints at the punctures, the $\{I_k\}$ labels encode only local information, and are measured by projective operators given by gluing cylinder fusion basis states $\psi_{\mathfrak{f}}^{\mathbb{S}_2}[\rho, I,I]$. 

The closed ribbon operators ${\cal K}_j[C_j,R_j]$ and the operation of gluing of the cylinder fusion basis states give together a maximal commuting set of observables characterizing the fusion basis.  
Coarse--graining in the fusion basis means that these observables determine the splitting into coarser and finer ones. 

In addition, there are conjugated observables, given by open ribbon operators extending from one puncture to another puncture. 
We leave it to future research to find a complete set of such independent operators. 
The coarse--graining scheme based on the fusion basis will also induce a splitting of the conjugated observables into a coarser and a finer set.
To deduce this splitting one needs to study in more detail the commutation relations or the corresponding symplectic structure in phase space.

The coarse--graining is now given by summing over the finer variables, just as usual.
Consider a density matrix defined using a fusion basis by 
\ba
\D \,=\, \sum_{\{\rho\},\{I\} }    \sum_{\{\tilde \rho\},\{\tilde I\} }    \d \Big[  \{\rho\},\{I\};  \{\tilde\rho\},\{\tilde I\}   \Big]   \,\,\,      \Big|  \psi_{\mathfrak{f}}[\{\rho\},\{I\}] \,\,   \Big \rangle  \Big \langle \psi_{\mathfrak{f}}[\{\tilde \rho\},\{\tilde I\} \Big|.
\ea

This density matrix is adapted to the intended coarse--graining (or fusion) of punctures into new `larger' punctures. That is  the $p$ punctures are partitioned into $p'\leq \tfrac12 p$ sets, each including at least two punctures.  The fusion tree needs then to include a subtree for each set that describes the fusion of the punctures in this set.  We label the variables attached to the subtrees with a super-index $f$, except for the pairs $(C_s,R_s)_{s=1}^{p'}$, which prescribe the excitations for the fused punctures.  We label these pairs and the remaining variables with a super-index $c$.

Working in the polarization given by the fusion basis variables, the coarser density matrix is then defined by

\ba
\d^c\Big[  \{\rho^c\},  \{\tilde \rho^c\} \Big]\,=\,   \sum_{\{\rho^f\}, \{I^f\}} \d\Big[  \{\rho^c\},\{\rho^f\}, \{I^f\};  \{\tilde\rho^c\},\{\rho^f\},\{ I^f\}   \Big]   .
\ea
In this scheme we get rid of all indices $I_k$, assuming that these are all classified as finer information. An alternative scheme introduces new indices $I^c_{k'}$ for the coarser punctures. This scheme is based on an extension of the Hilbert space (before coarse--graining).  We will explain this scheme%
\footnote{Note that for the definition of entanglement entropy based on subdividing the manifold into two regions the extension involves only a very limited set of additional degrees of freedom, which arises from cutting one edge of the fusion tree. In contrast, for the application to coarse--graining, one usually needs to cut many trees of the fusion tree. Correspondingly the extension would involve a much larger set of additional degrees of freedom.} 
in the forthcoming work \cite{Toappear}, where we will also discuss entanglement entropy.

\section{Discussion}\label{discussion}

In this work we introduced  the fusion basis for $(2+1)$ dimensional lattice gauge theories, in particular with non--Abelian structure groups. The basis is well adapted for the weak coupling regime and for describing topological BF theory with defects. The latter theory can also be taken as a description of $(2+1)$ dimensional gravity coupled to (possibly spinning) point particles. 

In contrast to the spin--network basis \cite{RovSmol}, the fusion basis is a multi--scale basis. It diagonalizes the traces of a certain multi--scale set of Wilson loop observables. This set does in itself not form a maximal set of commuting observables: for non--Abelian gauge groups one has rather to add further gauge--invariant observables describing  electric excitations.  Importantly, the electric (or torsion, in a gravitational context) excitation might emerge on larger scale even for gauge invariant states. This fact make it hard to control large scales in a spin--network basis.  For this reason, the fusion basis is ideally suited for coarse--graining  in lattice gauge theories and loop quantum gravity \cite{EteraCG,Clement,EteraTags,FiniteSF,Luca,BS,DittStein13,DecTNW}.

 More specifically, we have seen that the  fusion basis comes with a number of advantages. That we list and comment here below.
\begin{itemize}
\item The fusion basis incorporates the notion of basic excitations and their fusion to coarse grained excitations, hence making explicit the quasi--local structure of the excitations relative to the BF vacuum;
\item it makes transparent the Drinfel'd double algebra structure, which in past (loop quantum gravity) discussions was rather hidden in the algebra of constraints \cite{BonzomEtAl, PranzEtAl}; 
\item it incorporates a notion of cutting and gluing pieces of spatial manifolds along boundaries and thus comes automatically with a natural notion of local subsystems (see e.g. \cite{DonnFreid} for a different notion).
\end{itemize}
Moreover, in the context of $(2+1)$ dimensional loop quantum gravity coupled to point particles, the fusion basis
\begin{itemize}
\item  provides naturally and directly the physical states of the theory, even for states including spinning particles (i.e. states with torsion);
\item and diagonalizes the gauge and diffeomorphism invariant (Dirac) observables of the theory, which are given in terms of the closed ribbon operators.
\end{itemize}
This shows that the fusion basis is a convenient tool for describing the coupling of multiple particles to $(2+1)$ dimensional gravity. It would be of particular interest to consider a thermodynamic or continuum limit, possibly resulting in gravity coupled to a matter field, see also \cite{Karim2}. A further question in this direction is whether the resulting system can be described by a matter field propagating on an effective non--commutative space--time, as derived in a covariant framework by \cite{FreidelLivine}.

The use of the fusion basis emphasizes the Drinfel'd algebra or quantum double  structure of $(2+1)$ gravity coupled to point defects. This facilitates the comparison with other quantization schemes \cite{KarimCath}, such as the combinatorial quantization for Chern-Simons theory \cite{Rosly}.  Let us also point out the recent work \cite{MeusRecent}, which reformulates Kitaev models as a special case of combinatorial quantization via a reformulation of the latter in terms of a Hopf-algebra gauge theory.
~\\

Furthermore, in relation to coarse--graining, we emphasized that the fusion basis
\begin{itemize}
\item solves a deep problem related to coarse--graining in the spin--network basis: in non--Abelian gauge theories, coarse--graining generally leads to torsion degrees of freedom, even though these are not initially present, therefore the spin--network basis cannot be stable under coarse--graining. The fusion basis, on the other hand, incorporates torsion degrees of freedom from the onset, hence allowing for a consistent coarse--graining scheme;
\item moreover, the fusion basis can be naturally used to design multi--scale states, in the sense that it diagonalizes a set of operators defined at all available scales (cf. e.g. \cite{Ashley}), a fact that makes it ideal for discussing coarse--graining schemes.
\end{itemize}
We hope to make all this explicit within a new tensor network coarse--graining framework, by generalizing the recently developed schemes of \cite{DecTNW,Clement}. One of our principal aims is studying the continuum limit and coarse--graining of spin--foam models \cite{BD14,FiniteSF,Clement,BS}. In the context of $(2+1)$ dimensional gravity models, a particularly intriguing question is how to  flow via coarse--graining from models based on flat building blocks to models based curved building blocks, hence recovering in the quantum theory the classical result of \cite{NewRegge}.
More specifically, for spin--foam models one expects a transition from $\SU(2)$ to the quantum deformed $\SU(2)_q$. This requires besides a condensation of curvature degrees of freedom to a constant curvature state, also a condensation of torsion degrees of freedom. It is therefore important to have coarse--graining schemes which do not throw away the torsion degrees of freedom.
Notice also that the fusion basis is already available for $\SU(2)_q$, with $q$ root of unity,  \cite{KKR,DG16} and has in some aspects even a simpler structure than in the finite group case. (Even more so, if one compares with $\SU(2)$, since the corresponding fusion category is not finite.) The finiteness of $\SU(2)_q$ makes this choice particularly attractive for numerical approaches to coarse--graining, see e.g. \cite{DMS13}.\\

We have also anticipated that
\begin{itemize}
\item the fusion basis can be used to provide a new definition of entanglement entropy for non--Abelian gauge theories.  Let us add here, that in contrast to other proposals, this new definition leads to finite results, even in the continuum limit, for states describing a finite number of defects \cite{Toappear}.
\end{itemize}

Finally, we conclude with some remarks about open problems and further directions of investigation.

In this paper our analysis has been confined to lattice gauge theories with finite gauge groups. It is, however, an important point to generalize our analysis to Lie groups. For Lie groups there are two very different choices for the underlying topology of the state space and the related inner product. One possibility is to choose a discrete topology and measure on the gauge group, which is in fact necessary for the BF representation for {\it continuum} loop quantum gravity. In particular, this is needed for the BF vacuum to have a  finite norm \cite{BDG15}. Alternatively, if one is only interested in lattice gauge theory with a fixed lattice or with a fixed number of excitations (i.e. of punctures), one can also adopt the usual (continuous) Haar measure on the gauge group. Drinfel'd double representations and their tensor product, based on this choice, have been discussed in the case of $\SU(2)$ in \cite{koornw}. 

At last, a further interesting research direction consists of generalizing the fusion basis to $(3+1)$ dimensions. One possibility to achieve this is to use the idea formulated in \cite{DelDitt,Dittrich:2017nmq}, where it is proposed to formulate a $(3+1)$ dimensional theory of flat connections with defects, as a theory of flat connections in one less dimensions at the price of defining it on some topologically complicated Cauchy surface based on a Heegard splitting of the 3-dimensional hypersurface. The idea is then to use the machinery described in this paper to define a basis and operators for the  $(3+1)$ dimensional theory, in order to find a characterization of its basic excitations.  Also in $(3+1)$ dimensions, one expects the maximal set of commuting observables to include a multi--scale set of Wilson loops. Torsion degrees of freedom, on the other hand, are now captured by electric flux observables based on closed surfaces, as defined in \cite{DG14b,BDG15}. However, using the techniques of \cite{DelDitt}, these can be recast into ribbon operators acting around the non trivial cycles of the topologically complicated Cauchy surface designed to encode the $(3+1)$ dimensional theory.

\acknowledgements

We thank Laurent Freidel and Marc Geiller for insightful discussions,  as well as Etera Livine for many exchanges about the coarse--graining of spin--networks.

CD is supported by an NSERC grant awarded to BD.
This research was supported in part by Perimeter Institute for Theoretical Physics. Research at Perimeter Institute is supported by the Government of Canada through the Department of Innovation, Science and Economic Development Canada and by the Province of Ontario through the Ministry of Research, Innovation and Science.


\appendix


\newpage
\section{Inductive limit on $\{{\cal H}_p\}$\label{app_inductive}}

In this appendix, we explain how to construct an inductive limit of the Hilbert spaces ${\cal H}_p$ over the number of punctures $|p|$. Here, the index $p$ in $\mathbb{S}_p$ and ${\cal H}_p$ will not only denote the number of punctures (which we here denote by $|p|$) but also the embedding information of the punctures. Importantly, this embedding information includes not only the position of the punctures, which themselves are infinitesimally small disks removed from the manifold. But also a marked point on the boundary of each of these small disks. Alternatively a puncture can be described by a point in the manifold equipped with a tangent vector at this point \cite{Kir}. The open links of the graphs have to end at the marked points of the small disks (in one description) or to approach the punctures tangential to the associated vector (in the other description). 
This additional structure of the punctures allows the inclusion of torsion defects and is also crucial in order to make the gluing operation well defined. It however leads also to an entire family of continuum Hilbert spaces constructed via the inductive limit, as we will now explain.

We fix once and for all a coordinate atlas for the sphere $\mathbb{S}$. For the inductive limit we need to specify a partial ordering of a label set. This label set will be given by the punctures, including their embedding (and marked points) information. Given two punctured spheres $\S_p$ and $\S_{p'}$ we say that $\S_{p'}$ is a refinement of $\S_p$, denoted by $p \prec p'$, if all the punctures of $\S_p$ are punctures of $\S_{p'}$. That is there are $|p|$ punctures in $\S_{p'}$ whose positions and marked points agree with those of the punctures in $\S_p$.  

Thus we have a family of Hilbert spaces ${\cal H}_p$, labelled by the embedding information $p$ of a set of $|p|$ punctures. Theses labels are now equipped with a directed partial order $\prec$. What is needed for completing the definition of an inductive Hilbert space is the specification of (consistent) embedding maps $\iota_{pp'}: {\cal H}_p \rightarrow {\cal H}_{p'}$ for any pair $p \prec p'$. 

\begin{figure}[h]
	\includegraphics[scale =1]{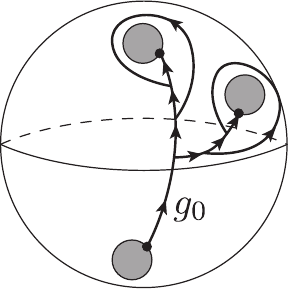}  \q \q \q  \q 
	\includegraphics[scale =1]{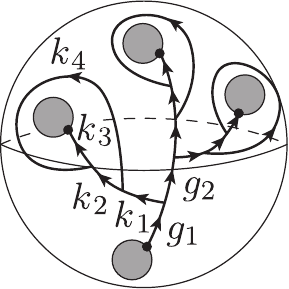} 
	\caption{ Adding a puncture and extending correspondingly the minimal graph. \label{figappA}}
\end{figure}

We construct such an embedding map as follows: given a state $\psi$ on ${\mathbb S}_p$ we can w.l.o.g. assume that this state is defined on a minimal graph $\Gamma_{p}$.  We add links and nodes to $\Gamma_p$ so that is becomes a minimal graph for $\Gamma_{p'}$.  This is done in $|p'|-|p|$ steps and we denote the graph after each step by $\Gamma_{p+i}, i=1,\ldots, |p'|-|p|$.  We label the new punctures with $i=1, \ldots,  |p'|-|p|$. For a new puncture added in the $i$--th step we do the following: we enclose  the new punctures with a cycle and furthermore add a path (composed out of links) that connects the marked point of the puncture to some (new) node on  the graph $\Gamma_{p+i-1}$, see Figure \ref{figappA}.

The embedding map from ${\cal H}_{p}$ to ${\cal H}_{p'}$ can be given as a composition of maps $\iota_{p_1p_2}$ with $|p_2|=|p_1|+1$.  For $|p'|=|p|+1$ the refinement of the graph adds the following additional holonomies:
\begin{itemize} \parskip -2mm
\item  One link $l_0$ of $\Gamma_p$ needs to be subdivided into two links $l=l_2 \circ l_1$ with a new two--valent node $n$. Correspondingly we will have two new holonomies $g_{l_1},g_{l_2}$ satisfying $ g_{l_0}=g_{l_2}g_{l_1}$. 
\item There will be a new path consisting of three new  links $l'_3 \circ l'_2 \circ l'_1$ starting from the new node $n$ to the marked point of the puncture. We will denote the associated holonomies by $k_1,k_2,k_3$. 
\item Furthermore there is one additional link $l'_4$ starting at the target node of $l'_1$ surrounding the new puncture and ending at the target node of $l'_2$. We denote the corresponding holonomy by $k_4$.
\end{itemize}

The embedding $\iota_{pp'}: {\cal H}_p \rightarrow {\cal H}_{p'}$ mapping a state $\psi(g_{l_0}, \{g_l\}_{l\neq l_0})$ is then given as
\ba\label{embeda}
\iota_{pp'}( \psi) (g_{l_1},g_{l_2}, \{g_l\}_{l\neq l_0}, k_1, \ldots ,k_4) \,=\, \sqrt{{\cal G}} \, \delta(   e,\, k_2^{-1} k_4) \,\, \psi( g_{l_2} g_{l_1} , \, \{ g_l\}_{l\neq l_0}) \q .
\ea  
 That is the refined state describes a trivial holonomy for the cycle around the new puncture, whereas it is totally spread on the holonomy $k=k_3 k_2 k_1$, that gives the parallel transport from the added node on the coarser graph to the new puncture. In summary the new puncture carries neither a curvature nor a torsion excitation. 
 
The embedding maps are consistent, that is reaching a given refinement via different smaller steps, leads to the same result. We have further chosen the pre--factor  in (\ref{embeda}) so that the embedding map is isometric. 

Consider now a family of Hilbert spaces $\{{\cal H}_p\}_{\{p \in {\cal P}\}}$, so that ${\cal P}$ carries a directed partial order.  The inductive limit Hilbert space ${\cal H}_{\cal P}$ is defined as the (closure) of the union over all Hilbert spaces ${\cal H}_p$ with the following equivalence relation imposed: two states $\psi \in {\cal H}_p$ and $\psi' \in {\cal H}_{p'}$ if they can be made equal under some refinement. That is, if there exist a $p''$ with $p\prec p''$ and $p\prec p'$ such that  
\ba
\iota_{pp''}(\psi) \,=\, \iota_{p'p''}(\psi')  \q .
\ea
On this inductive limit Hilbert space one can define an inner product  between as well as the addition of two states $\psi \in {\cal H}_p$ and $\psi' \in {\cal H}_{p'}$ by first embedding these two states in a common refinement Hilbert space ${\cal H}_{p''}$ with $p \prec p''$ and $p' \prec p''$.

Note that the inductive limit Hilbert space ${\cal H}_{\cal P}$ depends on the set of labels ${\cal P}$ and that we demanded that this set is directed. The latter property means that we can find for each two elements $p,p'$ a common refinement $p''$. By construction, this excludes the case that $p,p'$ have a puncture position in common, for which the marked points disagree, since in this case there is no common refinement.\footnote{Alternatively, one can introduce a refinement of the punctures itself, i.e. allow that an arbitrary number of open links end at a given puncture. We will discuss the consequences of this choice elsewhere.} Also, for ${\cal H}_{\cal P}$ to describe a sensible continuum limit one can demand for ${\cal P}$ to include some regular and infinite family of refinements.

\section{Properties of the irreducible representations of the Drinfel'd double \label{app_irrepsandsoon}}

\subsection{Defining property of the representations}
The irreducible representations of the Drinfel'd double are homomorphisms and as such they preserve the algebraic structure. The following shows how the irreducible representations are compatible with the definition of the star-product:

\begin{align} \nn
	&D^{C,R}_{i'M',iM}([\widetilde{G},\widetilde{H}]\star [G,H]) \\[0.5em] \nn &\q = \, D^{C,R}_{i'M',iM}([\widetilde{G}G,\widetilde{H}])\delta(\wG^{-1}\wH\wG,H) \\[0.5em] \nn
	&\q = \, \delta(\wG^{-1}\wH\wG,H)\delta(\wH,c_i')\delta(c_i,G^{-1}\wG^{-1}c_i'\wG G)D^R_{M'M}(q_{i'}^{-1}\wG G q_i) \\[0.5em] \nn
	&\q = \, \sum_{i'',M''}\delta(\wG^{-1}{\wH\wG,H})\delta(\wH,c_i')\delta(c_i,G^{-1}\wG^{-1}c_i'\wG G)\delta(H,c_i'') \\[-1em] \nn
			& \q \q \q \times D^R_{M'M''}(q_{i'}^{-1}\wG  q_i'')D^R_{M''M}(q_{i''}^{-1}Gq_i)  \\[0.5em] \nn
	&\q = \, \sum_{i'',M''}\delta(\wH,c_i')\delta(c_i'',\wG^{-1}c_i'\wG)D^R_{M'M''}(q_{i'}^{-1}\wG q_{i''})\delta(H,c_i'')\delta(c_i,G^{-1}c_i''G)D^R_{M''M}(q_{i''}^{-1}Gq_i) \\
&\q = \,  \sum_{i'',M''}D^{C,R}_{i'M',i''M''}([\wG,\wH])D^{C,R}_{i''M'',iM}([G,H])
\end{align}
where we have first used the linearity of the representations as well as the definition of the multiplication in $\mathcal{D}(\mathcal{G})$ together with the definition of the induced representations, and then the fact that the irreducible representations of the stabilizers preserve the group multiplication rule. 

\subsection{Orthogonality of the irreducible representations}

The space of functions on $\mG \times \mG$ is equipped with an inner product defined by
\be
	\langle f, f' \rangle = \frac{1}{|\mG|}\sum_{G,H \in \mG}f(G,H)\overline{f'(G,H)} \q.
\ee
Furthermore, the matrix elements of the  irreducible representations of the Drinfel'd double form an orthogonal set with respect to this inner product, i.e.
\begin{align}
	&\frac{1}{|\mathcal{G}|}\sum_{G,H \in \mathcal{G}}D^{C_1,R_1}_{i_1'M_1',i_1M_1}( [G,H])\overline{D^{C_2,R_2}_{i_2'M_2',i_2M_2}( [G,H])} \\ \nn
	&\q =\, \frac{1}{|\mathcal{G}|}\sum_{G,H \in \mathcal{G}}\delta(H,c_{i_1'})\delta(c_{i_1},G^{-1}c_{i_1'}G)\delta(H,c_{i_2'})\delta(c_{i_2},G^{-1}c_{i_2'}G)
	\\ \nn
	& \q \q \q \q \times D^{R_1}_{M_1'M_1}(q_{i_1'}^{-1}Gq_{i_1})\overline{D^{R_2}_{M_2'M_2}(q_{i_2'}^{-1}Gq_{i_2})} \\	\nn
	&\q = \, \frac{1}{|C_1|.|N_{C_1}|}\sum_{G \in \mathcal{G}}\delta_{C_1,C_2} \delta_{i_1',i_2'}\delta_{i_1,i_2}\delta(c_{i_1},G^{-1}c_{i_1}'G)
	D^{R_1}_{M_1'M_1}(q_{i_1'}^{-1}Gq_{i_1})\overline{D^{R_2}_{M_2'M_2}(q_{i_1'}^{-1}Gq_{i_1})} \\	
	& \q = \, \frac{\delta_{C_1,C_2}}{|C_1|}\delta_{i_1',i_2'}\delta_{i_1,i_2}\frac{\delta_{R_1,R_2}}{d_{R_1}}\delta_{M_1',M_2'}\delta_{M_1,M_2}. 
\end{align}
where we used between the third and last line the orthogonality of the irreducible representations of the stabilizer $N_{C_1}$. Note that for all $G$, the Kronecker delta $\delta(c_{i_1},G^{-1}c_{i_1'}G)$ ensures that $q_{i_1'}^{-1}Gq_{i_1} \in N_{C_1}$. However it does not mean that the summand is non-vanishing for all values of $G$. As a matter of fact only $N_{C_1}$ configurations are non-vanishing which explains why the summation over $G$ does not produce any additional factors.

\subsection{Completeness of the set of irreducible representations}
The irreducble representations form a complete set of representations, since they resolve the identity on ${\cal D}({\cal G})$:

\begin{align}
	&\sum_{C,R}\sum_{i',M' \atop i,M}d_{C,R} D^{C,R}_{i'M',iM}([G,H])\overline{D^{C,R}_{i'M',iM}([\widetilde{G},\widetilde{H}])} \\
	&\q  = \sum_{C,R} \sum_{i',M \atop i,M}d_{C,R} D^{C,R}_{i'M',iM}([G,H])D^{C,R}_{iM,i'M'}([\widetilde{G}^{-1},\widetilde{G}^{-1}\widetilde{H}\widetilde{G}]) \\
	&\q = \sum_{C,R}d_{C,R} \chi^{C,R}([G\widetilde{G}^{-1},H])\delta(G^{-1}HG,\widetilde{G}^{-1}\widetilde{H}\widetilde{G})\\ 
	&\q = \sum_{R}d_R. |C| \chi^R(q^{-1}_{k(H)}G \widetilde{G}^{-1}q_{k(H)})\delta(G\widetilde{G}^{-1}H,HG\widetilde{G}^{-1})
\delta(G^{-1}HG,\widetilde{G}^{-1}\widetilde{H}) \\ 
	&\q = |C|.|N_{C}|\delta(q^{-1}_{k(H)}G \widetilde{G}^{-1}q_{k(H)})
\delta(G\widetilde{G}^{-1}H,HG\widetilde{G}^{-1})
\delta(G^{-1}HG,\widetilde{G}^{-1}\widetilde{H}) \\ 
	&\q = |\mathcal{G}| \delta(G,\widetilde{G})\delta(H,\widetilde{H})
\end{align}
First, the formula (\ref{contra}) for the complex conjugate of the representation matrices is used together with the property (\ref{defining}); this allows us to obtain the character of the representation labeled by $C,R$. Then, we apply the definition (\ref{delta_stabilizer}) of the delta function for the stabilizer $N_C$ and use the fact that the identity in $N_C$ is the same as the identity in $\mG$.  Finally, we find that this resolves the identity in the following sense: 
\be
	\frac{1}{|\mG|}\sum_{G,H \in \mG} |\mathcal{G}| \delta(G,\widetilde{G})\delta(H,\widetilde{H}) f(\wG,\wH) = f(G,H)
\ee

\subsection{Diagonalization of the star-product \label{app_diagprod}}

As an algebra, the Drinfel'd double is semi-simple and therefore a decomposition into irreducible modules is provided via an idempotent decomposition. This means that the irreducible representations of the algebra are idempotent with respect to the multiplication. In other words, the irreducible representations diagonalize the star-product. Considering two basis states respectively labeled by the irreducible representations $\rho_1$ and $\rho_2$ as well as their corresponding matrix indices, we have indeed the following relation:

\begin{align} \nn
	& \psi^{\S_2}_{\rho_2,I'_2 I_2} \star  \psi^{\S_2}_{\rho_1,I'_1 I_1 } = \\
	&\q = \frac{\sqrt{d_{\rho_1}d_{\rho_2}}}{|\mG|^2}
	\sum_{G_1,H_1 \atop G_2,H_2}
	D^{\rho_2}_{I_2'I_2}([G_2,H_2]) D^{\rho_1}_{I_1'I_1}([G_1,H_1]) \psi^{\S_2}_{G_2,H_2} \star  \psi^{\S_2}_{G_1,H_1}\nn\\
	&\q=\frac{\sqrt{d_{\rho_1}d_{\rho_2}}}{|\mG|^2}
	\sum_{G_1,H_1 \atop G_2,H_2} \sum_{\rho_3}\sum_{I_3',I_3}
	D^{\rho_2}_{I_2'I_2}([G_2,H_2]) D^{\rho_1}_{I_1'I_1}([G_1,H_1]) \delta(G_2^{-1}H_2G_2,H_1) \nn \\
	&\q \q \q \q \q \q \q\times \sqrt{d_{\rho_3}}\psi^{\S_2}_{\rho_3,I_3'I_3}\overline{D^{\rho_3}_{I_3'I_3}([G_2G_1,H_2])}\nn\\
	&\q=\frac{\sqrt{d_{\rho_1}d_{\rho_2}}}{|\mG|^2}
	\sum_{G_1,H_1 \atop G_2,H_2} \sum_{\rho_3}\sum_{I_3'I_3} \sum_{I_3''}
	D^{\rho_2}_{I_2'I_2}([G_2,H_2]) D^{\rho_1}_{I_1'I_1}([G_1,H_1])\sqrt{d_{\rho_3}} \nn \\
	&\q \q \q \q \q \q \q\times \overline{D^{\rho_3}_{I_3'I_3''}([G_2,H_2])} \, \overline{D^{\rho_3}_{I_3''I_3}([G_1,H_1])}
	 \psi^{\S_2}_{\rho_3,I_3'I_3}\nn\\
	&\q=\sqrt{d_{\rho_1}d_{\rho_2}}
	\sum_{\rho_3}\sum_{I_3',I_3} \sum_{I_3''}\frac{1}{d_{\rho_3}^{3/2}}\delta_{\rho_2,\rho_3}\delta_{I_2',I_3'}\delta_{I_2,I_3''}\delta_{\rho_1,\rho_3}\delta_{I_1',I_3''}\delta_{I_1,I_3}
	\psi^{\S_2}_{\rho_3,I_3'I_3}\nn\\
 	&\q= \frac{\delta_{\rho_1,\rho_2}}{\sqrt{d_{\rho_1}}} \delta_{I_2,I_1'} \,  \psi^{\S_2}_{\rho_1,I_2'I_1}
\end{align}
\newpage
\subsection{Completeness of the Clebsch-Gordan coefficients \label{app_complete}}

The Clebsch-Gordan coefficients for the Drienfel'd double are unitary maps and as such they resolve the identity. Indeed, we have 
\begin{align}
	\sum_{\rho}\sum_I
\mathcal{C}^{\rho_1\rho_2\rho}_{I_1'I_2'I} \cdot
\overline{\mathcal{C}^{\rho_1\rho_2\rho}_{I_1I_2I} } &= 
\frac{1}{|\mathcal{G}|}\sum_{G \in \mathcal{G}}\sum_{H' \in C \atop H \in C_1}\sum_{\rho}
	d_{\rho} D^{\rho_1}_{I_1'I_1}([G,H]) D^{\rho_2}_{I_2'I_2}([G,H^{-1}H']) 
	\chi^{\rho}([G,H']) \\
 	&=\sum_{K \in \mathcal{G}}\sum_{H \in C_1}
	D^{\rho_1}_{I_1'I_1}([e,H]) D^{\rho_2}_{I_2'I_2}([e,H^{-1}K]) 
	\\
	&= \delta_{I_1'I_1}\delta_{I_2',I_2} 
\end{align}

where we made use between the first and the second of the following relation
\begin{align}
	\sum_{\rho} d_{\rho}\chi^{\rho}([G,H']) &= \sum_{\rho}d_{\rho}\delta(GH',H'G)\chi^R(q^{-1}_{k_C(H')}Gq_{k_C(H')}) \\ 
	&= |\mathcal{G}| \delta(GH',H'G)\delta(e,G)
\end{align}


\subsection{Invariance property of the Clebsch-Gordan coefficients \label{app_invar}}

We rewrite the defining equation for the Clebsch-Gordan coefficients (\ref{CG0})
\ba
D^{\rho_1}_{I_1'I_1} \otimes D^{\rho_2}_{I_2'I_2}(\Delta[G,H])
	\,=\,  \sum_{\rho_3}\sum_{I_3, I_3'}
	\mathcal{U}^{[\rho_1,\rho_2]}_{I_1'I_2',\rho_3I_3'}\,
	D^{\rho_3}_{I_3'I_3}([G,H])\,
	({\mathcal{U}^{[\rho_1,\rho_2] \dagger}) _{\rho_3I_3,I_1I_2}}
\ea
as
\ba\label{Inv2}
D^{\rho_1}_{I_1'I_1} \otimes D^{\rho_2}_{I_2'I_2}(\Delta[G,H])  \,\,\mathcal{U}^{[\rho_1,\rho_2] }_{I_1I_2,\rho_3I_3}
	&=&  \sum_{\rho_3}\sum_{I_3, I_3'}
	\mathcal{U}^{[\rho_1,\rho_2]}_{I_1'I_2',\rho_3I_3'}\,
	D^{\rho_3}_{I_3'I_3}([G,H]) \q .
\ea
We now make use of the equation
\ba
\sum_{I''_3}\sum_{H\in {\cal G}} D^{\rho_3}_{I_3'I''_3}([G,H])  D^{\rho_3}_{I_3'' I_3}([G^{-1},G^{-1}HG])  \,   \,=\, 
\sum_{H\in {\cal G}}       D^{\rho_3}_{I_3I''_3}([e,H])  \,=\, D^{\rho_3}_{I_3 I''_3}( \mathbb{I}) \,=\, \delta_{I_3 I_3''}  \q 
\ea
after multiplying (\ref{Inv2}) from the right with $D^{\rho_3}_{I_3'' I_3}([G^{-1},G^{-1}H,G])$ and summing over $H$. Resolving the co--product and using the property(\ref{contra}) of the matrix elements under complex conjugation we obtain the invariance property of the Clebsch-Gordan coefficients (remember that $\mathcal{C}^{\rho_1 \rho_2\rho_3}_{I_1 I_2 I_3} = 
	\mathcal{U}^{[\rho_1 ,\rho_2]}_{I_1 I_2,\rho_3I_3}$)
\ba
\sum_{H_1,H_2} \sum_{I_1,I_2,I_3} 
D^{\rho_1}_{I_1'I_1}([G,H_1])   D^{\rho_2}_{I_2'I_2}([G,H_2]) \overline{ D^{\rho_3}_{I'_3 I_3}([G, H_1H_2])  }
\,\,\mathcal{C}^{\rho_1 \rho_2\rho_3}_{I_1 I_2 I_3} &=&
\mathcal{C}^{\rho_1 \rho_2\rho_3}_{I'_1 I'_2 I'_3} \q .
\ea

\subsection{Orthonormality of the fusion basis states \label{app_orthostates}}
The basis states  $\{\psi^{\S_3}_{\frak f}\}$ defined on the three-punctured sphere are orthonormal with respect to the inner product of ${\cal H}_3$ (this is defined in (\ref{innprod1}) and the  following pages):
\begin{align}\nn
	&\Bigg\langle \psi^{\S_3}_{\mathfrak{f}}
	{ \scriptsize\begin{bmatrix}
		\rho_1,I_1' \\ \rho_2,I_2' \\ \rho_3,I_3
	\end{bmatrix}},
	\psi^{\S_3}_{\mathfrak{f}}
	{ \scriptsize\begin{bmatrix}
		\widetilde{\rho}_1\tilde{I}_1' \\ \widetilde{\rho}_2,\tilde{I}_2' \\ \widetilde{\rho}_3,\tilde{I}_3
	\end{bmatrix}} \Bigg\rangle = \\ \nn 
	&\q = \, \frac{1}{|\mathcal{G}|^{13}}\sum_{{g_0,g_1,\dots,g_4 \atop g_1',\dots,g_4'}}
	\sum_{{G_1,H_1,\widetilde{G}_1,\widetilde{H}_1 \atop G_2,H_2,\widetilde{G}_2,\widetilde{H}_2}}
	\sum_{{I_1,\tilde{I}_1 \atop I_2,\tilde{I}_2}}
	\overline{\psi^{\S_3}_{\{G\},\{H\}}(\{g_k\},\{g_k'\})}
	\psi^{\S_3}_{\{\widetilde{G}\},\{\widetilde{H}\}}(\{g_k\},\{g_k'\}) \\ \nn
	&\q \q \q \q \q   \times \prod_{k=1}^2\sqrt{d_{\rho_k}d_{\widetilde{\rho}_k}}
	\prod_{k=1}^2 \overline{D^{\rho_k}_{I_k'I_k}([G_k,H_k])}D^{\widetilde{\rho}_k}_{\tilde{I}_k'\tilde{I}_k}([\widetilde{G}_k,\widetilde{H}_k])
	\overline{ \mathcal{C}^{\rho_1\rho_2\rho_{3}}_{I_1I_2I_3}}\cdot 
	{\mathcal{C}^{\widetilde{\rho}_1\widetilde{\rho}_2\widetilde{\rho}_{3}}_{\tilde{I}_1\tilde{I}_2\tilde{I}_3}}\\ \nn
	&\q = \, \frac{1}{|\mathcal{G}|^{2}}
	\sum_{{G_1,H_1,\widetilde{G}_1,\widetilde{H}_1 \atop G_2,H_2,\widetilde{G}_2,\widetilde{H}_2}}
	\sum_{{I_1,\tilde{I}_1 \atop I_2,\tilde{I}_2}}
	\prod_{k=1}^2 \delta (G_k, \wG_k)\delta (H_k,\wH_k) \\ \nn
	&\q \q \q \q \q   \times \prod_{k=1}^2\sqrt{d_{\rho_k}d_{\widetilde{\rho}_k}}
	\prod_{k=1}^2 \overline{D^{\rho_k}_{I_k'I_k}([G_k,H_k])}D^{\widetilde{\rho}_k}_{\tilde{I}_k'\tilde{I}_k}([\widetilde{G}_k,\widetilde{H}_k])
	 \overline{\mathcal{C}^{\rho_1\rho_2\rho_{3}}_{I_1I_2I_3}}\cdot 
	{\mathcal{C}^{\widetilde{\rho}_1\widetilde{\rho}_2\widetilde{\rho}_{3}}_{\tilde{I}_1\tilde{I}_2\tilde{I}_3}}\\ \nn
	&\q = \, \frac{1}{|\mathcal{G}|^{2}}
	\sum_{{G_1,H_1 \atop G_2,H_2}}
	\sum_{{I_1,\tilde{I}_1 \atop I_2,\tilde{I}_2}} \prod_{k=1}^2\sqrt{d_{\rho_k}d_{\widetilde{\rho}_k}}
	\prod_{k=1}^2 \overline{D^{\rho_k}_{I_k'I_k}([G_k,H_k])}D^{\widetilde{\rho}_k}_{\tilde{I}_k'\tilde{I}_k}[{G}_k,{H}_k])
	 \overline{\mathcal{C}^{\rho_1,\rho_2,\rho_{3}}_{I_1,I_2,I_3}}\cdot 
	\mathcal{C}^{\widetilde{\rho}_1\widetilde{\rho}_2\widetilde{\rho}_{3}}_{\tilde{I}_1\tilde{I}_2\tilde{I}_3}\\ \nn
	& \q = \, \sum_{I_1,I_2}  \mathcal{C}^{\rho_1\rho_2\rho_{3}}_{I_1I_2I_3}\cdot \mathcal{C}^{\rho_1\rho_2\widetilde{\rho}_{3}}_{I_1I_2\tilde{I}_3}
	\, \delta_{I_1',\tilde{I}_1'}\delta_{I_2',\tilde{I}_2'}\delta_{\rho_1,\widetilde{\rho}_1}\delta_{\rho_2,\widetilde{\rho}_2}\\
	&\q = \, \delta_{I_3,\tilde{I}_3}\delta_{I_1',\tilde{I}_1'}\delta_{I_2',\tilde{I}_2'}\delta_{\rho_3,\widetilde{\rho}_3}\delta_{\rho_1,\widetilde{\rho}_1}\delta_{\rho_2,\widetilde{\rho}_2}.
\end{align} 
Here, we used the orthogonality of the irreducible representations together with the orthogonality relation (\ref{resolutionCC}) of the Clebsch-Gordan coefficients.
Similarly, the completeness relation for the fusion basis on the 3-punctured sphere follows from the completeness relation of the 2-punctured sphere together with the orthogonality relation (\ref{orthoW3J}). These properties generalize to the fusion basis states on the $p$-punctured sphere. For instance, the orthonormality is given by
\newpage
\begin{align} \nn
	&\Big\langle \psi_{\mathfrak{f}}^{\S_p} \big[ \{\rho_i\}_{i=1}^{2p-3},\{I'_k\}_{k=1}^{p-1},I_p \big] , \psi_{\mathfrak{f}}^{\S_p}\big[\{\tilde{\rho}_i\}_{i=1}^{2p-3},\{\tilde{I}'_k\}_{k=1}^{p-1},\tilde{I}_p \big] \Big\rangle \\ \nn
	&\q = \frac{1}{|\mG|^{5p-6}}.\frac{1}{|\mG|^{2p-2}}\sum_{\{g_i\}_{i=1}^{5p-6}}\sum_{\{I\},\{\tilde{I}\}} \sum_{\{G_i,H_i,\wG_i,\wH_i\}_{i=1}^{p-1}}
	\overline{\psi^{\S_p}_{\{G,H\}}(\{g\})}\psi^{\S_p}_{\{\wG,\wH\}}(\{g\}) \\ \nn
	& \q \q \q \times \overline{\mathcal{C}^{\rho_1\rho_2\rho_{(p+1)}}_{I_1I_2I_{(p+1)}}}\; \overline{\mathcal{C}^{\rho_{(2p-3)}\rho_{(p-1)}\rho_p}_{I_{(2p-3)}I_{(p-1)}I_p}}\;
	 \mathcal{C}^{\tilde{\rho}_1 \tilde{\rho}_2\tilde{\rho}_{(p+1)}}_{\tilde{I}_1\tilde{I}_2\tilde{I}_{(p+1)}}\;
	\mathcal{C}^{\tilde{\rho}_{(2p-3)}\tilde{\rho}_{(p-1)}\tilde{\rho}_p}_{\tilde{I}_{(2p-3)}\tilde{I}_{(p-1)}\tilde{I}_p}\\ \nn
	& \q \q \q \times \prod_{i=1}^{p-1}\sqrt{d_{\rho_i}d_{\tilde{\rho}_i}}\overline{D^{\rho_i}_{I_i'I_i}([G_i,H_i])}D^{\tilde{\rho}_i}_{\tilde{I}_i'\tilde{I}_i}([\wG_i,\wH_i])
	\prod_{i=p+1}^{2p-4}\overline{\mathcal{C}^{\rho_i\rho_{(i-p+2)}\rho_{(i+1)}}_{I_iI_{(i-p+2)}I_{(i+1)}}}
	\mathcal{C}^{\tilde{\rho}_i\tilde{\rho}_{(i-p+2)}\tilde{\rho}_{(i+1)}}_{\tilde{I}_i\tilde{I}_{(i-p+2)}\tilde{I}_{(i+1)}} \\ \nn
	&\q = \frac{1}{|\mG|^{p-1}}\sum_{\{I\},\{\tilde{I}\}}\sum_{\{G_i,H_i\}_{i=1}^{p-1}} \overline{\mathcal{C}^{\rho_1\rho_2\rho_{(p+1)}}_{I_1I_2I_{(p+1)}}}\; \overline{\mathcal{C}^{\rho_{(2p-3)}\rho_{(p-1)}\rho_p}_{I_{(2p-3)}I_{(p-1)}I_p}}\;
	 \mathcal{C}^{\tilde{\rho}_1 \tilde{\rho}_2\tilde{\rho}_{(p+1)}}_{\tilde{I}_1\tilde{I}_2\tilde{I}_{(p+1)}}\;
	\mathcal{C}^{\tilde{\rho}_{(2p-3)}\tilde{\rho}_{(p-1)}\tilde{\rho}_p}_{\tilde{I}_{(2p-3)}\tilde{I}_{(p-1)}\tilde{I}_p}\\ \nn
	& \q \q \q \times \prod_{i=1}^{p-1}\sqrt{d_{\rho_i}d_{\tilde{\rho}_i}}\overline{D^{\rho_i}_{I_i'I_i}([G_i,H_i])}D^{\tilde{\rho}_i}_{\tilde{I}_i'\tilde{I}_i}([G_i,H_i])
	\prod_{i=p+1}^{2p-4}\overline{\mathcal{C}^{\rho_i\rho_{(i-p+2)}\rho_{(i+1)}}_{I_iI_{(i-p+2)}I_{(i+1)}}}
	\mathcal{C}^{\tilde{\rho}_i\tilde{\rho}_{(i-p+2)}\tilde{\rho}_{(i+1)}}_{\tilde{I}_i\tilde{I}_{(i-p+2)}\tilde{I}_{(i+1)}} \\ \nn
	& \q = \sum_{\{I\},\{\tilde{I}\}} \overline{\mathcal{C}^{\rho_1\rho_2\rho_{(p+1)}}_{I_1I_2I_{(p+1)}}}\; 	\overline{\mathcal{C}^{\rho_{(2p-3)}\rho_{(p-1)}\rho_p}_{I_{(2p-3)}I_{(p-1)}I_p}}\;
	 \mathcal{C}^{{\rho}_1 {\rho}_2\tilde{\rho}_{(p+1)}}_{{I}_1{I}_2\tilde{I}_{(p+1)}}\;
	\mathcal{C}^{\tilde{\rho}_{(2p-3)}{\rho}_{(p-1)}\tilde{\rho}_p}_{\tilde{I}_{(2p-3)}{I}_{(p-1)}\tilde{I}_p}\; \prod_{i=1}^{p-1}\delta_{I_i'\tilde{I}_i'}\delta_{\rho_i\tilde{\rho}_i}\\ \nn
	& \q \q \q \times \prod_{i=p+1}^{2p-4}\overline{\mathcal{C}^{\rho_i\rho_{(i-p+2)}\rho_{(i+1)}}_{I_iI_{(i-p+2)}I_{(i+1)}}}
	\mathcal{C}^{\tilde{\rho}_i\tilde{\rho}_{(i-p+2)}\tilde{\rho}_{(i+1)}}_{\tilde{I}_i\tilde{I}_{(i-p+2)}\tilde{I}_{(i+1)}} \\ 
	&\q = \delta_{I_p\tilde{I}_p}\prod_{i=1}^{p-1}\delta_{I_i'\tilde{I}_i'}\prod_{i=1}^{2p-3}\delta_{\rho_i\tilde{\rho_i}}
\end{align}  
where we followed the same steps as for the case of the three-punctured sphere.


\subsection{Gauge invariant projection  of the fusion basis} \label{PfusionA}

Here we consider the Gau\ss~constraint projector ${\mathbb P}_{\rm gauge}$ applied to a fusion basis state on a cylinder:
\ba\label{PfA1}
{\mathbb P}_{\rm gauge}\psi^{\mathbb{S}_2}_{\mathfrak{f}}[CR,i'M',iM]&=&
|{\cal G}|^{-1/2} \sum_{h}\sum_{G,H} \sqrt{d_{CR}}\,\, \delta(H,c_{i'}) \, \delta(c_{i'}, Gc_iG^{-1}) \, D^R_{M'M}(q^{-1}_{i'}Gq_i)\, \nn\\&&
\q\q\q \; \times \delta(G,g_3g_2g_1h^{-1}) \delta(H,g_3g_4g_2^{-1}g_3^{-1}) \nn\\
&=&
|{\cal G}|^{-1/2} \sum_{h}\sum_{G,H} \sqrt{d_{CR}}\,\, \delta(H,c_{i'}) \, \delta(c_{i'}, Gh c_ih^{-1}G^{-1}) \, D^R_{M'M}(q^{-1}_{i'}Ghq_i)\, \nn\\&&
\q\q\q \; \times  \delta(G,g_3g_2g_1h) \delta(H,g_3g_4g_2^{-1}g_3^{-1}) \;.
\ea
One now writes $h$ as $h=q_j n q_i$ and splits the sum over $h \in {\cal G}$ into a sum over $n \in N_C$ and $q_j \in Q_C$. We have thus
\ba
hc_ih^{-1}=c_j \q,\q hq_i=q_j n  \;. 
\ea
Hence the summation variable $n$ only appears in $D^R_{M'M}(q^{-1}_{i'}Ghq_i)=D^R_{M'M}(q^{-1}_{i'}Gq_j n)$ which gives (Note that $q^{-1}_{i'}Gq_j$ is in $N_C$ due to the delta function in (\ref{PfA1}).)
\ba
\frac{1}{|N_C|}\sum_{n\in N_C} D^R_{M'M}(q^{-1}_{i'}Gq_j n) \,=\, \delta_{R,0 } \delta_{M'0} \delta_{M0} 
\,\, \delta(q^{-1}_{i'}Gq_j,e )   \; .
\ea
where we denote the trivial representation of $N_C$ with $R=0$.
We are left with the summation over $q_j$ which leads to the final result
\ba
{\mathbb P}_{\rm gauge}\psi^{\mathbb{S}_2}_{\mathfrak{f}}[CR,i'M',iM]&=&
|{\cal G}|^{1/2} \sum_{G,H} \sqrt{d_{C0}}\,\, \delta(H,c_{i'}) \,    \frac{1}{|Q_C|}\sum_{q_j} \delta(c_{i'}, Gq_j c_1 q_j^{-1}G^{-1}) \, \, \nn\\&&
\q\q\; \times \delta(G,g_3g_2g_1h^{-1}) \delta(H,g_3g_4g_2^{-1}g_3^{-1})\nn\\
&=&  \delta_{R,0 } \delta_{M'0} \delta_{M0}  \frac{1}{|Q_C|}\sum_{j}\psi^{\mathbb{S}_2}_{\mathfrak{f}}[C0,i'0,j0]  \; .
\ea

\section{Generalized fusion basis\label{app_mpuncture}}

Here we discuss an extension of the fusion basis, resulting from a generalization of the gluing procedure of section \ref{subsec_ocneanu}. 
There, using the fact that the gluing of two cylinders is another cylinder, we found that such a gluing procedure defines at the level of the states (defined modulo appropriate equivalence relations, as in section \ref{sec_alBF}) a multiplication mirroring that of the Drinfel'd algebra:
\ba\label{orgalgebra}
([\widetilde{G},\widetilde{H}],[G,H])\;\; &\longmapsto \;\;[\widetilde{G},\widetilde{H}]\star [G,H] = 
	\delta(\widetilde{H},\widetilde{G}H\widetilde{G}^{-1})[\widetilde{G}G,\widetilde{H}]  \; .
\ea

In defining the gluing procedure of cylinders and states, it was necessary to specify a marked point on the boundary of the punctures at which a link of the underlying graph terminates. This prescription can be readily generalized by introducing several, say $m\geq1$, marked points at a given puncture, prescribing now where $m$ different links can terminate. 
We will refer to such a puncture as a $m$--puncture. We can then consider the gluing of manifolds along two $m$-punctures. 

As an example consider a cylinder with two $2$--punctures denoted $a$ and $b$.  Labeling the marked points at $a$ by $a_1,a_2$ and at $b$ by $b_1,b_2$ we can associate four independent holonomies to such a cylinder: 
\ba
G=h(b_1,a_1), \, H=h(b_1,b_2) h(b_2,b_1), \, K'= h(b_2,b_1) \, ,  K= h(a_2,a_1),
\ea
where $h(y,x)$ denotes the holonomy from the marked point $x$ to the marked point $y$. All other holonomies between the marked points can be reconstructed using the flatness condition, {\it e.g.}
\be
	h(a_2,b_2)=     K' GK^{-1}  \;  . 
\ee
Considering a minimal embedded graph for the cylinder with two marked points at each puncture, we can define the following basis states via gauge fixing
\be
\psi^{\S_2}_{G,H;K'K} = \begin{array}{c}\includegraphics[scale =1]{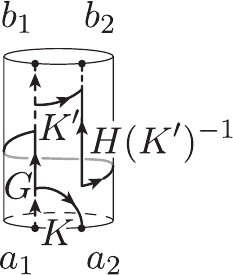}\end{array},
\ee
where the graphical notation is the same as in the main text. 
We can thus label a (basis) element of the algebra describing these cylinder states by $[G,H; K', K]$. 
Compared to the cylinder basis state with simple 1--punctures, these states contain additional information given by the holonomies between the two marked points at each of the punctures. In this sense, these are `refined' punctures.

Consider now gluing two such cylinders:
\begin{align}
	\label{Drinfeldbis}
	\begin{array}{c}\includegraphics[scale =1]{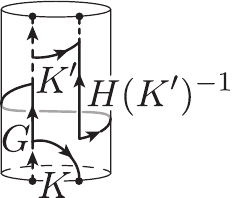}\end{array} \star
	\begin{array}{c}\includegraphics[scale =1]{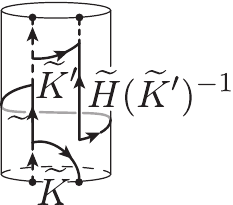}\end{array} &= \mathbb{P}_{\text{flat}} \circ \mathbb{P}_{\text{gauge}} 
	\Bigg( \; \; \begin{array}{c}\includegraphics[scale =1]{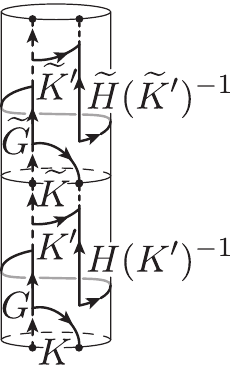}\end{array}  \Bigg) \\ &= 
	\delta(\wH,\wG H \wG^{-1}) \delta(\widetilde{K},K')
	\begin{array}{c}\includegraphics[scale =1]{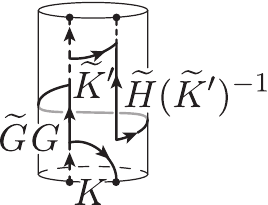}\end{array} 
\end{align}
That is, 
\ba\label{refalgebra}
[\widetilde{G},\widetilde{H}; \widetilde{K'}, \widetilde{K}] \star [G,H; K', K] &=& \delta(\widetilde{K},K' )\delta(\widetilde{H},\widetilde{G}H\widetilde{G}^{-1}) \, \,\, [\widetilde{G}G,\widetilde{H}; \widetilde{K'}, K]  \; .
\ea
Note that this new multiplication rule \eqref{Drinfeldbis} is essentially the same as the original Drinfel'd algebra multiplication \eqref{orgalgebra}: the $[G,H; \cdot,\cdot]$ part multiply precisely as in the Drinfel'd algebra, while the $[\cdot,\cdot, K',K]$ part functions as a factor with trivial multiplication rule
\ba
[\cdot,\cdot; \widetilde{K'}, \widetilde{K}] \star [\cdot,\cdot; K', K] &=&  \delta(\widetilde{K},K' )\,\, [\cdot,\cdot; \widetilde{K'}, K]    \; .
\ea 
In a way, this part already behaves like matrix indices in a representation. It is therefore not difficult to see, that the representations of this new algebra are given by a trivial extension of the Drinfel'd algebra representations  $V^{(C,R)}$ of section (\ref{sec_irreps}). It is indeed enough to extend the representation spaces to $V^{(C,R)}_{\rm ext}$ whose basis $|c_i,M,k\rangle$ is the tensor product of the basis $|c_i,M,\rangle$ of  $V^{(C,R)}$ with the basis $|k\rangle$ of the group algebra $\mathbb{C}[{\cal G}]$, $k\in \mG$.  This leads to the matrix elements
\ba
(D^{C,R}_{\rm ext})_{i'M'k',iMk}([G,H;K',K]) \,=\, \delta( k,K) \delta(k',K')  D^{C,R}_{i'M',iM}([G,H]) \; .
\ea
It is easy to see that the representation property holds, as well as the generalizations of the orthogonality and completeness relations (equations \eqref{orthoMat} and \eqref{completeness}, respectively). Thus, it follows that the basis states for the cylinder with ($m=2$)--punctures carry an additional label $k \in \mG$, which can again be absorbed into an extended multi-index $I$. This construction gives straightforwardly an extended fusion basis, which distinguishes itself from the original one only by its extended index structure associated to the punctures. 

Generalizations to $(m>2)$--punctures is obvious: for each additional marked point at a given puncture one obtains an additional index $k \in \mG$, describing the holonomy between two consecutive marked points. The generalization of the cylinder algebra and its representations is also obvious. In particular, the dimension of the corresponding extended representation is given by $|C|\times \dim(R) \times |{\cal G}|^{m-1}$. 

Therefore, we see that the physical content of the states, which is encoded in the set of charges $\rho=(C,R)$, is not altered at all by a refined puncture structure. This mirrors the Morita equivalence of tube algebras (with different number of open legs) discussed in \cite{Lan}. What changes is only the `amount of information' retained at the gluing interfaces. For a more thorough discussion of this point see the final sections of \cite{Toappear}, where the effect of refined punctures is discussed in relation to entanglement entropy.


\section{Properties of ribbon operators}

\subsection{Gluing of Ribbons} \label{gluingribbons}

Here we consider the gluing of two ribbons ${\cal R}_i[G_i,H_i]$, $i=1,2$ at a puncture. We assume that the state does not carry charges at this puncture.

Applying a ribbon ${\cal R}_1$ ending at a puncture $p$ and then a ribbon ${\cal R}_2$ starting at the puncture $p$ we obtain 
\begin{align}\label{4.8}
	{\cal R}_2[G_2,H_2]  {\cal R}_1[G_1,H_1] \psi 
	 = &\,\begin{array}{c}\includegraphics[scale =1]{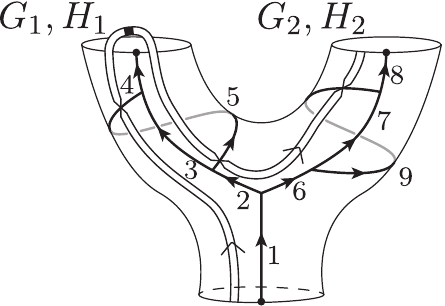}\end{array} \nn \\
	= &\,\delta(G_2, g_8g_7g_6g_2^{-1}g_3^{-1}g_4^{-1})\, \delta(G_1, g_4g_3g_2g_1)  \nn\\
	&\times \psi( \cdots, g_4^{-1} H_1^{-1} g_4 g_5 
	(g_8g_7g_6g_2^{-1})^{-1} H_2 g_8g_7g_6g_2^{-1}, \cdots, g_8^{-1} H_2^{-1} g_8 g_9)\nn\\
	=&\,\delta(G_2 G_1, g_8g_7g_6g_1)\delta(G_1, g_4g_3g_2g_1) \nn\\ 
	&\times \psi( \cdots, g_4^{-1} H_1^{-1} g_4 g_5 g_3^{-1} g_4^{-1}
	G_2^{-1} H_2 G_2g_4g_3, \cdots, g_8^{-1} H_2^{-1} g_8 g_9)\, .
\end{align}
Here ${\cal R}_1$ shifts $g_5$ and ${\cal R}_2$ shifts $g_9$ and $g_5^{-1}$.

A state without any charge at the puncture $p_2$ would have a $g_5$ dependence of the form $\delta(g_3^{-1} g_5 )$. For this factor to be left invariant we need $H_1= G_2^{-1} H_2 G_2$.  Thus the flatness projector ${\mathbb P}_\text{flat}$ at the puncture $p$ leads to the corresponding delta function. 
Also, we can now use that $   g_2^{-1}g_5^{-1}g_3 g_2=e$ in the first delta--function on the (last) RHS of (\ref{4.8}) so that for a state  $\psi=\delta(g_3^{-1} g_5 )\psi'$  we have
\begin{align}
	{\mathbb P}_\text{flat} \,{\cal R}_2[G_2,H_2]  {\cal R}_1[G_1,H_1] (\delta(g_3^{-1} g_5) \psi')
	&=
	\delta(H_1,G_2^{-1} H_2 G_2) \delta(G_2 G_1, g_8g_7g_6 g_2^{-1}g_5^{-1}g_3 g_2  g_1)\, \nn\\
	&\times \delta(G_1, g_4g_3g_2g_1)   \delta(g_3^{-1} g_5) \, \psi'(\cdots, g_8^{-1} H_2^{-1} g_8 g_9) . \; \;
\end{align}
We also apply a group averaging at the puncture
\begin{align}\label{A4}
	&|\mG|\,  {\mathbb P}_\text{gauge}  {\mathbb P}_\text{flat} \,{\cal R}_2[G_2,H_2]  {\cal R}_1[G_1,H_1] (\delta(g_3^{-1} g_5) \psi') \\
	& \q = 
	\sum_h \delta(H_1,G_2^{-1} H_2 G_2) \delta(G_2 G_1, g_8g_7g_6 g_2^{-1}g_5^{-1}g_3 g_2  g_1) \\[-0.7em]
	&\q \q \q \times \delta(G_1, h g_4g_3g_2g_1)   \delta(g_3^{-1} g_5) \, \psi'(\cdots, g_8^{-1} H_2^{-1} g_8 g_9)\nn\\
	& \q =\delta(H_1,G_2^{-1} H_2 G_2) \delta(G_2 G_1, g_8g_7g_6 g_2^{-1}g_5^{-1}g_3 g_2  g_1) \delta(g_3^{-1} g_5) \, \psi'(\cdots, g_8^{-1} H_2^{-1} g_8 g_9),
\end{align}
where we used that $\psi'$ is gauge invariant at the target node of $l_4$, i.e. cannot depend on $g_4$.

Now the RHS of (\ref{A4}) can be written as
\be
|\mG|\,  {\mathbb P}_\text{gauge}  {\mathbb P}_\text{flat} \,{\cal R}_2[G_2,H_2]  {\cal R}_1[G_1,H_1] (\delta(g_3^{-1} g_5) \psi')
= 
\delta(H_1,G_2^{-1} H_2 G_2)  {\cal R}_3[G_2 G_1,H_2]  (\delta(g_3^{-1} g_5) \psi'),
\ee
where the path underlying ${\cal R}_3$ is as follows
\begin{align*}
	\begin{array}{c}\includegraphics[scale =1]{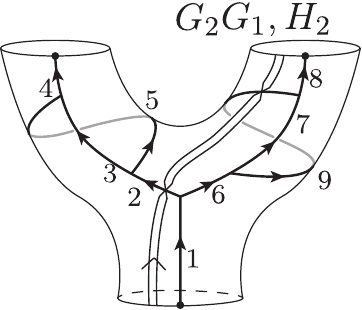}\end{array} \q .
\end{align*}
The ribbon does also cross the link $l_2$ and hence should shift the holonomy $g_2$. However we assumed that $\psi=\delta(g_3^{-1}g_5)\psi'$ does not carry charges at the puncture $p$ and that means that $\psi'$ does not depend on $g_2$ (and also not on $g_4$, as we used earlier). 

We conclude that for a gluing of ribbons at an (auxiliary) puncture which does not carrying any charge, we have
\ba\label{A6}
|\mG|\,  {\mathbb P}_\text{gauge}  {\mathbb P}_\text{flat} \,\, {\cal R}_2[G_2,H_2]  {\cal R}_1[G_1,H_1] 
&=& 
\delta(H_1,G_2^{-1} H_2 G_2)  {\cal R}_3[G_2 G_1,H_2]   .
\ea

\subsection{Gluing of charge ribbon operators}\label{GCR}

Using the previous result, equation \eqref{A6}, we can now repeat the same construction in the case of two charge ribbon operators
\begin{align}
	& |{\cal G}|\,{\mathbb P}_\text{gauge} {\mathbb P}_\text{flat} \, {\cal R}_2[C_2,R_2; i'_2M'_2,i_2M_2]   {\cal R}_2[C_1,R_1; i'_1M'_1,i_1M_1]\nn\\
	&\q =  \frac{d_{R_2}}{|N_{C_2}|} \frac{d_{R_1}}{|N_{C_1}|}
	\sum_{n_1,n_2}\!\!\!
	D^{R_2}_{M'_2M_2}(n_2) D^{R_1}_{M_1'M_1}(n_1) \, \delta( c^{(1)}_{i'_1} ,\,c^{(2)}_{i_2}  )\, 
	 {\cal R}[ q_{i'_2}n_2 q^{-1}_{i_2} q_{i'_1} n_1 q^{-1}_{i_1},c^{(2)}_{i'_2}]
	\nn\\
	&\q =
	\delta_{C_1,C_2}\frac{d_{R_2}}{|N_{C_2}|} \frac{d_{R_1}}{|N_{C_1}|} \sum_{n_1,n_2}\!\!
	D^{R_2}_{M'_2M_2}(n_2) D^{R_1}_{M_1'M_1}(n_1) \,
	 {\cal R}[ q_{i'_2}n_2  n_1 q^{-1}_{i_1},\, c^{(2)}_{i'_2}] \nn\\
	 &\q =
	\delta_{C_1,C_2} \frac{d_{R_2}}{|N_{C_2}|} \frac{d_{R_1}}{|N_{C_1}|}
	\sum_{n_1,n,M_2''}
	D^{R_2}_{M'_2M_2''}(n)  \overline{D^{R_2}_{M_2M''_2}(n_1)} D^{R_1}_{M_1'M_1}(n_1) \,
 	{\cal R}[ q_{i'_2}n q^{-1}_{i_1},\, c^{(2)}_{i'_2}]  \nn\\
 	&\q =\delta_{C_1,C_2}  \delta_{R_2,R_1}   \delta_{M_2 M_1'} \frac{d_{R_2}}{|N_{C_2}|}
    	\sum_{n}  D^{R_2}_{M'_2M_1}(n)  \,\,
 	{\cal R}[ q_{i'_2}n q^{-1}_{i_1},\, c^{(2)}_{i'_2}] \nn\\
	 &\q= \delta_{C_1,C_2} \delta_{R_2,R_1} \delta_{i_2 i'_1}  \delta_{M_2 M_1'}  \,   {\cal R}[C_2,R_2; i'_2M'_2,i_1M_1].
\end{align}
The delta function in the second line enforces $i_2=i'_1$, which we used in the third line. Furthermore, it requires that $C_2=C_1$. We then performed a variable transformation and  used orthogonality of the representation matrix elements.

\subsection{Lateral product of closed ribbons} \label{appOP2}

Here we consider the lateral product of two charge closed ribbons as defined in (\ref{defcl2}):
\ba
	{\cal K}[C , R]  {\cal K}[ C', R'] %
	& =& \frac{ d_R}{|N_C|}  \frac{d_{R'}}{|N_{C'}|}  \sum_{D,D'} {\chi^R(D)}  \, {\chi^{R'}(D')}
	\, \mathcal{K}[C,D] \mathcal{K}[C',D'] \\
	&  \stackrel{\eqref{opprodK}}{=}&\delta_{C,C'}  \frac{ d_Rd_{R'}}{|N_C|^2} \sum_{D,D'}\sum_{q \in Q_C}\sum_{d \in D \atop d' \in D'} {\chi^R(D)}  \, {\chi^{R'}(D')}
	\, \mathcal{R}[qc_1q^{-1},qdd'q^{-1}]\\
	& = & \delta_{C , C'}\frac{ d_R d_{R'} }{|N_C|^2}    \sum_{q\in Q_C} \sum_{d, d'\in N_C} {\chi^R(d)}  {\chi^{R'}(d')} \; {\cal R}[q c {q}^{-1} ,  q d d' {q}^{-1} ] \notag\\
	& = & \delta_{C , C'}\frac{ d_R d_{R'} }{|N_C|^2}   \sum_{q\in Q_C}  \sum_{d, d''\in N_C}  {\chi^R(d)}  \; {\chi^{R'} (d^{-1} d'')} \; {\cal R}[q c {q}^{-1} ,  q d'' {q}^{-1} ] \notag\\
	& = &\delta_{C , C'} \delta_{R, R'}\frac{ d_R  }{|N_C|}   \sum_{q\in Q_C}  \sum_{d''\in N_C} {\chi^R(d'')}  \; {\cal R}[q c {q}^{-1} ,  q d'' {q}^{-1} ] \notag\\
	& = &\delta_{C , C'} \delta_{R, R'} {\cal K}[C , R].
\ea
In the first step we used the result (\ref{opprodK}). We then rearranged the sums according to $\sum_D \sum_{d\in D} = \sum_{d\in N_C}$, together with the fact that $\chi^R(D) = \chi^R(d)$ for all $d\in D$. Finally, we performed a change of variables and used the orthogonality of the irreducible representations for the stabilizers. 

Thus, we conclude that the closed ribbons ${\cal K}[C,R]$ define a family of orthogonal projectors under the above lateral product.

\section{Action of closed ribbons on cylinder states}\label{app:clrc}

Here we consider the action of a closed ribbon on a fusion basis state on the cylinder. The minimal graph we use as well as the specific ribbon are indicated in figure \ref{fig:clrc}.

Using the expression (\ref{basiscylR}) for a fusion basis state on the cylinder and the definitions (\ref{defK}, \ref{defcl2}) for the closed ribbon we obtain
\begin{align}\label{clr01a}
	&{\cal K}[C',R'] \,\psi^{\mathbb{S}_2}_{\mathfrak f}[CR;i'M',iM] \nn\\
	&\q = |\mG|^{1/2}\sqrt{d_{R,C}} \frac{ d_{R'}} {|N_{C'}|} \sum_{n \in N_C}  \sum_{q \in Q_{C'}}\sum_{n'\in N_{C'}} \chi^{R'}(n') D^R_{M'M}(n) 
	 \delta(q c'_1 q^{-1},\, g_2^{-1} g_4) \, \delta( c_{i'},\, g_3g_4 g_2^{-1}g_3^{-1})  \nn\\
	& \q\q\q \q \q \q \q \q \times \delta( q_{i'} n q_{i}^{-1}, \, g_3 g_2 g_4^{-1} g_2 q (n')^{-1} q^{-1} g_2^{-1} g_4 g_1) .
\end{align}
To lighten the formulas we will evaluate the resulting wave function in a gauge-fixed form, $g_3=g_2=e$. Hence, 
\begin{align}\label{clr01}
	&\left({\cal K}[C',R'] \,\psi^{\mathbb{S}_2}_{\mathfrak f}[CR;i'M',iM] \right)_{|{\rm g.f.}} \nn\\
	&\q = |\mG|^{1/2}\sqrt{d_{R,C}} \frac{ d_{R'}} {|N_{C'}|} \sum_{n \in N_C}  \sum_{q \in Q_{C'}}\sum_{n'\in N_{C'}}\chi^{R'}(n') D^R_{M'M}(n) 
	\delta(q c'_1 q^{-1},\, g_4) \, \delta( c_{i'},\,g_4 ) \nn\\
	& \q\q\q \q \q \q \q \q \times \delta( q_{i'} n q_{i}^{-1}, \,  g_4^{-1}  q (n')^{-1} q^{-1}  g_4 g_1)\,. 
\end{align}
The first two delta functions on the RHS of \eqref{clr01}, enforce $C=C'$, and also allow us to determine $q$ from the condition
\ba
q c_1'q^{-1}\,=\, q c_1q^{-1}\,=\,  q_{i'} c_1 q_{i'}^{-1} ,
\ea
 which in turn follows from  the definition $c_{i'}=q_{i'} c_1 q_{i'}^{-1}$. Indeed, we find $q=q_{i'}$ (as now both $q,q_{i'}\in Q_C$).

Next, we turn to the last delta function in \eqref{clr01}. Using $g_4=qc_1'q^{-1}=qc_1q^{-1}$, one can show that $g_4$ and $q(n')^{-1}q^{-1}$ do commute (since $n' \in N_C$). Therefore, the condition enforced by the last delta function simplifies to
\ba
q_{i'}nq_i^{-1}\,=\, q (n')^{-1} q^{-1} g_1.
\ea
This can in turn be solved for $n'$:
\ba\label{clr03}
n'&=& q^{-1} g_1 q_i n^{-1} q_{i'}^{-1}  q 
\,\,=\,  q_{i'}^{-1}  g_1 q_i n^{-1}   .
\ea
Now, apart from determining $n \in N_C$,  equation \eqref{clr03} also requires that 
\ba
q_{i'}^{-1}  g_1 q_i \,\in \, N_C  .
\ea
We encode this into a characteristic function
\ba
\theta_{N_C}( g)\,=\, 1 \q \text{if} \, g\in N_C\, ,  \text{ and vanishing otherwise.}
\ea
Now, \eqref{clr01} can be written as
\begin{align}
	&\left({\cal K}[C',R']\, \,\psi^{\mathbb{S}_2}_{\mathfrak f}[CR;i'M',iM] \right)_{|{\rm gf}}\\
	&\q  = \delta_{C,C'} |\mG|^{1/2}\sqrt{d_{R,C}} \frac{ d_{R'}} {|N_{C'}|} \sum_{n \in N_C}  \chi_{R'}(  q_{i'}^{-1}  g_1 q_i n^{-1}    ) D^R_{M'M}(n)	
	\delta( c_{i'},\, g_4 ) \, \theta_{N_C} ( q_{i'}^{-1}  g_1 q_i )\nn\\
	&\q =
 \delta_{C,C'} \delta_{R,R'} |\mG|^{1/2}\sqrt{d_{R,C}}  \,\,D^R_{M'M}( q_{i'}^{-1}  g_1 q_i  ) \, \theta_{N_C} ( q_{i'}^{-1}  g_1 q_i )\, \delta( c_{i'},\, g_4 )\nn\\
 &\q =  \delta_{C,C'} \delta_{R,R'} \,\,\psi^{\mathbb{S}_2}_{\mathfrak f}[CR;i'M',iM]_{|{\rm gf}}  .
\end{align}
 Thus the closed ribbon operator ${\cal K}[C,R]$ projects onto states $\psi[CR;i'M',iM]$. In particular, the projective cylinder states are eigenstates for the closed ribbons ${\cal K}$.

\section{The S--matrix}\label{SmatrixR}

The S--matrix can be defined as \cite{Pasquier,Verlinde}
\ba\label{SmatrixV2}
{\bf S}_{CR,C'R'} \,=\, \frac{1}{|\cal G|}  \sum_{h_i  \in C, h'_j \in C'}  \delta( h_i h'_j, \, h'_j h_i) \, \, \overline{\chi^R}( q_i^{-1}h'_jq_i) \,\, \overline{ \chi^{R'}} ( (q'_j)^{-1} h_i q'_j)  .
\ea
with $h_i:= q_i c_1 q_i^{-1}$ and $h'_j:= q_j c'_1 q_j^{-1}$ where $c_1 \in C, \, c'_1 \in C'$ and $q_i \in Q_C, \, q'_j \in Q_{C'}$. 

As $h_j'$ commutes with $h_i$, it has to be of the form
\ba
q_j'c'_1 (q'_j)^{-1}\,=\,h_j'=  \, q_i  n q_i^{-1} \, \q \text{with} \q n \in N_C    .
\ea
Here, $n$ is given by 
\ba
n \,=\, q_i^{-1} q_j c'_1 (q'_j)^{-1} q_i \,=\, {}_nq_{k} \, c'_1\, {}_nq_{k}^{-1} \q .
\ea
The second equation comes from \eqref{3.39}  and defines ${}_nq_k$. Note that we use $c_1'={}_nc_1$. Thus $(q'_j)^{-1} h_i q'_j=  {}_nq_{k}^{-1} c_1 \, {}_nq_{k} \in D_{n,c_1}$. This shows that the summation over $h_i \in C$  is superfluous, and we can write
\ba\label{3.45}
{\bf S}_{CR,C'R'} &=& \frac{1}{|\cal G|}  \sum_{h_i  \in C}  \sum_{n \in N_C} \delta_{C',C_n}   \, \, \overline{\chi^R}( n) \,\, \overline{ \chi^{R'}} ( D_{n,c_1}) \nn\\
 &=& \frac{1}{|N_C|}  \,  \sum_{n \in N_C} \delta_{C',C_n}   \, \, \overline{\chi^R}( n) \,\, \overline{\chi^{R'}}(D_{n,c_1}) .
\ea

\section{Constructing the fusion basis via charge ribbon operators}\label{RibbonsTOFusion}

\begin{figure}[h]
	\includegraphics[scale =1]{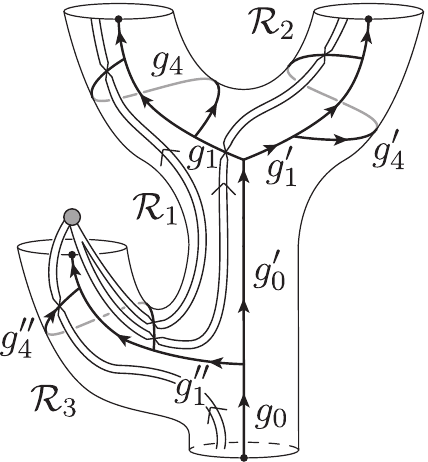} 
	\caption{Construction of the fusion basis states on the three-punctured sphere using charge ribbon operators. An auxiliary puncture is introduced at which the three ribbons are fused via a Clebsch-Gordan coefficient. \label{F_fig3S}}
\end{figure}

Here we construct the fusion basis on the three--punctured sphere by applying three charge ribbon operators, ending at an auxiliary puncture, see figure \ref{F_fig3S}. 

We start with a vacuum on the three punctured sphere, which in gauge fixed form is given by
\ba
{\psi_0^{\mathbb{S}_3}}_{|\rm g.f.}\,=\,  \delta(g_4,e) \delta(g_4',e)   \; .
\ea
We now introduce an auxiliary puncture, using the embedding map detailed in (\ref{embeda}):
\ba
{\psi_0^{3\rightarrow4}}_{|\rm g.f.}:={(\iota_{pp'}(\psi_0^{\mathbb{S}_3}))}_{|\rm g.f.}\,=\,  |\mG|^{1/2} \delta(g_4,e) \delta(g_4',e) \delta(g_4'',e) \; .
\ea
This allows us to apply three ribbon operators, as indicated in figure \ref{F_fig3S}:
\begin{align}\label{AF2}
	&\left( {\cal R}_1[G_1,H_1]{\cal R}_2[G_2,H_2]{\cal R}_3[G_3,H_3] {\psi_0^{3\rightarrow4}}\right)_{|\rm g.f.} \nn\\
	&\q = |\mG|^{1/2}   \delta(g_1''g_0 ,G_3) \delta(   g''_4  G_1^{-1} H_1 G_1 G_2^{-1} H_2 G_2  , H_3) \, \, \delta(g'_1 (g_1'')^{-1},G_2 ) \delta(g'_4,H_2) \,\, \nn\\
	&\q \q \times \;\delta(g_1 (g_1'')^{-1},G_1 ) \delta(  g_4,  H_1) \; .
\end{align}

To glue the ribbons together we apply the flatness projector  $\mathbb{P}_{\rm flat}$ and the gauge averaging $|\mG| \mathbb{P}_{\rm gauge}$ at the auxiliary puncture.  The flatness projector leads to an additional delta function $\delta(g_4'',e)$. We use its solution in the second delta function factor on the RHS in (\ref{AF2}). The gauge averaging leads in this gauge fixed setting to a summation over $g_1''$, that is we can solve e.g. the first  delta function in (\ref{AF2}) for $g''_1$. This results in
\begin{align}
&|\mG| \mathbb{P}_{\rm gauge} \mathbb{P}_{\rm flat}\left( {\cal R}_1[G_1,H_1]{\cal R}_2[G_2,H_2]{\cal R}_3[G_3,H_3] {\psi_0^{3\rightarrow4}}\right)_{|\rm g.f.} \nn\\
&\q = |\mG|^{1/2} \delta(    G_1^{-1} H_1 G_1 G_2^{-1} H_2 G_2  , H_3) \, \delta(g_4'',e)
\, \, \delta(g'_1 g_0,G_2 G_3) \delta(g'_4,H_2) \,\, \nn\\
&\q \q \times \;\delta(g_1 g_0,G_1 G_3 ) \delta(  g_4,  H_1)   \; .
\end{align}
We have now projected away any charge contend at the auxiliary puncture, and can therefore use the equivalence relations in section \ref{sec_alBF}, to express the state on a minimal graph for the 3--punctured sphere. (The  gauge fixing is the same as in (\ref{3pstate1})). We write this as
\begin{align}
&\left(\left(   |\mG| \mathbb{P}_{\rm gauge} \mathbb{P}_{\rm flat} ({\cal R}_1[G_1,H_1]{\cal R}_2[G_2,H_2]{\cal R}_3[G_3,H_3] )\right) \psi^{{\mathbb S}_3}_0\right)_{|\rm g.f.} \nn\\
&\q = \delta(    G_1^{-1} H_1 G_1 G_2^{-1} H_2 G_2  , H_3)\, \, \delta(g'_1 ,G_2 G_3) \delta(g'_4,H_2) \delta(g_1,G_1 G_3 ) \delta(  g_4,  H_1)  \; .
\end{align}

We now transform the three ribbons to charge ribbons and contract the appropriate $I$--indices with a Clebsch-Gordan coefficient. This gives
\begin{align}
	&\Bigg( \bigg(   |\mG| \mathbb{P}_{\rm gauge} \mathbb{P}_{\rm flat} \sum_{I_1,I_2,I'_3}  {\cal C}^{\rho_1\rho_2\rho_3}_{I_1I_2I_3'}  {\cal R}_1[\rho_1,I_1',I_1]{\cal R}_2[\rho_2,I'_2,I_2]{\cal R}_3[\rho_3,I_3',I_3] \bigg) \psi^{{\mathbb S}_3}_0\Bigg)_{|\rm g.f.} \nn\\
	&\q =\sum_{G_1,G_2,G_3,\atop H_1,H_2,H_3}  \sum_{I_1,I_2,I'_3}
	\left( \prod_{\alpha=1}^3 \frac{d_{\rho_\alpha}}{|\mG|} \,D^{\rho_\alpha}_{I'_\alpha I_\alpha}([G_\alpha,H_\alpha]) \right) {\cal C}^{\rho_1\rho_2\rho_3}_{I_1I_2I_3'} \nn\\
	&\q \q \q \times \delta(    G_1^{-1} H_1 G_1 G_2^{-1} H_2 G_2  , H_3)\, \, \delta(g'_1 g_0,G_2 G_3) \delta(g'_4,H_2)\delta(g_1 g_0,G_1 G_3 ) \delta(  g_4,  H_1)  \nn\\
	& \q =
	\sum_{G_1,G_2,G_3,\atop H_1,H_2}  \sum_{I_1,I_2,I'_3} \frac{ d_{\rho_1} d_{\rho_2} d_{\rho_3}}{ |\mG|^3} 	
	D^{\rho_1}_{I'_1 I_1}([G_1 G_3^{-1},H_1])\, D^{\rho_2}_{I'_2 I_2}([G_2 G_3^{-1},H_2]) \, {\cal C}^{\rho_1\rho_2\rho_3}_{I_1I_2I_3'} \nn\\
	& \q \q \q \times D^{\rho_3}_{I'_3 I_3}([G_3 , G_3G_1^{-1} H_1 G_1 G_2^{-1} H_2 G_2G_3^{-1} ])
 	\, \delta(g'_1 g_0,G_2 ) \delta(g'_4,H_2)\delta(g_1 g_0,G_1  ) \delta(  g_4,  H_1)  \, .
\end{align}
For the second equation we solved the delta function for $H_3$, and furthermore translated the summation variables $G_1 \rightarrow G_1 G_3^{-1}$ and $G_2 \rightarrow G_2 G_3^{-1}$.  We proceed by using equation (\ref{CGconstraint2})
\begin{align}\label{CGconstraint2a}
&\sum_{I_1,I_2,I'_3} 
D^{\rho_1}_{I_1'I_1}([G_1G_3^{-1},H_1])   D^{\rho_2}_{I_2'I_2}([G_2G_3^{-1},H_2]) 
\,\mathcal{C}^{\rho_1 \rho_2\rho_3}_{I_1 I_2 I'_3}  \, 
D^{\rho_3}_{I'_3 I_3}([G_3 ,  G_3 G_1^{-1} H_1 G_1 G_2^{-1} H_2 G_2 G_3^{-1}])  \nn\\
&\q =
\sum_{I_1,I_2} 
D^{\rho_1}_{I_1'I_1}([G_1,H_1])   D^{\rho_2}_{I_2'I_2}([G_2,H_2]) 
\,\mathcal{C}^{\rho_1 \rho_2\rho_3}_{I_1 I_2 I_3}   \; ,
\end{align}
and performing the (now trivial) summation over $G_3$:
\begin{align}
&\Bigg( \bigg(   |\mG| \mathbb{P}_{\rm gauge} \mathbb{P}_{\rm flat} \sum_{I_1,I_2,I'_3}  {\cal C}^{\rho_1\rho_2\rho_3}_{I_1I_2I_3'}  {\cal R}_1[\rho_1,I_1',I_1]{\cal R}_2[\rho_2,I'_2,I_2]{\cal R}_3[\rho_3,I_3',I_3] \bigg) \psi^{{\mathbb S}_3}_0\Bigg)_{|\rm g.f.} \nn\\
& \q =
\sum_{G_1,G_2\atop H_1,H_2}  \sum_{I_1,I_2,I'_3} \frac{ d_{\rho_1} d_{\rho_2} d_{\rho_3}}{ |\mG|^2} 
D^{\rho_1}_{I'_1 I_1}([G_1,H_1])\, D^{\rho_2}_{I'_2 I_2}([G_2 ,H_2]) \, {\cal C}^{\rho_1\rho_2\rho_3}_{I_1I_2I_3} \nn\\
&\q\q\q \times   \delta(g'_1 g_0,G_2 ) \delta(g'_4,H_2)\delta(g_1 g_0,G_1  ) \delta(  g_4,  H_1) \; .\q\q
\end{align}
The right hand side of this equation can be compared with the definition of the fusion basis state in (\ref{3pstate1}) and (\ref{deffus48}). (Note that $\psi^{\S_3}_{G_1,H_1;G_2,H_2} = |\mG|^3  \delta(g'_1 g_0,G_2 ) \delta(g'_4,H_2)\delta(g_1 g_0,G_1  ) \delta(  g_4,  H_1)$.)

We finally obtain
\begin{align}
&\bigg(    \sum_{I_1,I_2,I'_3}  \Big({\cal R}_1[\rho_1,I_1',I_1]{\cal R}_2[\rho_2,I'_2,I_2]  {\cal C}^{\rho_1\rho_2\rho_3}_{I_1I_2I_3'} \Big) \star{\cal R}_3[\rho_3,I_3',I_3]    )  \bigg) \psi^{{\mathbb S}_3}_0
\nn\\
&\q :=\bigg(   |\mG| \mathbb{P}_{\rm gauge} \mathbb{P}_{\rm flat} \sum_{I_1,I_2,I'_3}  {\cal C}^{\rho_1\rho_2\rho_3}_{I_1I_2I_3'}  {\cal R}_1[\rho_1,I_1',I_1]{\cal R}_2[\rho_2,I'_2,I_2]{\cal R}_3[\rho_3,I_3',I_3] \bigg) \psi^{{\mathbb S}_3}_0 \nn\\
&\q =    \frac{ \sqrt{d_{\rho_1}d_{\rho_2}} d_{\rho_3}}  {|\mG|^3} \,\,\,
 \psi^{\S_3}_{\frak f}
	 { \scriptsize\begin{bmatrix} \rho_1,I_1'\\ \rho_2,I_2'\\ \rho_3,I_3\end{bmatrix}}  \; .
\end{align}


\newpage

\end{document}